\documentclass[sn-mathphys-num]{sn-jnl}

\usepackage{graphicx}%
\usepackage{multirow}%
\usepackage{amsmath,amssymb,amsfonts}%
\usepackage{amsthm}%
\usepackage{mathrsfs}%
\usepackage[title]{appendix}%
\usepackage{xcolor}%
\usepackage{textcomp}%
\usepackage{manyfoot}%
\usepackage{booktabs}%
\usepackage{algorithm}%
\usepackage{algorithmicx}%
\usepackage{algpseudocode}%
\usepackage{listings}%
\usepackage{lmodern}

\def\e{\epsilon}

\begin{document}

\title[Article Title]{On Modeling Tear Breakup Dynamics with a Nematic Lipid Layer}

\author[1]{\fnm{M.J.} \sur{Taranchuk}}\email{mjgocken@udel.edu}

\author*[1]{\fnm{R.J.} \sur{Braun}}\email{rjbraun@udel.edu}

\affil[1]{\orgdiv{Department of Mathematical Sciences}, \orgname{University of Delaware}, \orgaddress{\street{210 South College Avenue}, \city{Newark}, \postcode{19716}, \state{DE}, \country{USA}}}


\abstract{
One of the main roles of the lipid layer (LL) of the tear film (TF) is to help prevent evaporation of the aqueous layer (AL). The LL thickness, composition, and structure all  contribute to its barrier function. It is believed that the lipid layer is primarily nonpolar with a layer of polar lipids at the LL/AL interface. There is evidence that the nonpolar region of the LL may have liquid crystalline characteristics. We investigate the structure and function of the LL via a model of the tear film with two layers, using extensional flow of a nematic liquid crystal for the LL and shear-dominated flow of a Newtonian AL. Evaporation is taken into account, and is affected by the LL thickness, internal arrangement of its rod-like molecules, and external conditions. We conduct a detailed parameter study with a focus on the evaporative resistance parameter, the Marangoni number, and primary liquid crystal parameters including the Leslie viscosities and director angle. This new model responds similarly to previous Newtonian models in some respects; however, incorporating internal structure via the orientation of the liquid crystal molecules affects both evaporation and flow. As a result, we see new effects on TF dynamics and breakup.}

\keywords{Tear film, Nematic liquid crystal, Lipid layer, Marangoni effect}



\maketitle

\section{Introduction}\label{sec1}

With each eye blink, a thin liquid tear film is established that helps protect and nourish the ocular surface, and helps provide a good smooth surface for clear sight.  The tear film (TF) is a multi-layer film, and each layer plays a role in TF dynamics \cite{BronTiffRev04}.  Despite the importance of the TF, there is still much that is not understood.  This paper develops a new model for how the TF may fail, which is often called tear breakup (TBU). 
The purpose is to better understand the function and structure of the outermost layer of the TF, called the lipid layer (LL). A prime motivation for understanding TF dynamics and TBU is that these phenomena are thought to contribute to the initiation and progression of dry eye disease (DED), which affects millions of people \cite{DEWSepi, stapletonDEWSIIepi2017}.  For comprehensive views of clinical aspects of DED and the tear film see Refs.\ \cite{lempReportNational1995,DEWSdef,craig2017tfos,willcox2017tfos,BaudouinEtal13}.  Reviews related to the tear film include Refs.\ \cite{tsubota1998tearfilm,BronTiffRev04,braun2015dynamics}. In this introduction, we now describe the overall structure of the LL.  We then introduce previous work on properties, proceed to mathematical models of the tear film and its dynamics, and finally discuss details of the LL and our goals for this paper.

\subsection{Overall structure of the tear film}
At the anterior interface with air is an oily LL that retards evaporation \cite{MishimaMaurice61,svitova2021barrier}, and helps to retain a smooth, well-functioning tear film \cite{Norn79}. 
The LL also contains polar lipids that are surface active and lower the surface tension between the LL and the aqueous layer (AL), which is posterior to the LL \cite{berger74,craster2009}.  The AL consists mostly of water \cite{Holly73}, but it also contains various ions and proteins \cite{MishimaMaurice61}.
At the corneal surface, which is posterior to the AL, there is a region with transmembrane mucins protruding from the epithelial cells of the cornea or conjunctiva.  This forest of glycosolated mucins is called the glycocalyx.
It is generally agreed that the presence of the healthy hydrophilic glycocalyx prevents the tear film from dewetting the ocular surface \cite{tiffany1990,tiffany1990b,Gipson04}. 

The overall thickness of the TF is a few microns \cite{King-SmithFink04}, while the average thickness of the LL is on the order of tens to 100 nanometers when the eye is open \cite{Norn79,Yokoi96,GotoTseng03,KSOculSurfRev11,braun2015dynamics}; the thickness of the glycocalyx is a few tenths of a micron \cite{Gipson04,Gipson10}.
The overall structure is reformed on the order of a second after each blink in a
healthy tear film \cite{braun2015dynamics,berger74,OwPh01,KNNFB08}.

TBU occurs when a thinned region forms in the TF.  
Clinically, this is defined as the first dark area that is observed in the fluorescent TF following instillation of fluorescein dye \cite{norn1969-1}.
Various mechanisms are thought to cause different types of TBU, or to combine in various ways: evaporation of water from the AL \cite{lemp2007,willcox2017tfos,king2018}, Marangoni-driven tangential flow  (from non-uniform polar lipid surface concentrations) \cite{ZhongBMB19}, or dewetting-driven tangential flow (from uneven ocular surface conditions) \cite{ZhangMatar03,ZhangMatar04,yokoi2019TForiented}. Here, we use the phrase tangential flow to mean flow along the eye.  
The evaporation of water from the AL causes relatively slow thinning \cite{king2010application}, while one may expect rapid thinning to be explained by Marangoni-driven tangential flow \cite{ZhongBMB19} or dewetting-driven tangential flow. 
Concentration gradients from surface active polar lipids in the LL induce shear stress at the LL/AL interface and drive outward tangential flow via the Marangoni effect \cite{berger74,craster2009}.
Dewetting from irregularities in a corneal surface region has been hypothesized to generate outward tangential flow from internal pressure gradients inside the AL due to van der Waals type forces \cite{SharmRuck85,ShaKhaRei99,ZhangMatar03}.
A related term is full-thickness tear film breakup (full thickness-TBU), which is when the AL has thinned to the point where the LL and glycocalyx touch \cite{king2018,begleyQuantitativeAnalysis2013}.  It is our belief that the strongest evidence to date regarding mechanisms for TBU are evaporative and Marangoni-driven thinning and their interaction \cite{King-SmithOS13,lukeParameterEstimationMixedMechanism2021a}; unambiguous evidence of van der Waals-driven dewetting is not yet in hand in our opinion \cite{king2018}.  We now discuss studies of these mechanisms in more detail.

The aqueous part of tear fluid is primarily supplied from the lacrimal gland near the temporal canthus (corner) and the excess is drained through the puncta near the nasal canthus \cite{Doane81}.
Water lost from the tear film due to evaporation into air is an important process  \cite{MishimaMaurice61,TomDoaneMcF09,kimball2010}.  We believe that this is the primary
mechanism by which the osmolarity (essentially, the total concentration of ions on a molar basis) is
increased in the tear film \cite{braun2015dynamics}.
Some water is supplied from the ocular epithelia via osmosis when the osmolarity increases above the isotonic
level (the so-called hyperosmolar state)
\cite{Braun12,CerretaniEtal14,braun2015dynamics,TomDoaneMcF09,stahl2012}.
According to the Dry Eye WorkShop (DEWS) reports, the osmolarity is important because it plays an
important etiological role in the development of dry eye disease \cite{lemp2007}.
An accepted view in the DEWS reports is that repeated exposure to sufficient hyperosmolarity in tears, as may occur in TBU, can result in inflammation and ocular surface damage in a vicious cycle that may lead to DED.  
There is still considerable ongoing research to more fully understand the causes and etiology of DED; some mathematical modeling results discussed in this paper help to quantify the relationship between tear breakup and hyperosmolarity.

\subsection{Measurements and properties}
Both TF instability and hyperosmolarity are important to study because they are proposed as etiological causes of dry eye disease \cite{craig2017tfos, willcox2017tfos}. 
An osmolarity difference between the corneal epithelium and AL induces osmotic flow from the cornea to the TF \cite{peng2014evaporation,braun2015dynamics}.
TF osmolarity may be measured in the inferior meniscus clinically \cite{lemp2011}; the healthy range is 296-302 mOsM  \cite{lemp2011,tomlinson2006,versura2010}. Dry eye measurements in the same location can reach 316-360 mOsM \cite{GilbardFarris78,tomlinson2006,sullivan2010,dartt2013} but estimates for the TF over the cornea reach 900 mOsM or higher \cite{liu2009link,braun2015dynamics,peng2014evaporation,lukeParameterEstimation2020}. High levels of TF osmolarity are associated with pain, inflammation and cellular changes \cite{pflugfelder2011,belmonte2015,liu2009link}. In support of these potentially high levels of TF osmolarity over the cornea, mathematical models without spatial variation have estimated peak osmolarities up to ten times the isotonic concentration \cite{Braun12,braun2015dynamics}. The modeling work of \citet{peng2014evaporation} found that evaporation elevates osmolarity in breakup regions.

TF thinning rates have been measured experimentally or estimated in many studies. A few experimental methods include spectral interferometry \cite{nichols2005,kimball2010,king2010application}, an open chamber \cite{hamano1981}, an evaporimeter \cite{peng2014b}, and averaging pure water and functioning LL rates over the cornea obtained by heat transfer analysis and thermal imaging \cite{DurschFLandThermal2017}.
Parameter estimation schemes were developed in \citet{lukeParameterEstimation2020,lukeParameterEstimationMixedMechanism2021a} for fitting partial differential equation (PDE) models to experimental fluorescent intensity distributions \cite{BraunDrisTBU17}.  They found evaporative thinning rates ranging from 36.9 to $-4.91$ $\mu$m/min (negative indicates thickening) and overall TF thinning rates ranging from 23.5 to 1.85 $\mu$m/min. These estimates were quite similar to previous experimental spot-thinning rates that did not specifically seek TBU instances \cite{nichols2005}.

\citet{TiffDart81} measured values of the viscosity of human meibum (the material expressed from the meibomian glands inside the lids) using a capillary tube
and they estimated values from 9.7 to 19.5 Pa s at 308 K.  This is a representative value
for the temperature on the surface of the tear film \citep{PurWol05}.  
In a series of papers by Tiffany and coworkers, the human tear film has been shown to have non-Newtonian properties \cite{Tiffany94,PanNagBT99,tiffany1991viscosity}.
\citet{tiffany1991viscosity} found that whole tears, that is, tear sampled directly from the eye,
are shear thinning.  The shear thinning has been fit with empirical models, namely Cross \cite{tiffany1991viscosity} and Ellis models \cite{JossicLefevre09}.
\citet{Tiffany94} also found weak elastic effects; \citet{TiffNag02} found that removing all lipids from tears caused tear fluid to become Newtonian.
The rheology of the LL depends on temperature
\citep{TiffRev87}; \textsl{in vitro} observations showed temperature dependence of meibum on saline for monolayers  \cite{Leiske11,Leiske12} and for bulk samples \cite{rosenfeld2013structural}.
The elastic contributions to viscous response was small above 305 K for these measurements \cite{Leiske11}.
\citet{KSOculSurfRev11} used high-resolution microscopy to observe LL dynamics, and they saw fine structure of the LL during the interblink period that included islands of lipids.

The surface tension of the tear/air interface was measured to be 46.2 mN/m by \citet{Miller69} \textsl{in vivo} using a special contact lens; this is about 2/3 that of an air/pure-water interface.  
\citet{PanNagBT99} measured the surface tension of whole tears with a capillary tube and obtained $43$ mN/m for unstimulated tears and $46$ mN/m for stimulated tears; \citet{Miller69} likely measured stimulated tears.  A value of $45$ mN/m is often used in theoretical studies.  The components responsible for lowering the surface tension from that of pure water has been the subject of a number of papers \citep{NagyTiff99,TiffNag02,MudTorMil06,MudMil08}. 
A challenge for this measurement is the dependence of surface tension on surface active components like phospholipids and other lipids \cite{butovich2013tearfilmlipids,rantamaki2017phospolipids,galor2022meibumaltered} and still other species \cite{georgiev2017strucfun}. Detailed equations of state for studying the Marangoni effect from these species is not available for the tear film \textsl{in vivo} to our knowledge, and this is an ongoing area of research.

\subsection{Mathematical modeling}
A variety of mathematical models have incorporated various important effects of tear film dynamics as has been recently reviewed  \cite{Braun12,braun2019,braun2015dynamics}.
The most common assumptions for these models are a Newtonian tear fluid and a flat cornea \cite{berger74,BraunUsha12}.  
One category of models treats the flow over an open eye-shaped domain; these models have used lubrication theory with two space dimensions and time as independent variables \cite{MakiBraun10a,MakiBraun10b,LiBraun14,LiBraun16,LiBraun18}. These papers were on a stationary domain, and emphasized the effects of supply and drainage, osmolarity in various locations in the tear film, and the overall appearance of fluorescein imaging of the tear film.  Some initial studies have examined flow on a blinking domain \cite{driscollSimulationParabolic2018,maki2019blinkingeye,braun2019}, but they have yet to incorporate the same level of effects as in the static boundary papers.  
There have also been useful compartment models developed for tear film dynamics \cite{GaffEtal10,CerretaniEtal14}.  These models use compartments of the tear film as dependent variables in systems of ordinary differential equations (ODE) and thus avoid the computational difficulties of solving PDEs on 2D domains.  The models can be solved over many blinks and give insight into the osmolarity and other variables at various operating conditions.

There are quite a few models that have been developed using a single variable in space and time for independent variables.  There are two main categories here: (1) a spatial domain that extends across the eye opening vertically, with or without a moving end, and (2) a shorter domain that is a local model for TF dynamics.  We first discuss category (1). First, a number of papers considered a domain that extended across the open eye with fixed ends which typically called post-blink drainage models \cite{SharmaTiwari98,wong1996deposition,MilPolRad02a,BraunFitt03}.  These models typically imposed large menisci at the domain ends that drained fluid from the relatively flat interior of the TF, and that drove the dynamics.  Notable among these papers is that evaporation was included to find that it could drive thinning to TBU \cite{BraunFitt03}, and that a moving end, representing the upstroke of the blink, was treated as a coating flow problem \cite{wong1996deposition}.  Both of those aspects would find many applications in later papers.  Heat transfer and evaporation were considered in \citet{LiBraun12}, and later treated together with thermal imaging in \citet{DurschFLandThermal2017}.  A counterpoint to these papers Berger and Corrsin's treatment of the Marangoni effect in post blink spreading \cite{berger74}, which provided a basis for using this effect in many later papers.  

Using a moving end on a one-dimensional domain to simulate the upper eyelid during a blink has been employed in a number of papers.  The first to do this was by \citet{wong1996deposition}; they treated the problem as a coating flow analysis and developed a closed form equation for the thickness of the deposited film as a function of end (lid) speed.  Later papers extended this analysis by computing the thickness and other TF variables numerically.  These included a combined TF formation and drainage model \cite{JonesEtal05}, and then adding surfactant transport on the surface of the TF \cite{JonesEtal06}. Additional papers using the eye opening and interblink together included gravity and reflex tearing \cite{MakiBraun08}; using optimization varying viscosity to simulate and design eyedrops \cite{JossicLefevre09}; a detailed look at insoluble surfactant \cite{AydemirBreward2010}; a (Newtonian) two-layer model of the tear film with insoluble surfactant between them \cite{bruna2014influence}; the effect of (constant) globe curvature on tear film thickness \cite{Allouche17}; and a shear-thinning (power law) model for TF thickness on a curved substrate \cite{mehdaoui2021numerical}.  Some papers also treated the moving end for complete blink cycles.  Simple sinusoidal motion was treated in \citet{BrKS07}, while more realistic motion was treated in \citet{heryudono2007single}, and insoluble surfactant was added in \citet{zubkov2012coupling}.  Heat transfer was later considered with either full \cite{deng2014heat} or partial \cite{deng2013model} blink cycles.  

We next discuss category (2) which are local one-dimensional models for TF dynamics.  These models are along a short line segment and are designed to be local models of TF flow.  
The effects of evaporation and the Marangoni effect on TBU have been studied in a number of papers.  Evaporation drove all the dynamics in the models in \cite{peng2014evaporation,braun2015dynamics,BraunDrisTBU17}, though the treatments were different. In \cite{peng2014evaporation}, a stationary LL with variable thickness caused increased localized thinning; also included were osmolarity transport in the AL and osmosis across the AL/cornea interface.  An important result from that work is that diffusion of osmolarity out of the region of evaporation prevents osmosis from stopping thinning as it would in spatially uniform models \cite{Braun12,braun2015dynamics}.  Simpler evaporation models were used in \cite{braun2015dynamics,BraunDrisTBU17}, but they included transport of fluorescein so that they could explain what was seen in TF visualization experiments.  \citet{ZhongBMB19} developed a PDE model with one spatial variable that incorporated both mechanisms.  Including fluorescein dye transport and fluorescence enabled fitting of models to \textsl{in vivo} fluorescence data within TBU regions to estimate parameters that are not possible to directly measure there at the time of writing \cite{lukeParameterEstimation2020,lukeParameterEstimationMixedMechanism2021a}.  Recently, ODE models with no space dependence have been successfully fit to fluorescence data in small TBU spots and streaks \cite{lukeParameterEstimationMixedMechanism2021b}.  Those models have been coupled to a neural-network based data extraction system to greatly expand the amount of TBU instances that may be studied \cite{driscoll2023fittingMAIO}.

Other effects have been included in local models of TBU \cite{ZhangMatar03,ZhangMatar04,peng2014evaporation,BraunDrisTBU17,ZhongEtal18,ZhongJMO18,ZhongBMB19,deyContinuousMucinProfiles2019,deyContinuousMucinCorrection2020,choudhuryMembraneMucin2021}.  
Some models incorporated the effect of soluble mucins in the AL necessitating departure from more standard lubrication models \cite{deyContinuousMucinProfiles2019,deyContinuousMucinCorrection2020}.  Localized non-wettability of the ocular surface was modeled in \cite{choudhuryMembraneMucin2021}.
Other models have provided fundamental analyses on evaporation \cite{ji2016finitetimerupture,ji2020ssdynnon-conserved} and instability \cite{shi2021instabilitysoliddome,shi2022instabilitybubble}.  

\subsection{LL structure and the model in this paper}

The composition and structure of the LL, and their relation to function, are the subject of much ongoing research \cite{butovich2013tearfilmlipids,pucker2015lipidreview,butovich2017meibogenesis}.  The complexity of the composition affects the determination of structure and function, and has led to a variety of perspectives in this area \cite{georgiev2017strucfun,georgiev2019lipidsat}.  It is generally thought that the LL has an outermost nonpolar LL with a layer of polar lipids and other surface active components at the LL/AL interface \cite{mcculley1997llmodel,BronTiffRev04}.  Measuring all but the most basic properties can be a challenge, however.  Microscopy has shown some example forms that the LL may take \cite{KSOculSurfRev11}; the shapes observed there suggest non-Newtonian behavior of the LL \cite{Leiske12,rosenfeld2013structural}.  For a number of years, it was very difficult to reproduce significant resistance to evaporation using \textsl{in vitro} meibum or LL films   \cite{cerretani2013water,sledge2016nobarrier}, even though resistance to evaporation has been thought to be an important property \textsl{in vivo} \cite{MishimaMaurice61,king2010application}.  However, recent progress on \textsl{in vitro} films with many components has been made \cite{svitova2021barrier,xu2023barrier}.  The barrier function does depend on thickness \cite{king2010application,King-SmithIOVS13a}, but clearly there is an effect from internal structure as well \cite{fenner2015moretostable,king-smith2015moretostable}.

Molecular dynamics simulations, together with Langmuir trough experiments, have also given insight into the structure and function of the LL or model systems for it.  Coarse grained models were initially employed \cite{wizert2014organization,cwiklik2016molecular} and those models suggest that the LL may not be so as well organized as layers of nonpolar lipids atop a polar lipid monolayer.  Later, all-atom models have suggested that the specific polar lipids \cite{bland2019oahfa,viitaja2021tfll-and-oahfa} and wax esters \cite{paananen2019waxesters} may form ordered states to slow evaporation of water from the AL.  Molecular dynamics simulations are likely to continue to give important insights into LL structure and function, but they currently can only be solved at small space and time scales.  A link to bigger scales has been made via Langmuir trough observations with Brewster angle microscopy, e.g. \cite{bland2019oahfa,viitaja2021tfll-and-oahfa} for equilibrium states. Those insights need to be converted to larger scale dynamic models to address other situations such as TBU.  In this paper, we develop a two-layer continuum model that attempts to incorporate some internal structure of the LL that may be expected to produce evaporation resistance \cite{King-SmithOS13,king2018} that depends on that structure.  We use a nematic liquid crystal structure as a model for the LL, which can show birefringence as has been observed for meibum \cite{butovich2014biophysical}.

Adapting two-layer models of tear dynamics \cite{bruna2014influence, peng2014evaporation} to dynamic models of local TBU regions \cite{stapf2017duplex, taranchuk24} has made use of extensional flow for the LL and shear dominated flow for the AL.  These models have a common basis from the approach of \citet{matar2002surfactant}.  Of particular interest here is including extensional flow of a nematical liquid crystal \cite{cummings2014extensional,taranchuk2023extensional} for the LL with a Newtonian shear-dominated AL layer.  That extensional flow modeling used the Ericksen-Leslie equations and multiple scale perturbation methods to reduce the equations to a small PDE system.  
A detailed derivation and preliminary results for the combined model were developed in \citet{taranchuk24}. With this new model, they found dependence of TF dynamics on the director angle $\theta_B$ with all other parameters fixed, and we continue this work with a detailed parametric study.  We find new ways the effects combine to affect TF dynamics and TBU.

The paper is organized as follows. In Section \ref{sec:Model} we describe the problem formulation. In Section \ref{sec:results} we present our results. These include profiles of the film thicknesses, velocities and pressures that lead to different outcomes depending on the internal orientation of the liquid crystal molecules. Finally, in Section \ref{sec:discussion} we discuss the results and outline our conclusions.

\section{Model}\label{sec:Model}
\begin{figure}[htbp]%
\centering
\includegraphics[width=0.8\textwidth]{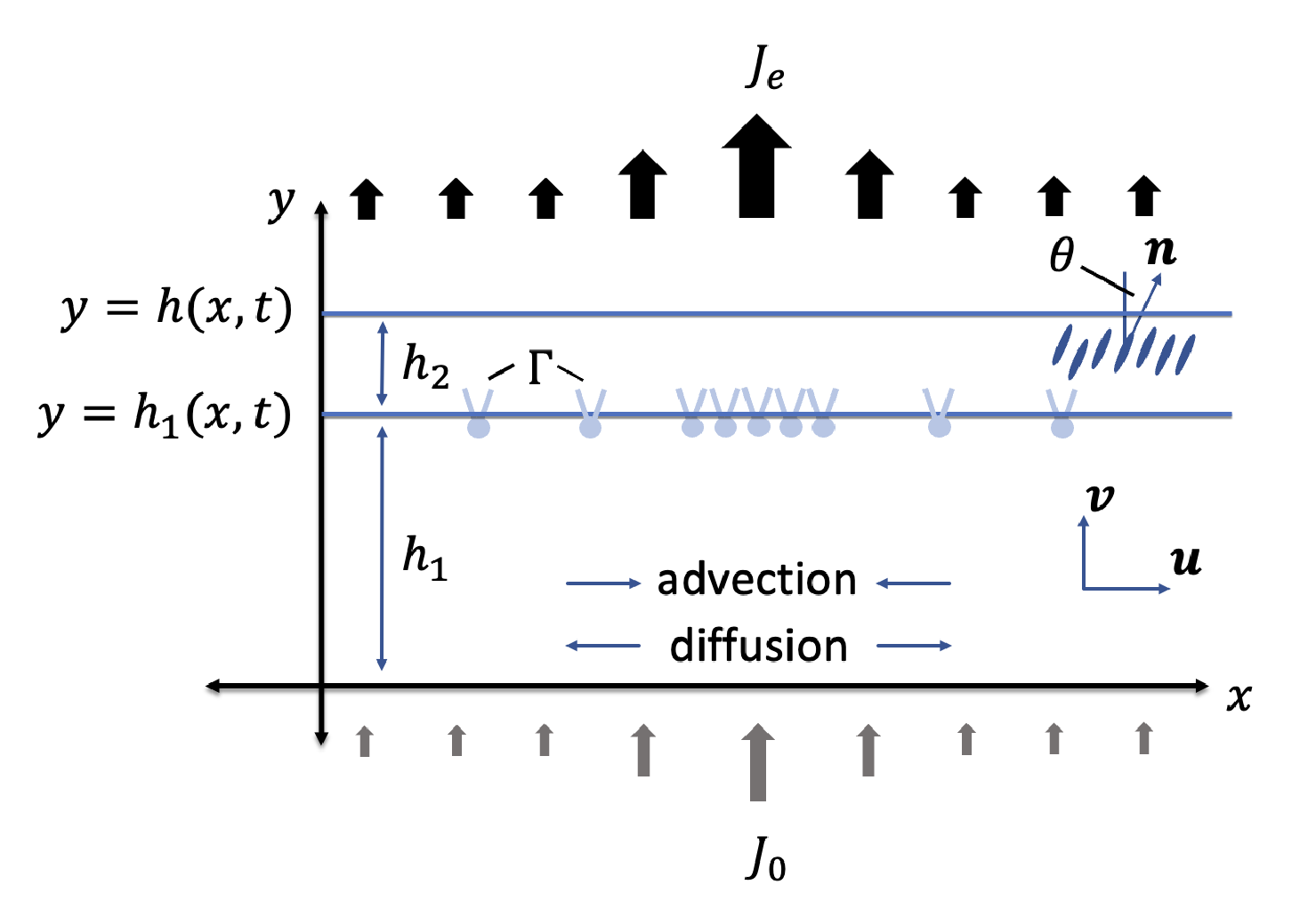}
\caption{Schematic of the two-layer tear film model.}\label{fig:system}
\end{figure}

Our two-layer model of the tear film lies on top of the corneal surface, which we take to be flat. It is composed of a Newtonian AL containing various solutes and a liquid crystal LL containing non polar lipids. At the interface of these two fluid layers is a monolayer of surfactant comprising primarily polar lipids. Water in the AL is lost through evaporation, $J_e$, and supplied from the corneal surface through osmosis, $J_o$. Figure \ref{fig:system} shows a sketch of the model. In the AL, we solve for the thickness $h_1$ and the osmolarity concentration $c$. In the LL we solve for the following quantities: thickness $h_2$, axial velocity $u_2$, angle of the director $\theta$, and surface concentration of surfactant $\Gamma$. 

We model the AL using the Navier-Stokes equations, and the LL using the Ericksen-Leslie equations for nematic liquid crystals. We use the following scalings, where primes denote dimensional quantities. The coordinates $(x',y')$ and velocity components $(u_i',v_i')$ correspond to the axial and transverse directions respectively, $t'$ is time, and $p'_i$ is pressure. The subscript $i$ indicates the fluid layer under consideration; $i=1$ corresponds to the AL, and $i=2$ corresponds to the LL. 
\begin{align}
    x'&=Lx, \qquad\qquad  h_1'=\e L h_1,\qquad\qquad h_2'=\e \delta L h_2,\nonumber\\
y'&=\e Ly,\hspace{36pt}      u_1'=Uu_1, \hspace{44pt}      u_2'= Uu_2,  \nonumber\\
h'&=\e L h, \hspace{36pt}     v_1'=\e U v_1,\hspace{42pt} v_2'=\e U v_2,  \nonumber\\
t'&=\frac{L}{U}t,\hspace{38pt} p_1'= \frac{\mu_1 U}{\e^2L}p_1,\hspace{33pt} p_2'= \frac{\mu_2 U}{L}p_2, \label{eq:scalings} \\
 \Gamma'& = \Gamma_0\Gamma,\hspace{38pt} c'=Cc, \hspace{50pt}      \nonumber\\
 J_o'& = \e U J_o,\hspace{30pt}\nonumber J_e'=\e\rho_1 UJ_e.
\end{align}

Here the total thickness of the tear film $h'=h'_1+h'_2\implies h=h_1+\delta h_2$. The scalings used in the model, chosen with the dimension of the human eye in mind, are defined in Table \ref{tab:scalings}. The LL is the thinnest layer in the tear film, and has an average thickness of tens to 100 nm during the interblink \cite{Norn79, KSHinNic10, king2011high, goto2003kinetic}. The AL is much thicker, averaging about 3 $\mu$m \cite{king2004thickness}. The glycocalyx is less than a micron thick \cite{Gipson04, Gipson10}; while it does not appear in the model explicitly, we account for its height by defining TBU as an AL thickness of 0.5 $\mu$m or less. We consider a domain of length 2 mm. 

\begin{table}[htbp]
\caption{Parameters used in dependent variable scalings} \label{table:scalings}%
\begin{tabular}{@{}lllll@{}}
\toprule
	  Parameter
		&  Value 
		&  Units
		&  Description
		&  Reference  \\ 
\midrule
	 $L=H_1(\gamma_1+\gamma_2)/(\mu_1 E_0)^{1/4}$
		& $3.1779 \times 10^{-4}$
		& m
		& Length scale
		&   \cite{stapf2017duplex} \\
	 $H_1$
		& $3.5 \times 10^{-6}$
		& m
		& AL thickness 
            & \cite{king2004thickness}\\
	 $H_2$
		& $5 \times 10^{-8}$
		& m
		& LL thickness
		&  \cite{KSHinNic10}  \\
	$V = E_0$
            & $5.093\times 10^{-7}$ 
            & m/s
            & Thinning rate
            &\cite{stapf2017duplex}\\
	$U=V/\e $
		& $4.624\times 10^{-5}$
		& m/s
		& Horizontal velocity
		& \cite{stapf2017duplex}   \\
	 $C$
		& $300$
		& mOsM
		& Reference osmolarity
		& \cite{peng2014evaporation}   \\
	 $\Gamma_0$
		& $4 \times 10^{-7}$
		& mol/$\text{m}^2$
		& Surfactant concentration
		& \cite{AydemirBreward2010,bruna2014influence}  \\
\botrule
\end{tabular}
\label{tab:scalings}
\end{table}

The nondimensional quantities that result from scalings are given in Table \ref{tab:nondim}, and the physical parameters are listed in Table \ref{tab:physparam}. We note that while experiments show evidence of liquid crystal structure in the LL of the TF, the parameter values for the LL have not been measured. For liquid crystals, we return to a Newtonian fluid by setting the Leslie viscosities $\alpha'_i=0,\,i=1,\cdots,6,\,i\neq4$, and $\alpha'_4=2\mu_2$. The dynamic viscosity of the LL, $\mu_2$, has previously been taken to be 0.1 Pa s \cite{bruna2014influence, stapf2017duplex}. Thus, to match this viscosity in the Newtonian limit, as default values we choose our Leslie viscosities to be three times the properties of 5CB, a liquid crystal that has been relatively well studied near physiological temperatures \cite{stewart2019static}. We explore a larger range of values in Section \ref{sec:LCparams}.

\begin{table}[htbp]
\caption{Nondimensional Parameters} 
\begin{tabular}{@{}llll@{}}
\toprule
	Parameter
		&  Formula
		&  Value 
		&  Description  \\  
\midrule
	$\epsilon$
		&  $H_1 / L$
		&  $0.0110$
		&  Aspect ratio of AL  \\
	$\delta$
		&  $H_2 / H_1$
		&  $0.0143$
		&  Lipid to aqueous thickness ratio \\
	$\Upsilon$
		&  $\epsilon^2 \mu_2 / \mu_1$
		&  $0.0091$
		&  Reduced viscosity ratio  \\
	$\mathcal{C}_1$
		&  $\epsilon^3 (\gamma_1+\gamma_2) / (\mu_1 U)$
		&  $1.0001$
		&  Reduced aqueous capillary number  \\
	$\mathcal{C}_2$
		&  $\epsilon \gamma_2 / (\mu_2 U)$
		&  $43.8371$
		&  Reduced lipid capillary number  \\
	$\mathcal{M}$
		&  $\epsilon R T_0 \Gamma_0 / (\mu_1 U)$
		&  $187.7687$
		&  Reduced Marangoni number  \\
	$\mathcal{E}$
		&  $ E_0 / (\epsilon\rho_1  U)$
		&  $1.0001$
		&  Evaporation parameter  \\
	$\mathcal{R}_0$
		&  $k_m H_2 / (Dk)$
		&  $17.68$
		&  Evaporative resistance parameter \\
	$\mathcal{P}$
		&  $C P_c / (\epsilon U)$
		&  $0.1355$
		&  Osmosis parameter  \\
	$\textrm{Pe}_1$
		& $UL/D_1$
		&  $9.1842$
		&  P\'{e}clet number for osmolarity diffusion  \\
	$\textrm{Pe}_s$
		&  $UL/D_s$
		&  $0.4898$
		&  P\'{e}clet number for surfactant diffusion  \\
\botrule
\end{tabular}
\label{tab:nondim}
\end{table}

\begin{table}[htbp]
\caption{Physical Parameters}
\label{table:constants}%
\begin{tabular}{@{}lllll@{}}
\toprule
	Constant
		&  Value 
		&  Units
		&  Description
		&  Reference  \\  
\midrule
	$\rho_1$
		&  1000
		&  kg/$\text{m}^3$
		&  Density of liquid water
		&  \cite{CRC77} \\
	$\rho_2$
		&  900
		&  kg/$\text{m}^3$
		&  Density of lipid
		&  \cite{bruna2014influence} \\
	$\mu_1$
		&  $1.3 \times 10^{-3}$
		&  Pa s
		&  Dynamic viscosity of water
		&  \cite{bruna2014influence,tiffany1991viscosity}  \\
	$\mu_2=\alpha_4'/2$
		&  $0.0978$
		&  Pa s
		&  Dynamic viscosity of lipid
		&  \cite{stewart2019static} \\
	$\gamma_1$
		&  $0.027$
		&  N/m
		&  Surface tension of water
		&	 \cite{tiffany1987lipid}	\\
	$\gamma_2$
		&  $0.018$
		&  N/m
		&  Surface tension of lipid
		&	 \cite{nagyova1999components}	\\
	$R$
		&  $8.3145$
		&  J/(K mol)
		&  Ideal gas constant
		&  \cite{CRC77}	\\
  	$D_1$
		&  $1.6 \times 10^{-9}$
		&  $\text{m}^2$/s
		&  Osmolarity diffusion constant
		&  \cite{riquelme2007interferometric}  \\
	$D_s$
		&  $3 \times 10^{-8}$
		&  $\text{m}^2$/s
		&  Surfactant diffusion constant
		&  \cite{bruna2014influence}  \\
	$k_m$
		&  $0.0182$
		&  $ \text{m} / \text{s}$
		&  Mass transfer coefficient
		&  \cite{stapf2017duplex}  \\
	$Dk$
		&  $ 5.136  \times 10^{-11}$
		&  $ \text{m}^2 / \text{s}$
		&  Lipid permeability
		&  \cite{stapf2017duplex}  \\
	$P_c$
		&  $2.3 \times 10^{-10}$
		&  $\text{kg}/(\text{mOsM} \ \text{m}^2 \ \text{s})$
		&  Corneal osmosis coefficient
		&  \cite{peng2014evaporation}  \\
	$E_0$
		&  $5.0928 \times 10^{-7}$
		&  $ \text{m} / \text{s}$
		&  Evaporative thinning rate
		&  \cite{stapf2017duplex}  \\
        $T_0$
        & 308.15
        & K
        & Eye surface temperature
        &\cite{peng2014evaporation} \\
        $\alpha_1'$
            & $-0.0180$
            & Pa s
            & Leslie viscosity 
            & \cite{stewart2019static}\\
        $\alpha_2'$
            & $-0.2436$
            & Pa s
            & Leslie viscosity
            & \cite{stewart2019static}\\
        $\alpha_3'$
            & $-0.0108$
            & Pa s
            & Leslie viscosity
            & \cite{stewart2019static}\\
        $\alpha_4'$
            & $0.1956$
            & Pa s
            & Leslie viscosity
            & \cite{stewart2019static}\\
        $\alpha_5'$
            & $0.1920$
            & Pa s
            & Leslie viscosity
            & \cite{stewart2019static}\\
        $\alpha_6'$
            & $-0.0624$
            & Pa s
            & Leslie viscosity
            & \cite{stewart2019static}\\
        $\theta_B$
            &$0$ - $\pi/2$
            & radians
            & Director angle
            & \cite{stewart2019static}\\
\botrule
\end{tabular}
\label{tab:physparam}
\end{table}

\subsection{Governing equations}\label{sec:governingeqs}
After nondimensionalizing, we asymptotically expand all of the dependent variables in powers of the small parameter $\epsilon$. We then collect like powers of $\e$, and solve to find a closed system of leading order equations. In the process, we find that the director angle is constant at leading order, $\theta=\theta_B$. We also determine the leading order AL pressure $p_1$, and the AL depth averaged axial velocity $\Bar{u}_1$. This process is described in more detail in the appendix \ref{sec:appendix_derivation}, and derived thoroughly in \citet{taranchuk24}. We find the closed leading order system to be
\begin{align}
    h_{10t}+(\Bar{u}_{10}h_{10})_x=J_o-J_e,\label{eq:ALconsmass}\\
    h_{20t}+\left(u_{20}h_{20}\right)_x=0,\label{eq:LLconsmass}\\
    -\frac{A(\theta_B)}{B(\theta_B)} \delta\Upsilon (u_{20x} h_{20})_x -\frac{u_{20}}{h_{10}}=\frac{h_{10}}{2} \bigg(-\mathcal{C}_1 h_{10xxx}- \mathcal{C}_2 \Upsilon h_{0xxx}\bigg)\nonumber\\
- \mathcal{C}_2\delta\Upsilon h_{20} h_{0xxx} +\mathcal{M} \Gamma_{0x},\label{eq:LLforcebalance}\\
\Gamma_{0t}+ (\Gamma_0 u_{20})_x=\frac{1}{Pe_s}\Gamma_{0xx},\label{eq:surfconc}\\
(c_0h_{10})_t =\frac{1}{Pe_1}(c_{0x}h_{10})_x-\left(c_0\Bar{u}_{10}h_{10}\right)_x,\label{eq:osmconc}
\end{align}
where
\begin{align}
    h_0=h_{10}+\delta h_{20},\\
    J_o=\mathcal{P}\left(c-\frac{c_0}{C}\right),\\
    J_e= \frac{\mathcal{E}}{1+\mathcal{R}(\theta_B)h_{20}},\label{eq:je}\\
    \mathcal{R}(\theta_B)=\mathcal{R}_0(0.1+0.9\sin \theta_B),\label{eq:r0}\\
    p_{10}=-\Upsilon\mathcal{C}_2h_{0xx} - \mathcal{C}_1 h_{10xx},\\
    \Bar{u}_{10}=-p_{10x}\frac{h_{10}^2}{12}+\frac{u_{20}}{2},
\end{align}
and
\begin{align}
    A(\theta_B)=&\;2 \cos (2 \theta_B) (\alpha_1+\alpha_5+\alpha_6+4) (\alpha_2+\alpha_3-\alpha_5+\alpha_6)\nonumber\\
&+\cos (4 \theta_B) (\alpha_1 [\alpha_2-\alpha_3]+[\alpha_2+\alpha_3] [\alpha_5-\alpha_6])+\alpha_1 \alpha_2-\alpha_1 \alpha_3-2 \alpha_1 \alpha_5\nonumber\\
&-2 \alpha_1 \alpha_6-8 \alpha_1+\alpha_2 \alpha_5+3 \alpha_2 \alpha_6+8 \alpha_2-3 \alpha_3 \alpha_5-\alpha_3 \alpha_6-8 \alpha_3-2 \alpha_5^2\nonumber\\
&-4 \alpha_5 \alpha_6-16 \alpha_5-2 \alpha_6^2-16 \alpha_6-32,\\
B(\theta_B)=&-\alpha_1 \cos (4 \theta_B)+\alpha_1-2 \cos (2 \theta_B) (\alpha_2+\alpha_3-\alpha_5+\alpha_6)\nonumber\\
&-2 \alpha_2+2 \alpha_3+2 \alpha_5+2 \alpha_6+8.
\end{align}
Equation (\ref{eq:ALconsmass}) and (\ref{eq:LLconsmass}) represent conservation of mass (of solvent) in the AL and LL, respectively. Equation (\ref{eq:LLforcebalance}) is the axial force balance in the LL. Equation (\ref{eq:surfconc}) conserves mass of surfactant, and Equation (\ref{eq:osmconc}) conserves osmolarity in the AL. 

The evaporation function in Equation (\ref{eq:je}) depends on the evaporative resistance in the LL which is described by Equation (\ref{eq:r0}). We vary the evaporative resistance function $\mathcal{R}(\theta_B)$ by changing both $\mathcal{R}_0$ and $\theta_B$. Various director angles are considered throughout the paper, and Sections \ref{sec:r0_gaussian} and \ref{sec:r0_surf} focus on the effect of $\mathcal{R}_0$ for two sets of initial conditions. 

\subsection{Initial and boundary conditions}
\label{sec:icandbc}
\subsubsection{Evaporation-driven flow}\label{sec:evap_ic}
To illustrate flow driven by evaporation, we choose initial conditions in such away that valley, or thin spot, in the LL drives all the dynamics of the tear film toward TBU. We choose a Gaussian perturbation of the LL 
\begin{align}
    h_{2}(x,0)=1-0.9e^{-x^2/(2\sigma^2)},
\end{align}
where $2\sigma^2=1/9$.
 We take $h_1=\Gamma=c=1$ initially. We compute $p_1(x,0)$ from Equation~(\ref{eq:pressure1}), and solve the discrete version of the axial force balance Equation~(\ref{eq:axforcebal}) using the other initial values to obtain $u_2(x,y,0)$. This initial condition corresponds to the Gaussian disturbance described in \citet{stapf2017duplex}; we refer to it as a valley here. 
\subsubsection{Marangoni effect}\label{sec:maragoni_ic}
To illustrate surfactant gradient driven flow, we choose initial conditions in which all variables are uniform initially ($h_1=h_2=c=1$), except for a local excess of surfactant in the center of the domain. We compute $p_1(x,0)=0$ and $u_2(x,y,0)$ as described above. 
The initial condition for $\Gamma$ is
\begin{align}
    \Gamma(x,0)=1+s-s\left(\frac{1+0.5\tanh\left(\frac{x-x_0}{x_w}\right)+0.5\tanh\left(\frac{-x-x_0}{x_w}\right)}{n}\right)\label{eq:surfactant_IC}
\end{align}
where the height of the region of excess surfactant is $s=1.8$, the inflection point is $x_0=1$, and the width of the region is $x_w=0.36$, and normalization factor $n=0.9915$.

\subsubsection{Combining LL and surfactant perturbations}

Having considered perturbations in the LL and surfactant surface concentration individually, we now consider two scenarios which combine perturbations in both. Namely, we consider a bump in $h_2$ combined with a local excess of $\Gamma$, and a valley in $h_2$ combined with a local excess of $\Gamma$.

Imaging of the LL has shown instances of regions of thicker lipid \cite{GotoTseng03}. It is reasonable to assume that when the LL is thicker, there is more surfactant. To illustrate this, we consider a thin LL with a central bump and central increased $\Gamma$ in that area. We take $h_1(x,0)=c(x,0)=1$, and we calculate $p_1(x,0)$ and $u_2(x,0)$. The initial condition for surfactant is Equation (\ref{eq:surfactant_IC}), and for LL thickness
\begin{align}
    h_2(x,0)=1+s-s\left(\frac{1+0.5\tanh\left(\frac{x-x_0}{x_w}\right)+0.5\tanh\left(\frac{-x-x_0}{x_w}\right)}{n}\right)\label{eq:h2_bump}
\end{align}
where the height of the bump is $s=1$, the inflection point is $x_0=1$, and the width of the bump is $x_w=0.36$, and normalization factor $n=0.9915$. This corresponds dimensionally to a LL with background thickness of 25 nm, and a bump of lipid of height 50 nm in the center of the domain.

We also consider a pathological scenario in which a valley in the LL is combined with a local excess of $\Gamma$. Keeping all other initial conditions as described above, we modify only Equation (\ref{eq:h2_bump}) by changing $s=-0.5$, creating a valley in the LL. This corresponds dimensionally to a LL with thickness 50 nm, and a valley of depth 25 nm in the center of the domain.

\subsubsection{Boundary conditions}
At both boundaries we impose the following no flux boundary conditions,
\begin{align}
    h_{1x}=h_{2x}=p_{1x}=\Gamma_{x}=c_{x}=0,
\end{align}
and specify $u_2=0$ for the LL velocity.

\subsection{Numerical solution}
Before solving numerically, we make two change of variables. In order reduce the order of the system, we define $p_{10}$ as a dependent variable (Equation (\ref{eq:pressure1})) and substitute into Equation (\ref{eq:LLforcebalance}). We also use the substitution $m_0=h_{10}c_0$ in Equation (\ref{eq:osmconc}), where $m_0$ is the mass. This yields the following closed system which we wish to solve
\begin{align}
    h_{10t}+(\Bar{u}_{10}h_{10})_x=J_o-J_e,\\
    h_{20t}+\left(u_{20}h_{20}\right)_x=0,\\
    p_{10}=-\Upsilon\mathcal{C}_2h_{0xx} - \mathcal{C}_1 h_{10xx},\label{eq:pressure1}\\
    -\frac{A(\theta_B)}{B(\theta_B)} \delta\Upsilon (u_{20x} h_{20})_x -\frac{u_{20}}{h_{10}}=\frac{h_{10}}{2} p_{10x}
- \mathcal{C}_2\delta\Upsilon h_{20} h_{0xxx} +\mathcal{M} \Gamma_{0x},\label{eq:axforcebal}\\
\Gamma_{0t}+ (\Gamma_0 u_{20})_x=\frac{1}{\mbox{Pe}_s}\Gamma_{0xx},\\
m_{0t} =\frac{1}{\mbox{Pe}_1}\left(m_{0x}-\frac{m_0h_{10x}}{h_{10}}\right)_x-\left(m_0\Bar{u}_{10}\right)_x,
\end{align}
where
\begin{align}
    h_0=h_{10}+\delta h_{20},\\
    J_0=\mathcal{P}\left(\frac{m_0}{h_{10}}-1\right),\\
    J_e= \frac{\mathcal{E}}{1+\mathcal{R}(\theta_B)h_{20}},\label{eq:evap}\\
    \Bar{u}_{10}=-p_{10x}\frac{h_{10}^2}{12}+\frac{u_{20}}{2}.\label{eq:ubar}
\end{align}

We then use the method of lines to solve the system numerically subject to the initial and boundary conditions of Section \ref{sec:icandbc}. Spatial derivatives are approximated with second-order finite differences on a uniform grid of $512$ points. This results in a system of differential algebraic equations that we solve forward in time in \textsc{Matlab} (MathWorks, Natick, MA, USA) using \texttt{ode15s}. We found that the solutions were converged using this number of grid points, and that the mass of water and osmolarity were conserved to the tolerances of the computation.  

We integrate the system until either of the following happens: (i) a final time for 60 s is reached, or (ii) TBU occurs. We define TBU as the point at which the AL thickness reaches a minimum value of $0.5\mu$m, and use event detection to stop the simulation if this happens. When breakup occurs, the TBUT becomes the final time.

\section{Results}\label{sec:results}

\subsection{Evaporation-driven thinning}
\begin{figure}[htbp]%
\centering
\includegraphics[width=0.49\textwidth]{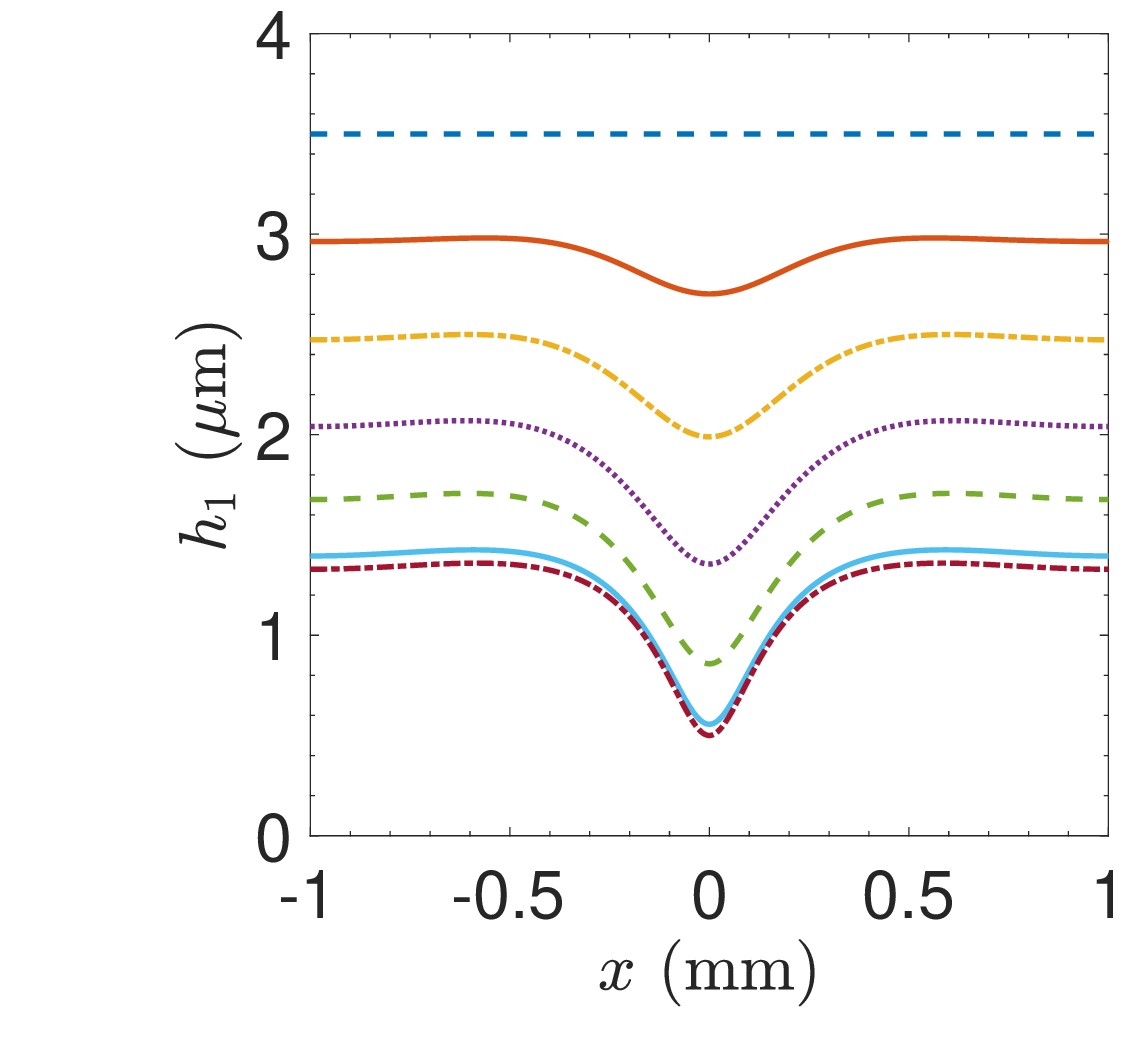}
\includegraphics[width=0.49\textwidth]{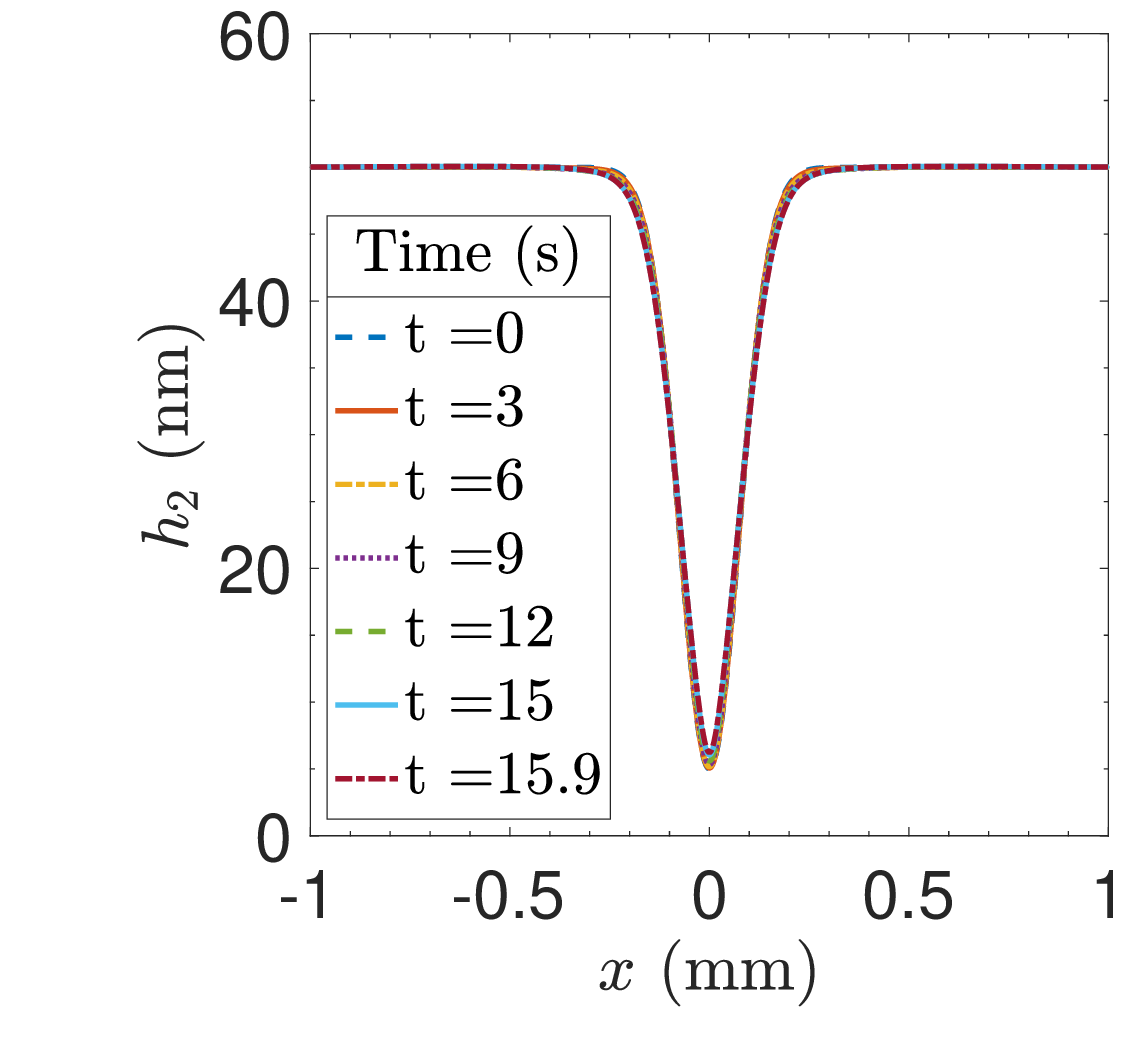}
\includegraphics[width=0.49\textwidth]{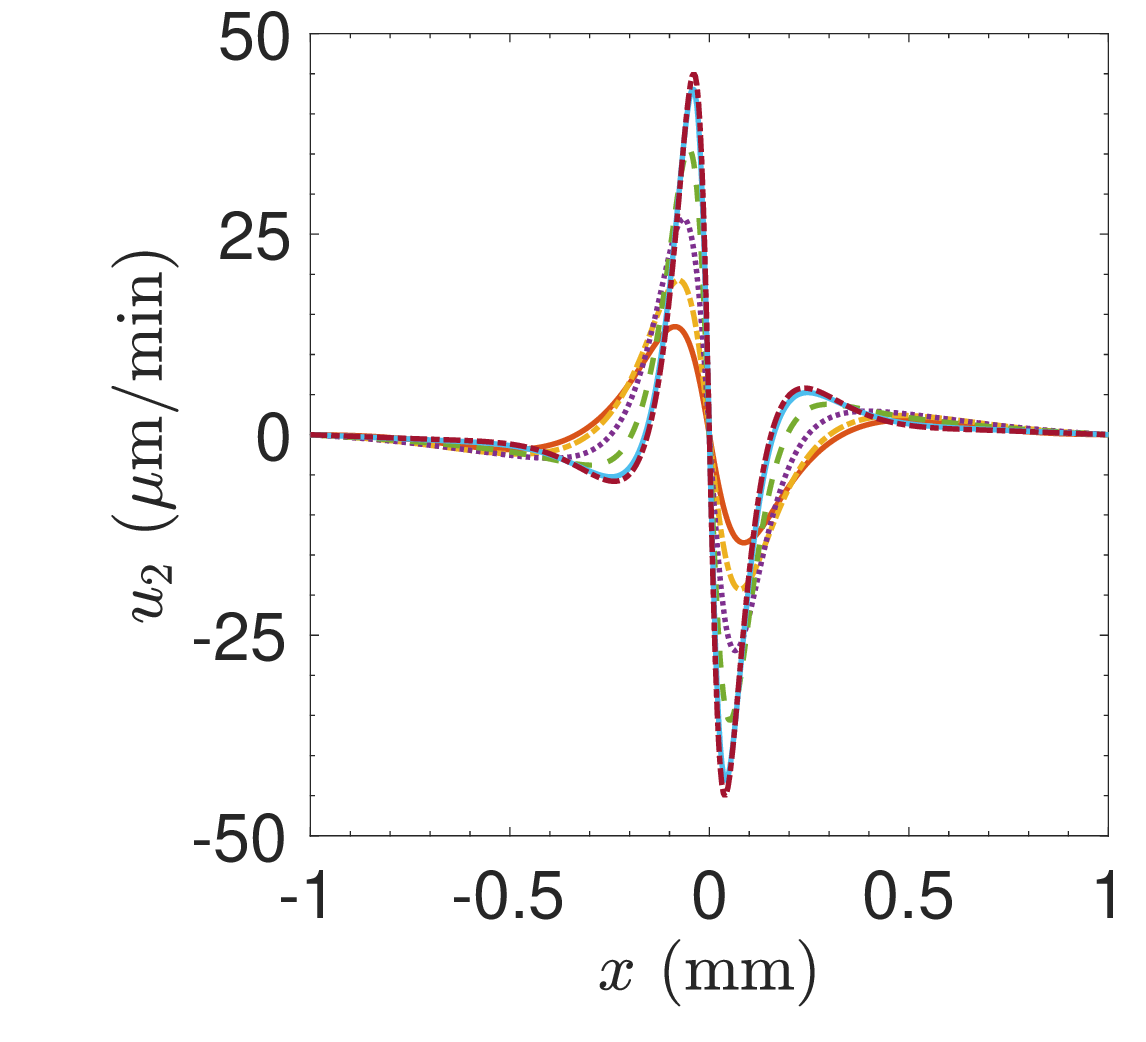}
\includegraphics[width=0.49\textwidth]{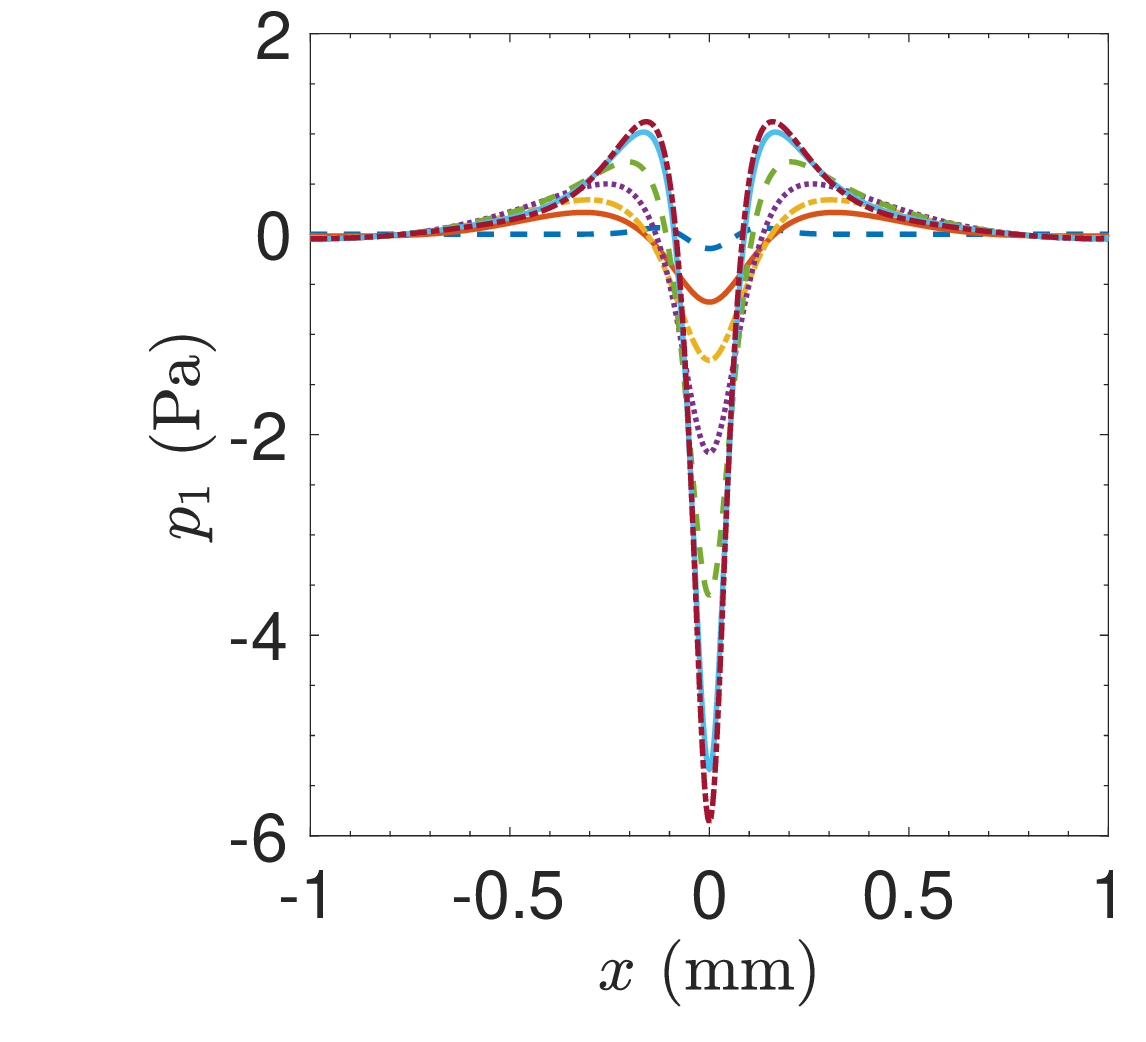}
\includegraphics[width=0.49\textwidth]{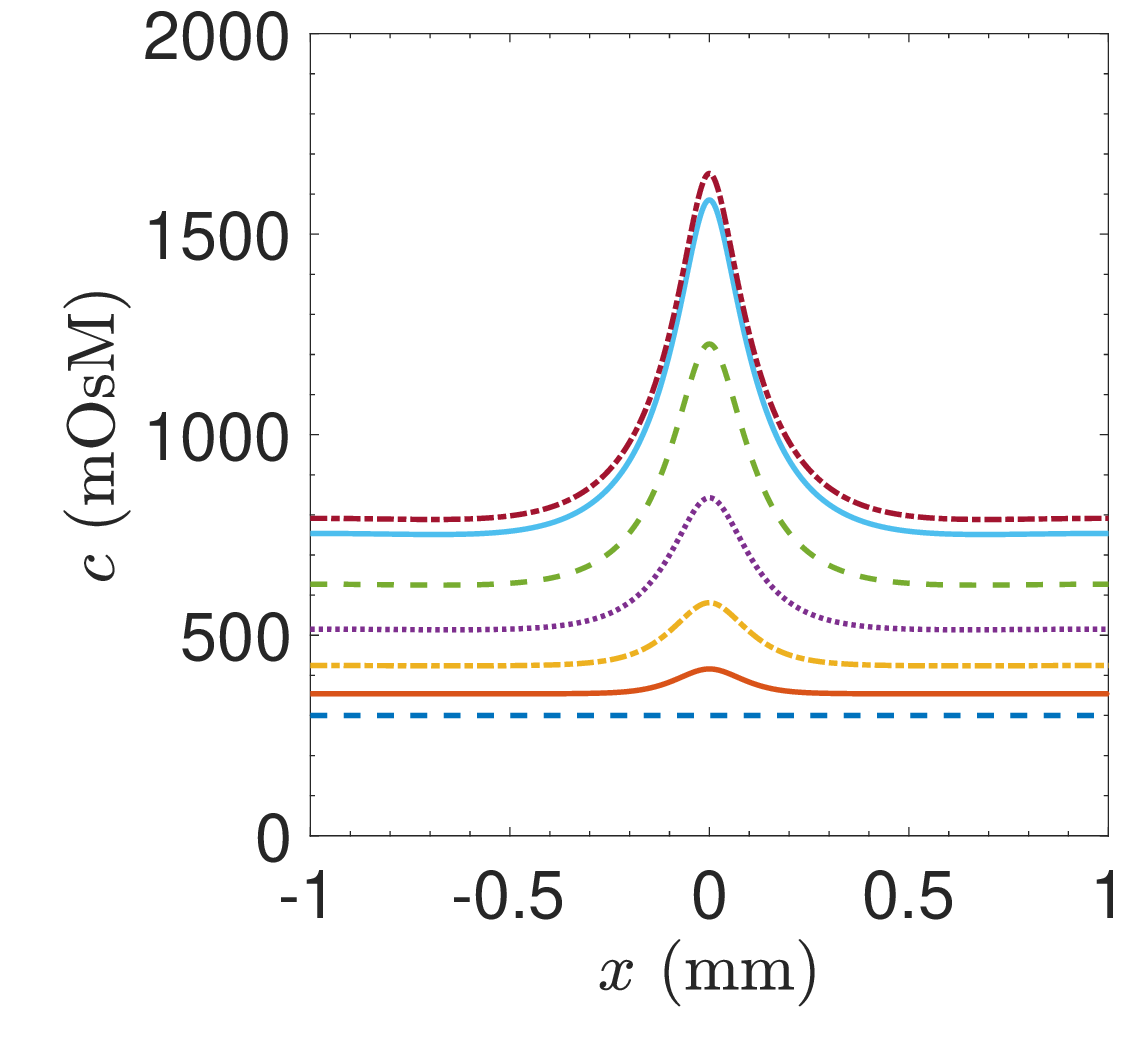}
\includegraphics[width=0.49\textwidth]{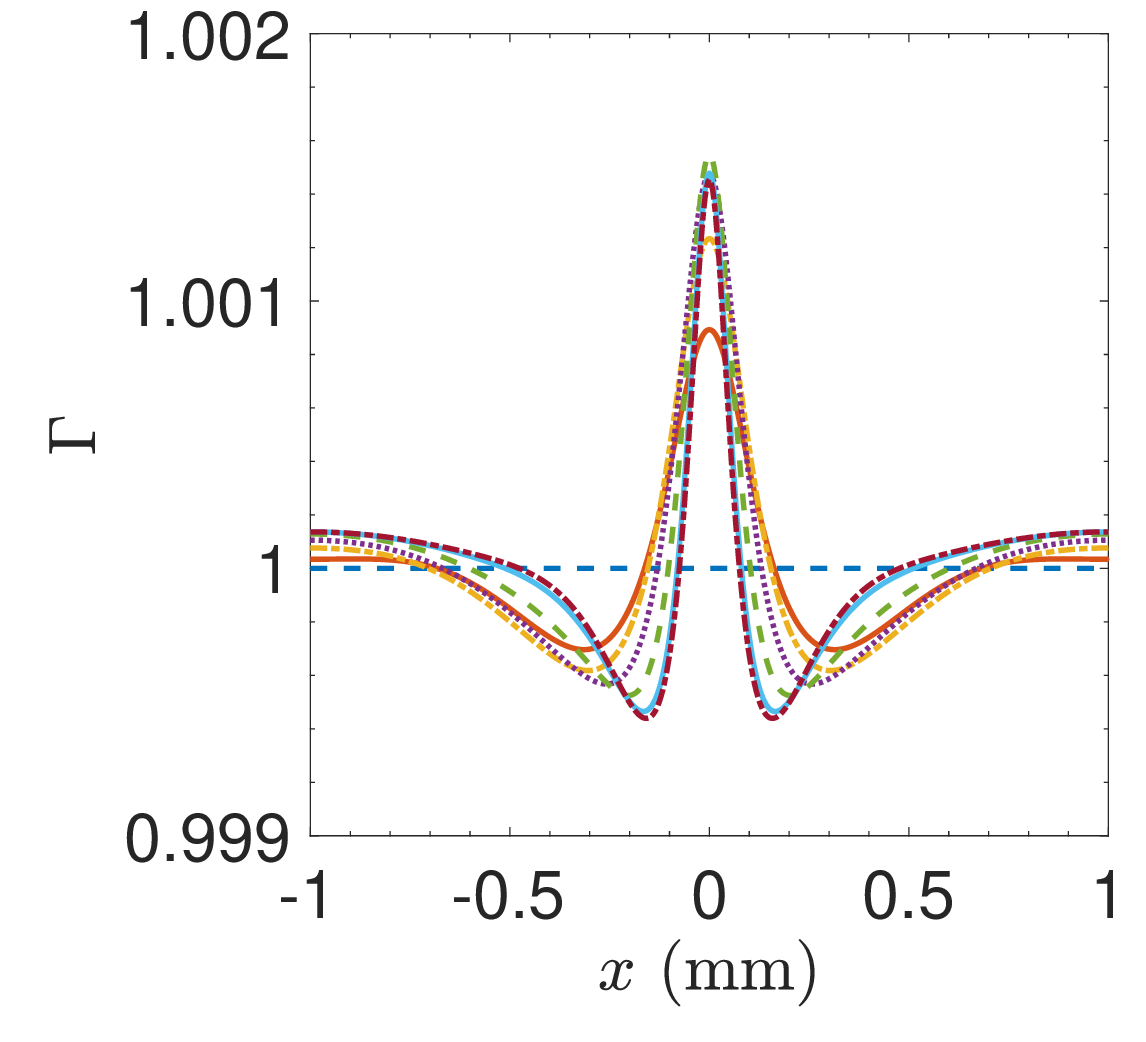}
\caption{Solutions for spatially-uniform initial conditions except for a Gaussian disturbance in the LL thickness.  The director angle is $\theta_B=0$.  This figure illustrates typical dynamics of the dependent variables in space and time for the case of TBU driven by increased evaporation from the valley in the LL.}\label{fig:gaussian_BU}
\end{figure}

As an example of evaporation-driven thinning, we consider the initial conditions described in Section \ref{sec:evap_ic} with a director angle of $\theta_B=0$. This choice maximizes the evaporation rate for our chosen evaporation function.
Evaporative resistance in the LL is maximized for $\mathcal{R}(\pi/2)$; $\mathcal{R}(0)$ reduces that resistance by a factor of 10. 

Solutions are shown in Figure \ref{fig:gaussian_BU}. The LL changes very little over time, maintaining its original shape with a valley in the center of the domain. The initial velocity in the LL, which ranges between $\pm 50\,\mu$m, is moderate compared to the dynamics of other scenarios considered later. The thinning in the LL leads to increased evaporation of the AL, which quickly thins, reaching breakup conditions in the center of the domain by 15.9 s. AL pressure begins to drop in the center of the domain; in general, this would help to restore fluid to the area; however TBU occurs because that capillarity-driven contribution is too weak. 

Osmolarity $c$ increases over time, peaking in the center of the domain as a result of increased evaporation in that area. By the final time, osmolarity in the center of the domain is 1651 mOsM, more than five times the initial value; and near the ends it is 789 mOsM, over two and a half times the the initial value. 

The surface surfactant concentration $\Gamma$, shown relative to its initial value, builds up slightly in the center of the domain. We explore Marangoni-driven flow in detail in Section \ref{sec:marangoni}; however in this example, the increase in surfactant concentration is not large enough to contribute significantly to the flow. 

These solutions are representative of evaporation-driven thinning that leads to breakup before the final time of 60 s. With a larger director angle, evaporation resistance of the LL is increased, and the AL thins more slowly. When $\theta_B\lesssim\pi/17$, TBU is not reached within the simulation time of 60 s. Because evaporation is slower, the valley that develops in the AL is less pronounced, and the resulting peak in osmolarity is smaller; see \citet{taranchuk24} for more details. 

\subsubsection{Varying the Marangoni number}\label{sec:gaussian_marangoni}

\begin{figure}[htbp]%
\centering
\includegraphics[width=0.32\textwidth]{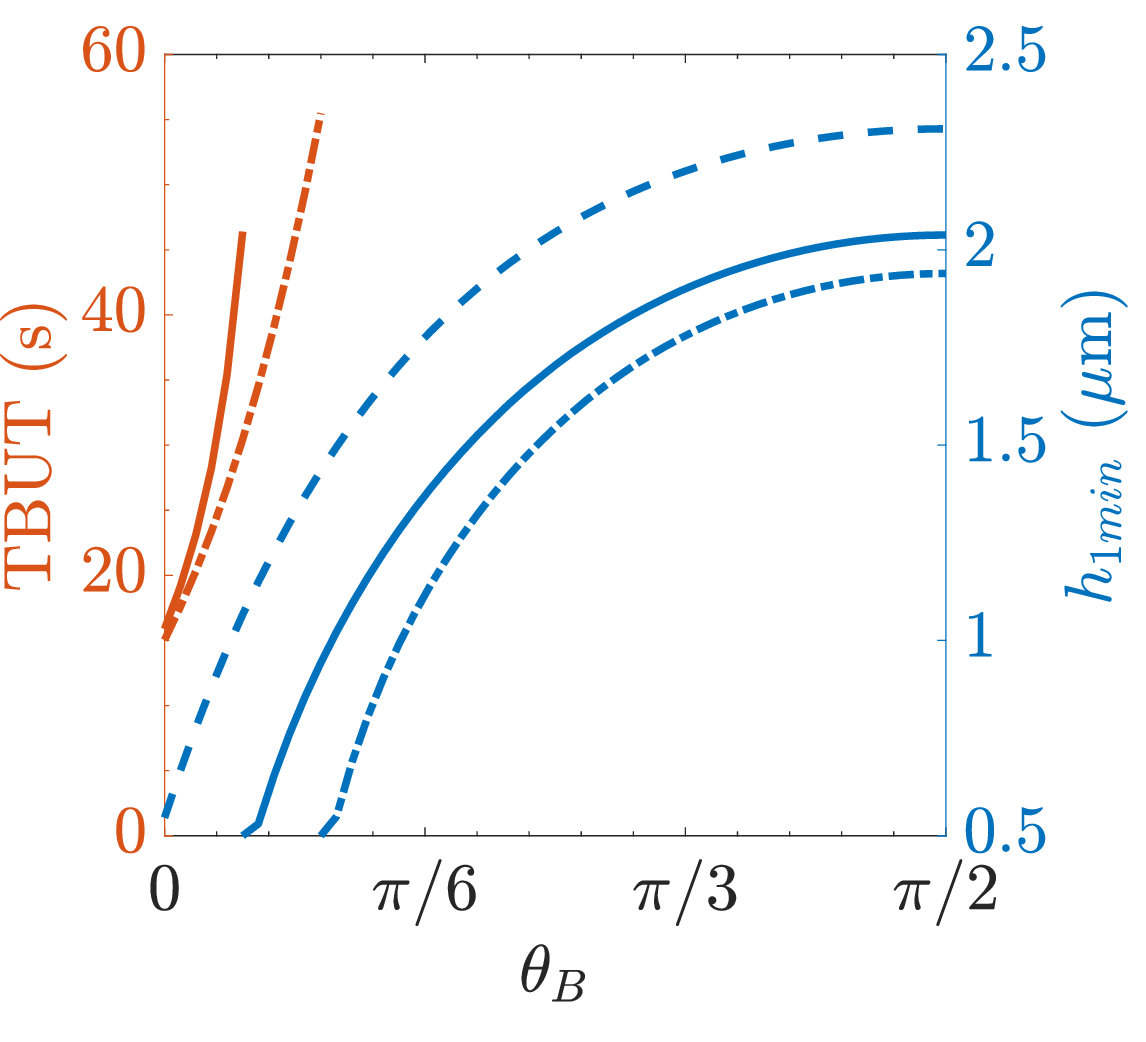}
\includegraphics[width=0.32\textwidth]{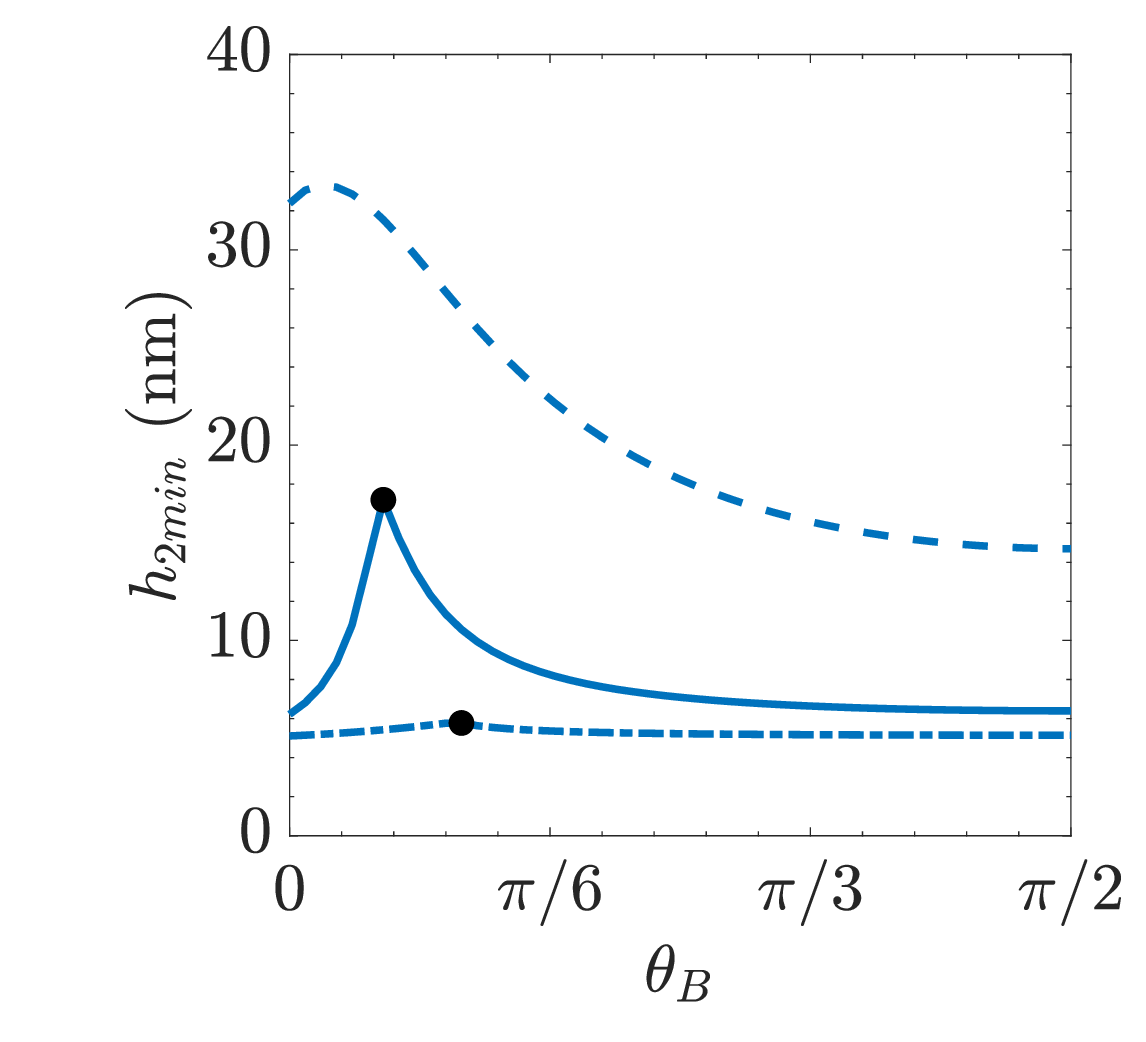}
\includegraphics[width=0.32\textwidth]{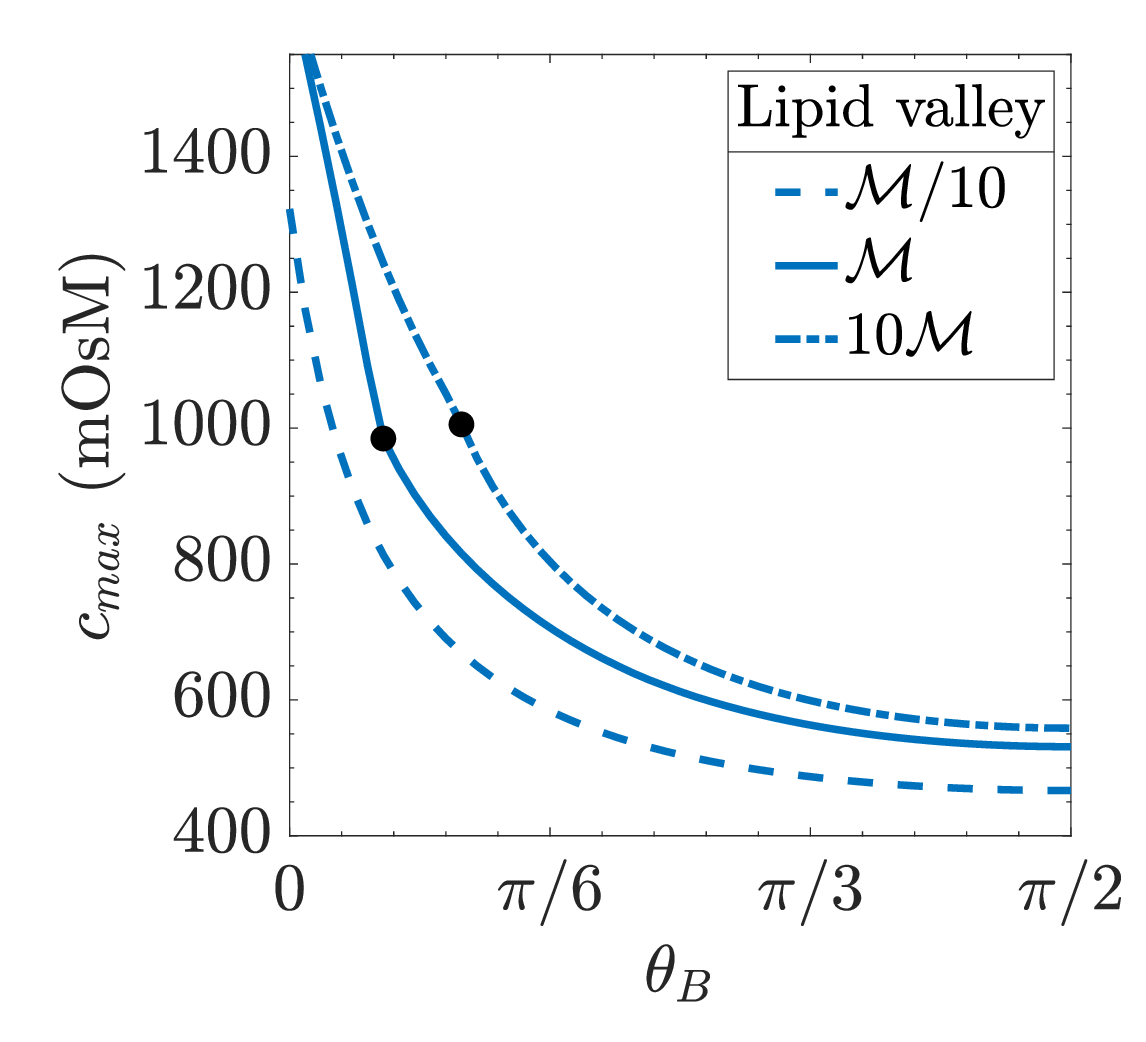}
\includegraphics[width=0.32\textwidth]{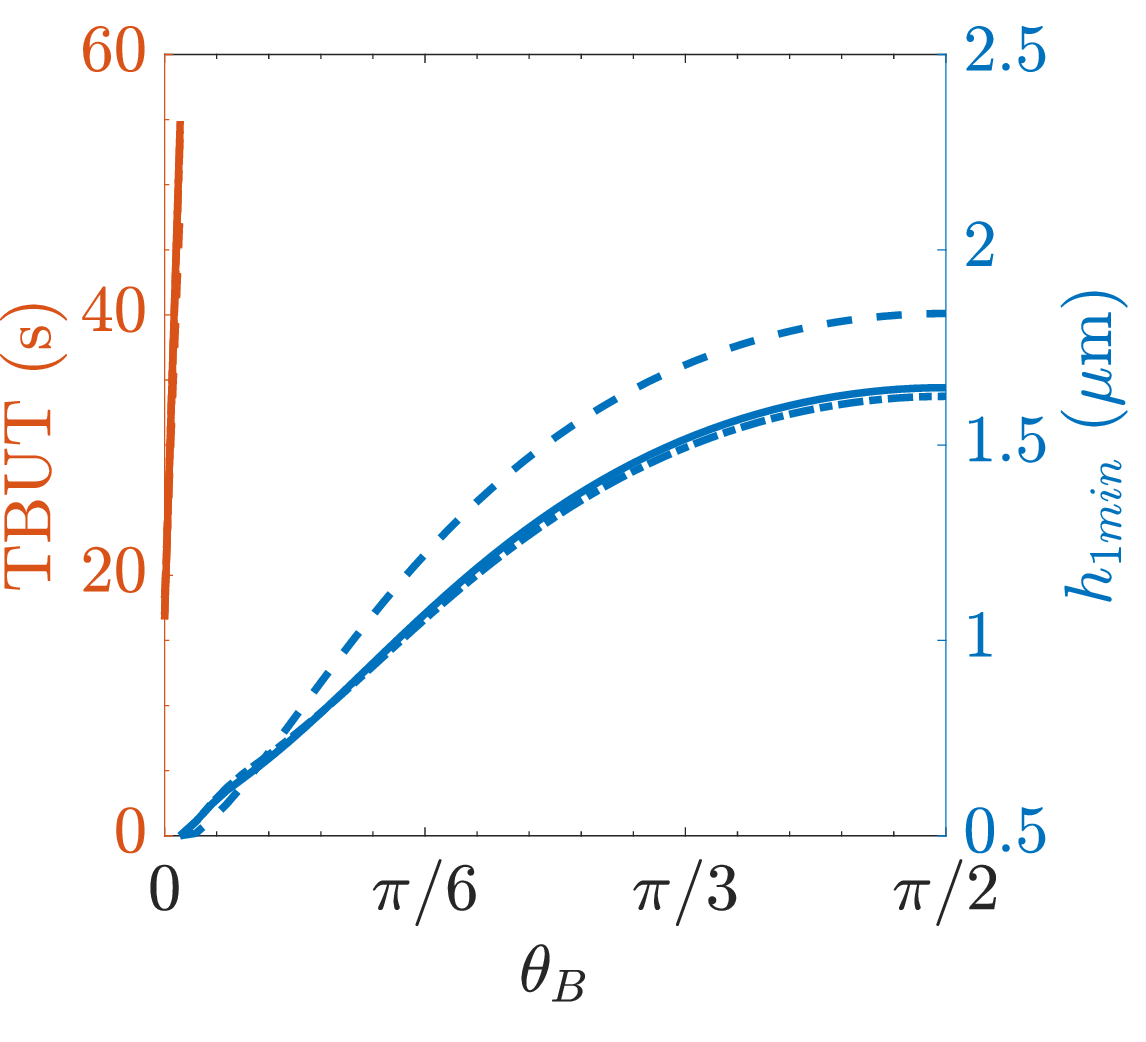}
\includegraphics[width=0.32\textwidth]{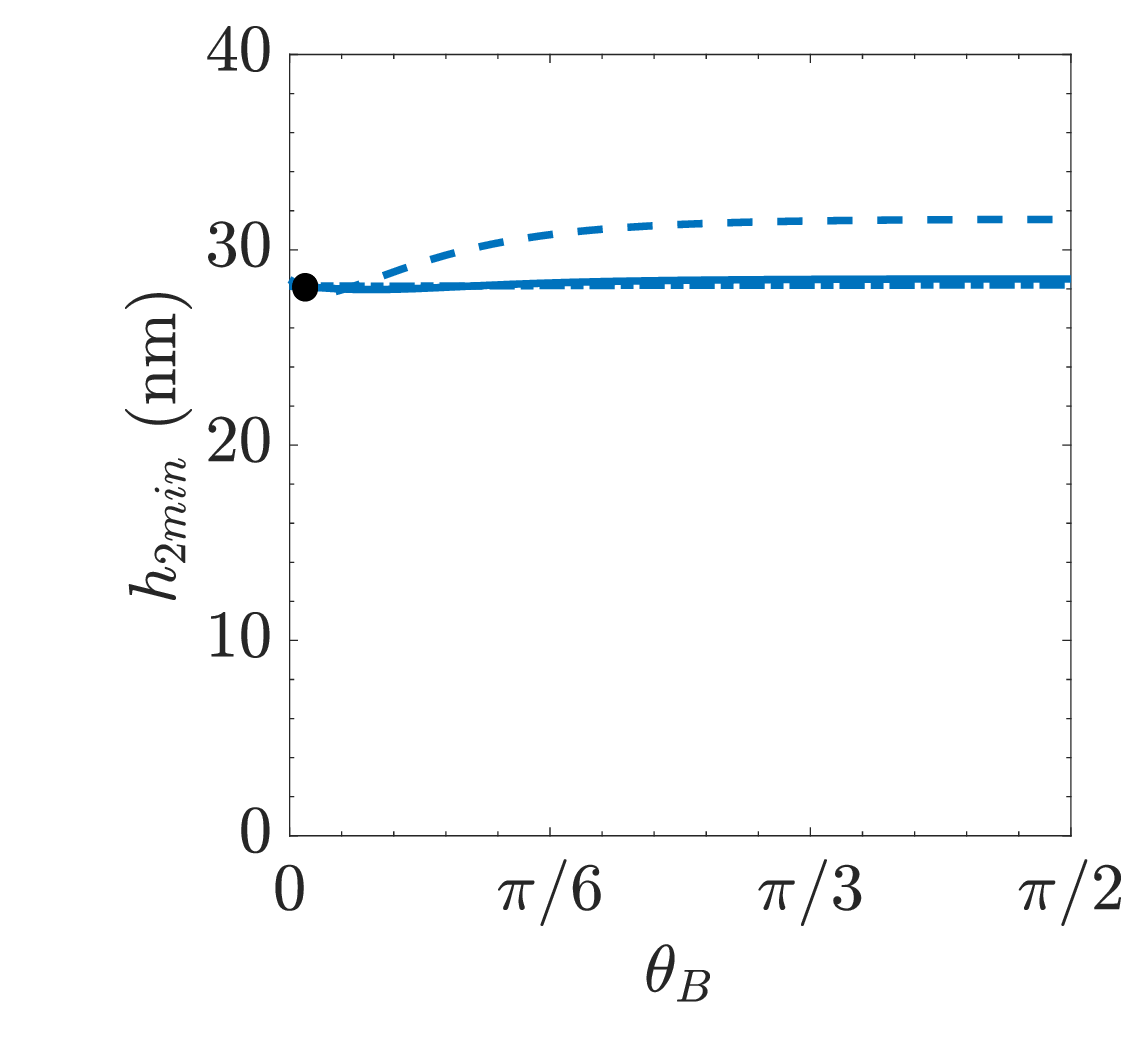}
\includegraphics[width=0.32\textwidth]{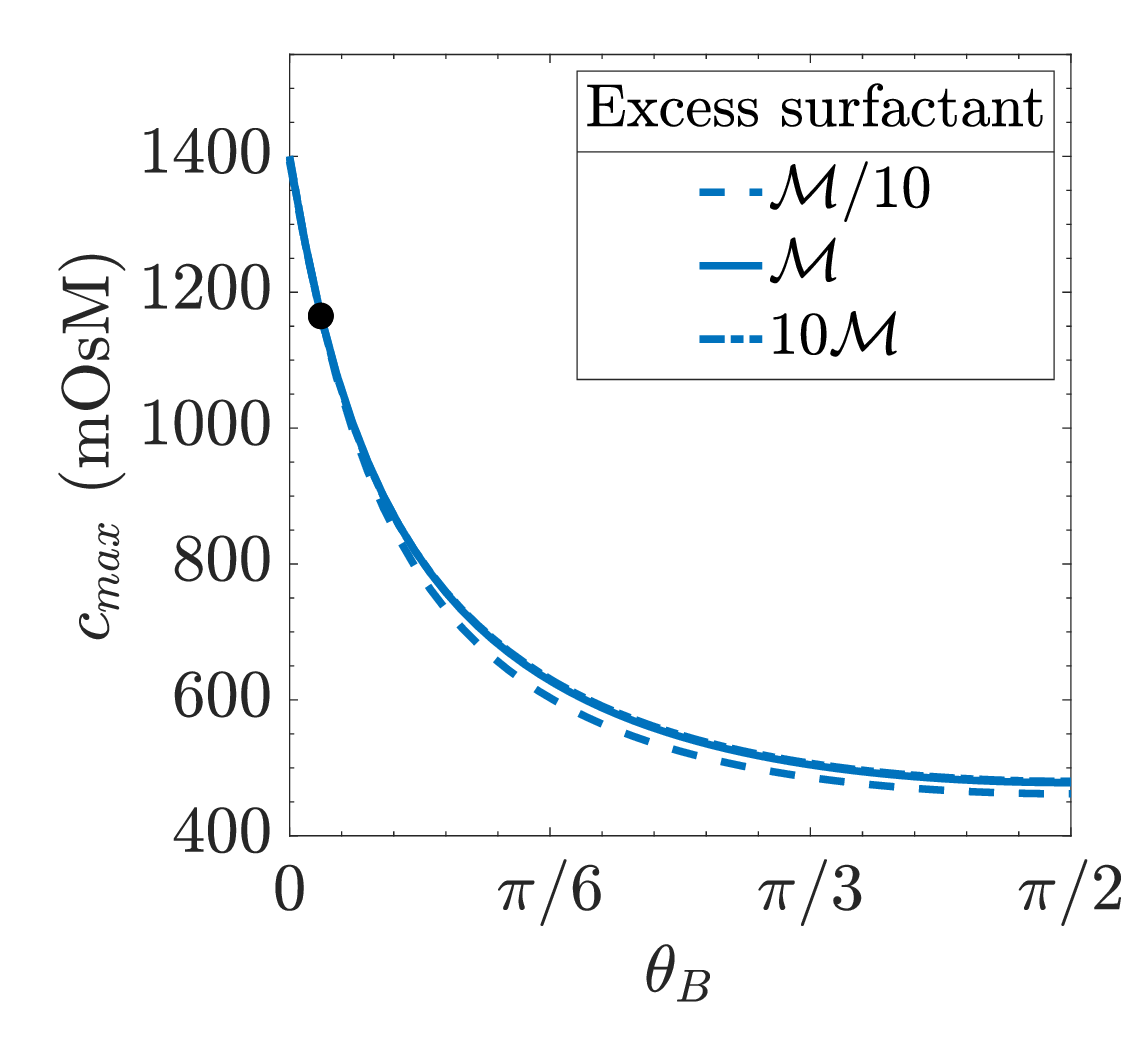}
\caption{Extreme values of AL thickness, osmolarity, and LL thickness at the final time as $\theta_B$ is varied from $0$ to $\pi/2$, for three values of the Marangoni number $\mathcal{M}$. To the left of the black dots, TBUT occurs before 60 s; to the right, the simulations end at 60 s. The first row corresponds to the initial condition with a valley in the LL described in Section \ref{sec:evap_ic}. The second row corresponds to the initial condition with a local excess of surfactant described in Section \ref{sec:maragoni_ic}. }\label{fig:thetaB_varyM}
\end{figure}

We now consider the effect of the Marangoni number on tear film dynamics as the director angle $\theta_B$ varies from 0 to $\pi/2$. Figure \ref{fig:thetaB_varyM} shows the extreme values of AL thickness, osmolarity, and LL thickness for three values of the Marangoni number: $\mathcal{M}$ (the default value of 187.7), $\mathcal{M}/10$, and $10\mathcal{M}$. Solutions are also shown for the initial conditions of Section \ref{sec:maragoni_ic}, the local excess of surfactant; these are described in further detail below in Section \ref{sec:maragoni_surf}. 

We see that when $\mathcal{M}\rightarrow\mathcal{M}/10$, the Marangoni effect is weakened, and there are no cases of TBU within 60 s. The LL is the most mobile in this case, as healing flow (driven by capillarity) allows the valley in the LL to thicken from its initial minimum of 5 nm, to local minima of 14.7 nm to 33.3 nm, depending on the angle of the director. Osmolarity decreases monotonically from 1323 mOsM to 467 mOsM. 

For the default Marangoni number, TBU occurs before 60 s for $\theta_B \lesssim \pi/17$. Around this critical director angle the LL has some mobility, with the lipid valley thickening to a maximum of 17.2 nm; however, for $\theta_B\rightarrow\pi/2$, the LL moves very little from its initial profile.

When $\mathcal{M}\rightarrow10\mathcal{M}$, TBU occurs before 60 s for $\theta_B\lesssim\pi/9$. The LL remains tangentially immobile for all angles of the director. Osmolarity decreases over the range of the $\theta_B$, but its decline is slower than for the default Marangoni number.  

\subsubsection{Varying the evaporative resistance parameter}\label{sec:r0_gaussian}

\begin{figure}[htbp]%
\centering
\includegraphics[width=0.8\textwidth]{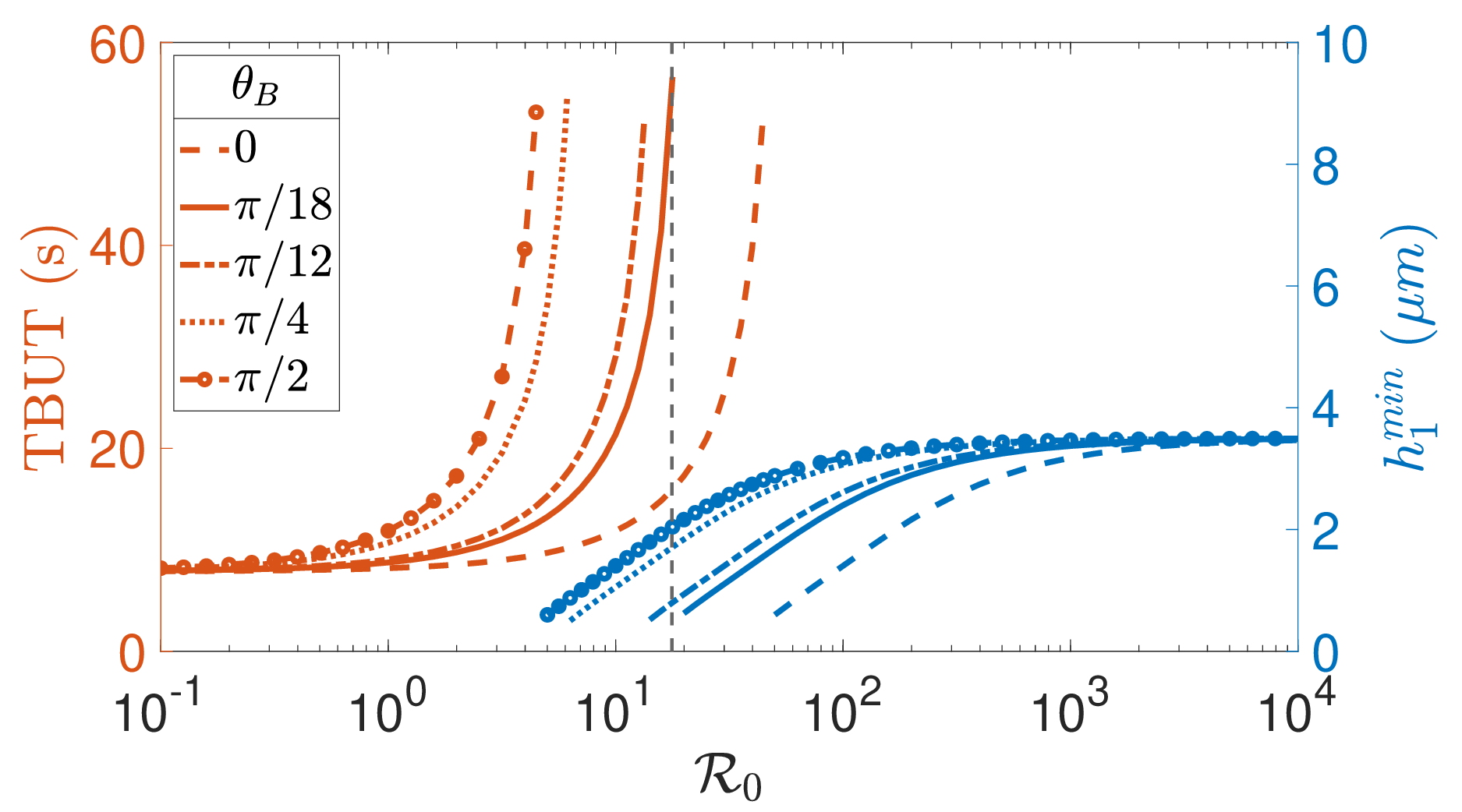}
\caption{Tear film breakup time or the minimum thickness of the AL after 60 s as a function of the evaporation resistance parameter $\mathcal{R}_0$ for the initial condition with a valley in the LL. The black dashed vertical line denotes the default value of $\mathcal{R}_0=17.68$.} \label{fig:varyR_gaussian}
\end{figure}

Figure \ref{fig:varyR_gaussian} shows the effect of varying $\mathcal{R}_0$, the evaporative resistance parameter of the LL. The default value $\mathcal{R}_0=17.68$ is based on a nonlinear least squares fit to lipid thickness and permeability measurements performed by \citet{stapf2017duplex}; this is shown with the vertical dashed line. With this level of resistance, only a director angle of $\theta_B=0$ results in a tear breakup time (TBUT) before 60 s. However, if $\mathcal{R}_0$ is decreased by a factor of 10, then TBU occurs in under 15 s for all values of the director angle. As the evaporative resistance parameter is increased, the minimum AL thickness tends toward the initial value of 3.5 $\mu$m. 

\subsubsection{The effect of liquid crystal viscosities on thinning}\label{sec:LCparams}

\begin{figure}[htbp]%
\centering
\includegraphics[width=0.8\textwidth]{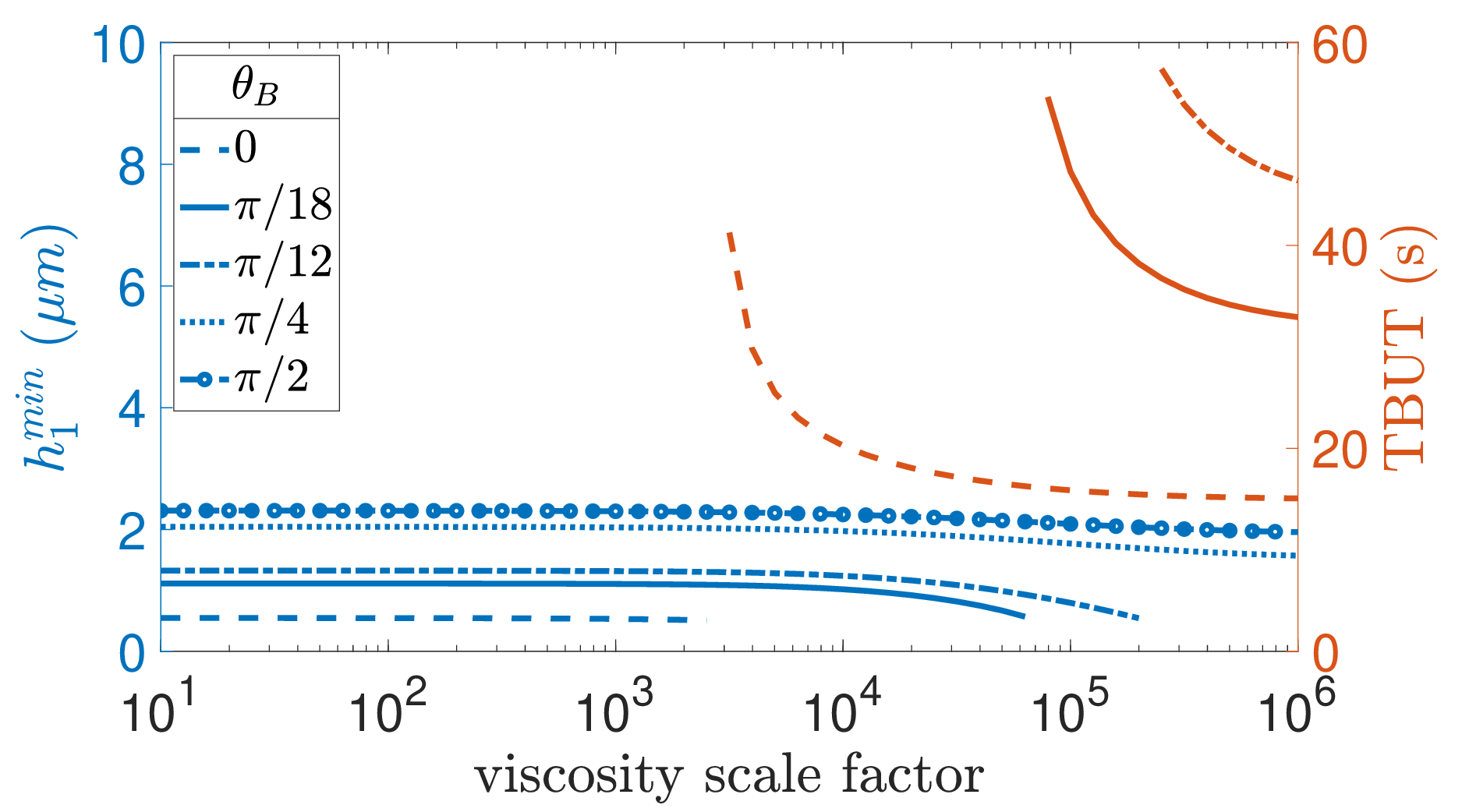}
\includegraphics[width=0.8\textwidth]{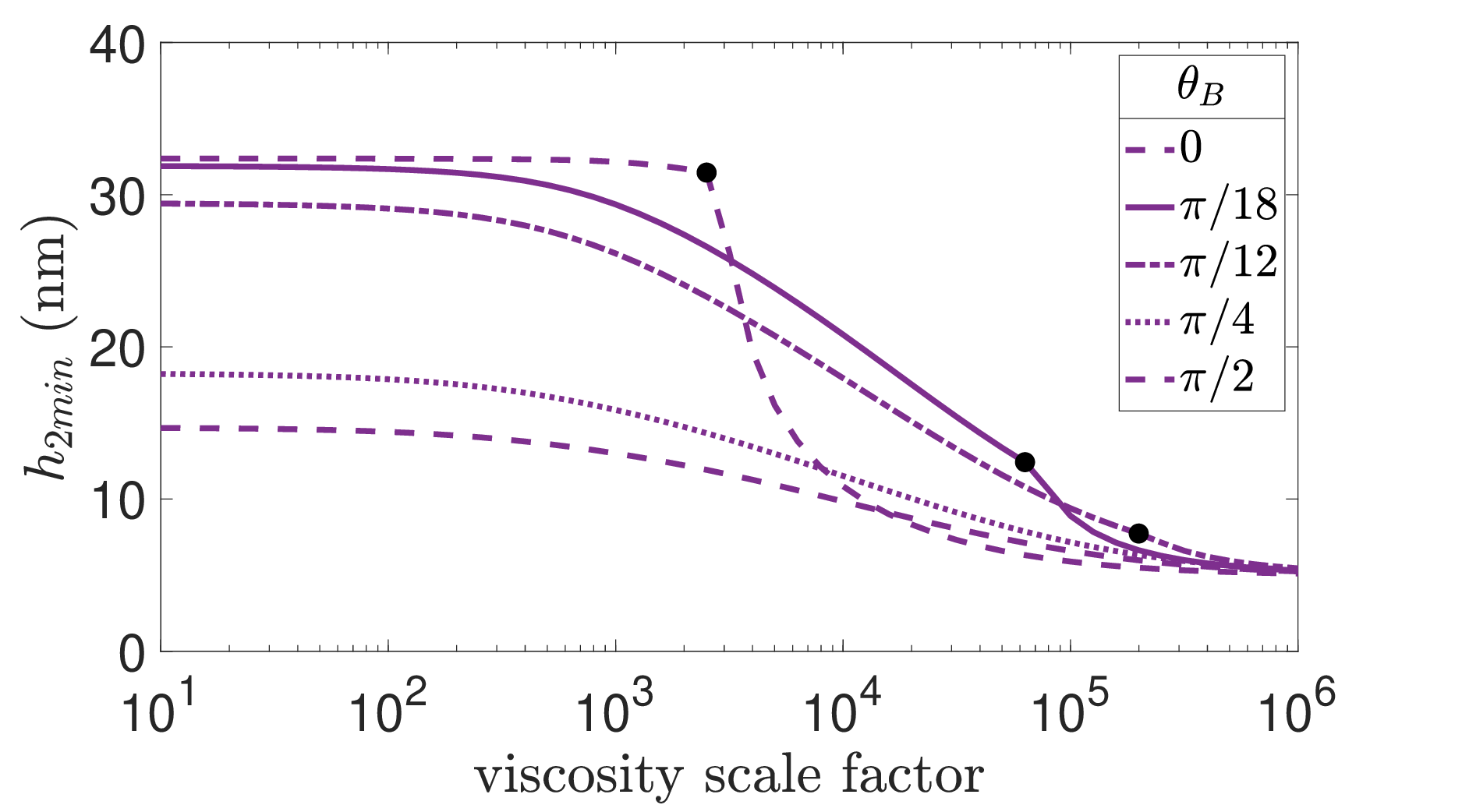}
\caption{Minimum thickness of the AL after 60 s or TBUT as a function of the scale factor applied to the 5CB liquid crystal parameters. To the left of the black dots, TBUT occurs before 60 s; to the right, the simulations end at 60 s.} \label{fig:varymu2_gaussian}
\end{figure}

We have chosen to set the Leslie viscosities to three times the values of 5CB, a relatively well studied liquid crystal. By doing this, we match the dynamic viscosity of the LL, $\mu_2$, to previously used value of 0.1 Pa s \cite{bruna2014influence,stapf2017duplex}. However, there is much uncertainty around the values for the Leslie viscosities as they have not been measured for the LL, and using a scale factor of 3 is only one possible choice. Thus, in Figure \ref{fig:varymu2_gaussian}, we consider the effect of a wide range of scalings on the minimum values for both the aqueous and LL. Note that because we are scaling the dimensional quantities, in the nondimensional form we are effectively scaling $\mu_2$. Here we reduce the Marangoni number by a factor of 10 ($\mathcal{M}\rightarrow\mathcal{M}/10$) in order to isolate the effect of the liquid crystal viscosities on tear film thinning and breakup. 

For values of $\mu_2$ that are increased by less than a factor of 1000, there is little change in either the minimum AL thickness or LL thickness at the final time. For values of $\mu_2$ more than a factor of 1000 larger than default, we begin to see breakup within 60 s for small values of $\theta_B$. However, for larger director angles, the AL is fairly unaffected. On the other hand, for small values of $\mu_2$ we observe a much more mobile LL for all orientations of the director.  As $\mu_2$ increases, the LL becomes  immobile, and the minimum thickness remains close to the initial value of 5 nm. 

\begin{figure}[htbp]
    \centering
    \begin{minipage}{.45\textwidth}
        \centering
        \includegraphics[width=\linewidth]{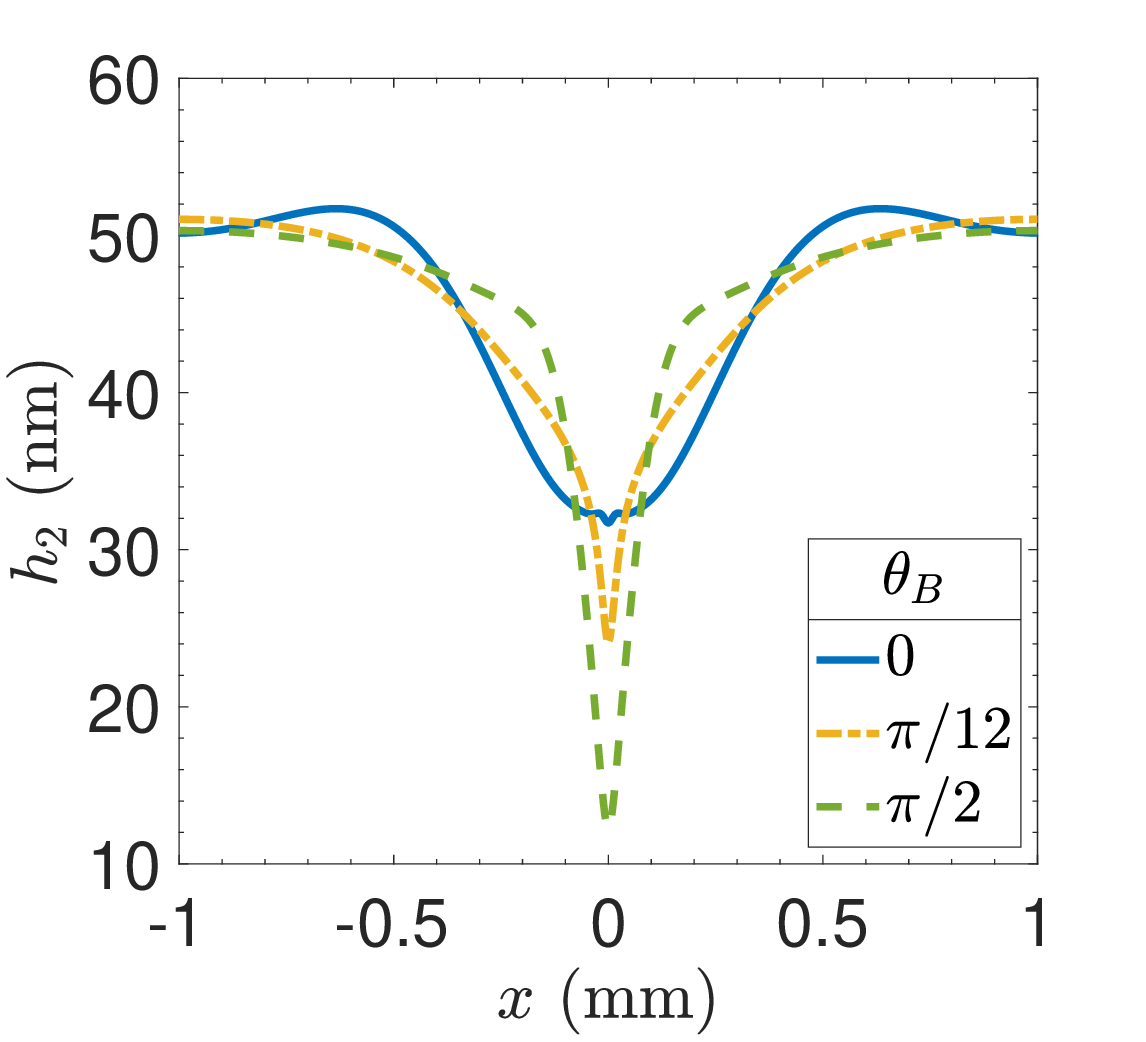}
        \caption{Solutions for the LL at 60 s when the initial condition has a valley in the LL and $\mu_2\rightarrow2000\,\mu_2$, and $\mathcal{M}\rightarrow\mathcal{M}/10$.}
    \label{fig:varymu2_finaltime}
    \end{minipage}
    \hfill
    \begin{minipage}{0.45\textwidth}
        \centering
\includegraphics[width=\linewidth]{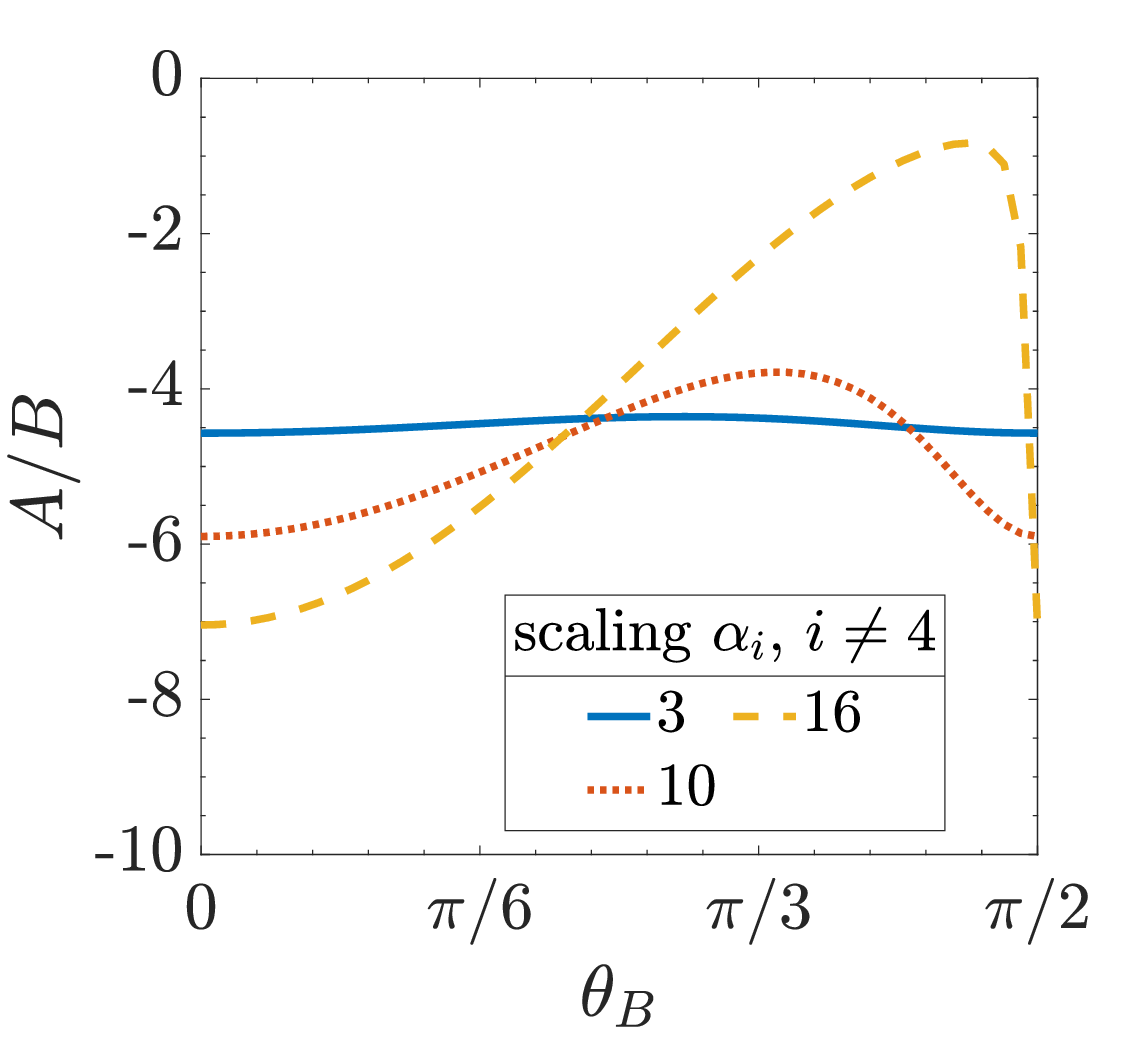}
\caption{The coefficient $A/B$ as $\theta_B$ varies for scalings of 3, 10, and 16 on the Leslie viscosities $\alpha_i,\,i=1,\cdots,6,\,i\neq4$  for the liquid crystal 5CB. } \label{fig:scaling}
    \end{minipage}
\end{figure}

Figure \ref{fig:varymu2_finaltime} shows solutions at the final time when $\mu_2$ is increased by a factor of 2000. At this value, we see the most variation in LL thickness with respect to the director angle. For clarity, we omit $\theta_B=\pi/18$ and $\pi/4$. When $\theta_B=\pi/2$, the minimum LL thickness is just over 12 nm, while when $\theta_B=0$, the effects of capillarity result in healing flow, which increases the minimum thickness to 31.7 nm. For the intermediate value of $\theta_B=\pi/12$, the LL experiences some healing flow with a minimum of 24.1 nm, but mobility is reduced.

It is natural to explore an alternative scenario: keeping a scaling of three on $\alpha'_4$, which maintains $\mu_2=0.1$ Pa s, while scaling the other Leslie viscosities, $\alpha'_i,\,i=1,\cdots,6,\,i\neq4$. By doing this, we are changing the coefficient $A/B$ on the extensional term in Equation (\ref{eq:LLforcebalance}). Note that as $\alpha'_i\rightarrow0,\,i=1,\cdots,6,\,i\neq4$, we return to the Newtonian limit, where $A/B=4$. However, we find that if we increase the scaling too much (more than a factor of 16), the coefficient $A/B$ becomes positive for some values of $\theta_B$, making the problem ill-posed. In addition, the effect of the scalings in this feasible range is negligible for all variables. If we look at the coefficient $A/B$ for scalings, shown in Figure \ref{fig:scaling}, we see that the change on the coefficient is very little, even with the maximum scale factor of 16. 

\clearpage
\subsection{Marangoni-driven thinning} \label{sec:marangoni}
\begin{figure}[htbp]%
\centering
\includegraphics[width=0.49\textwidth]{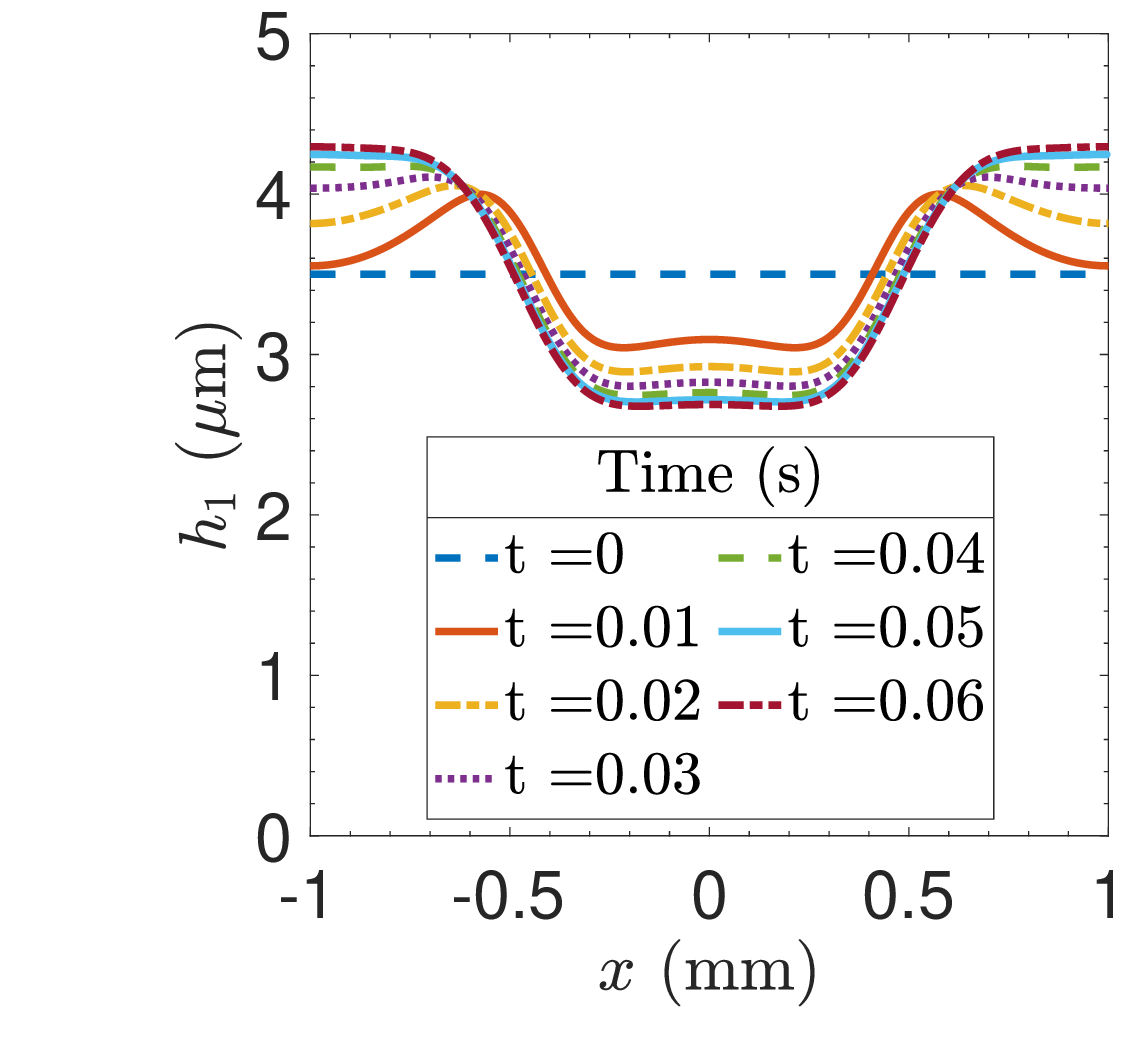}
\includegraphics[width=0.49\textwidth]{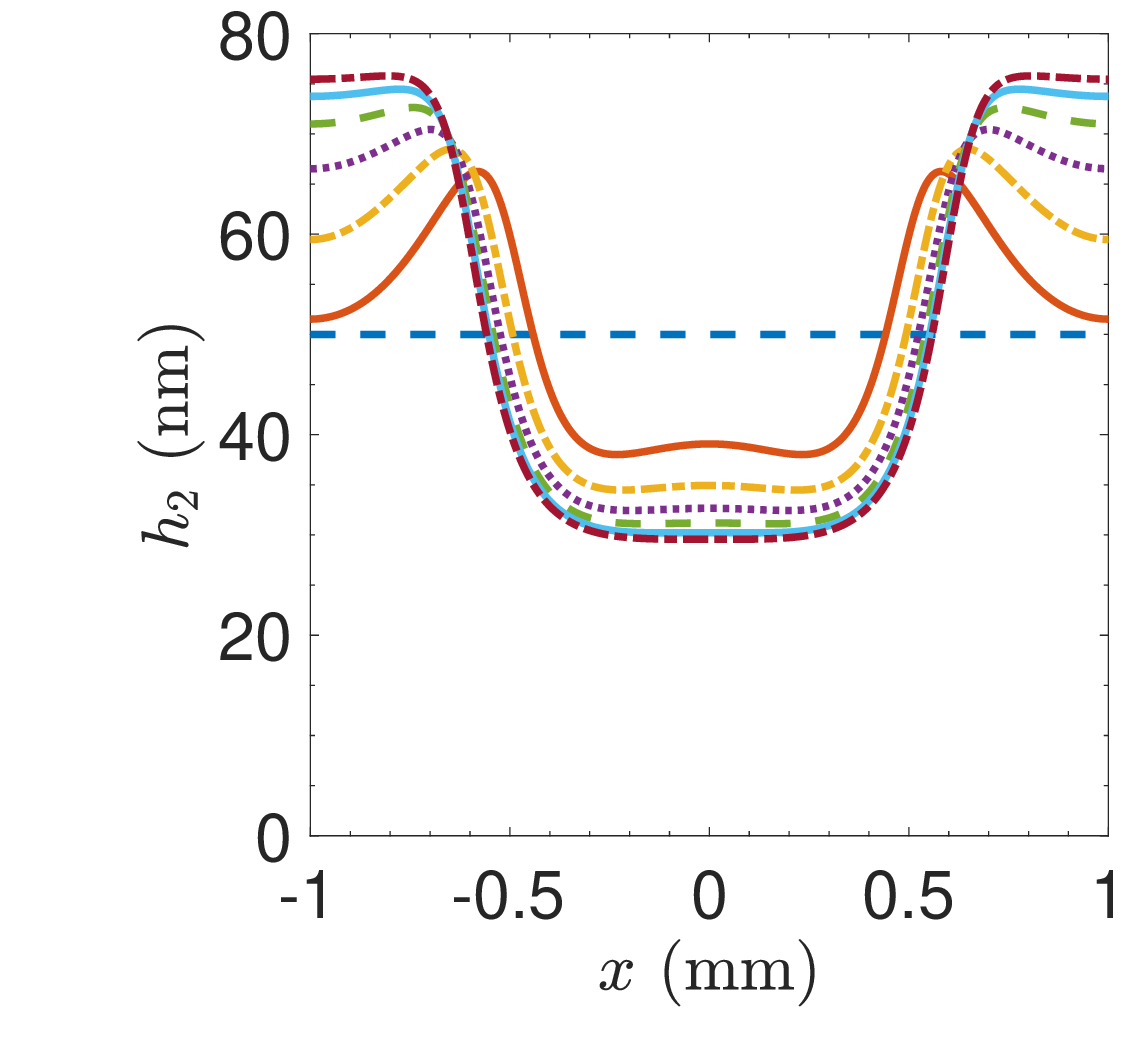}
\includegraphics[width=0.49\textwidth]{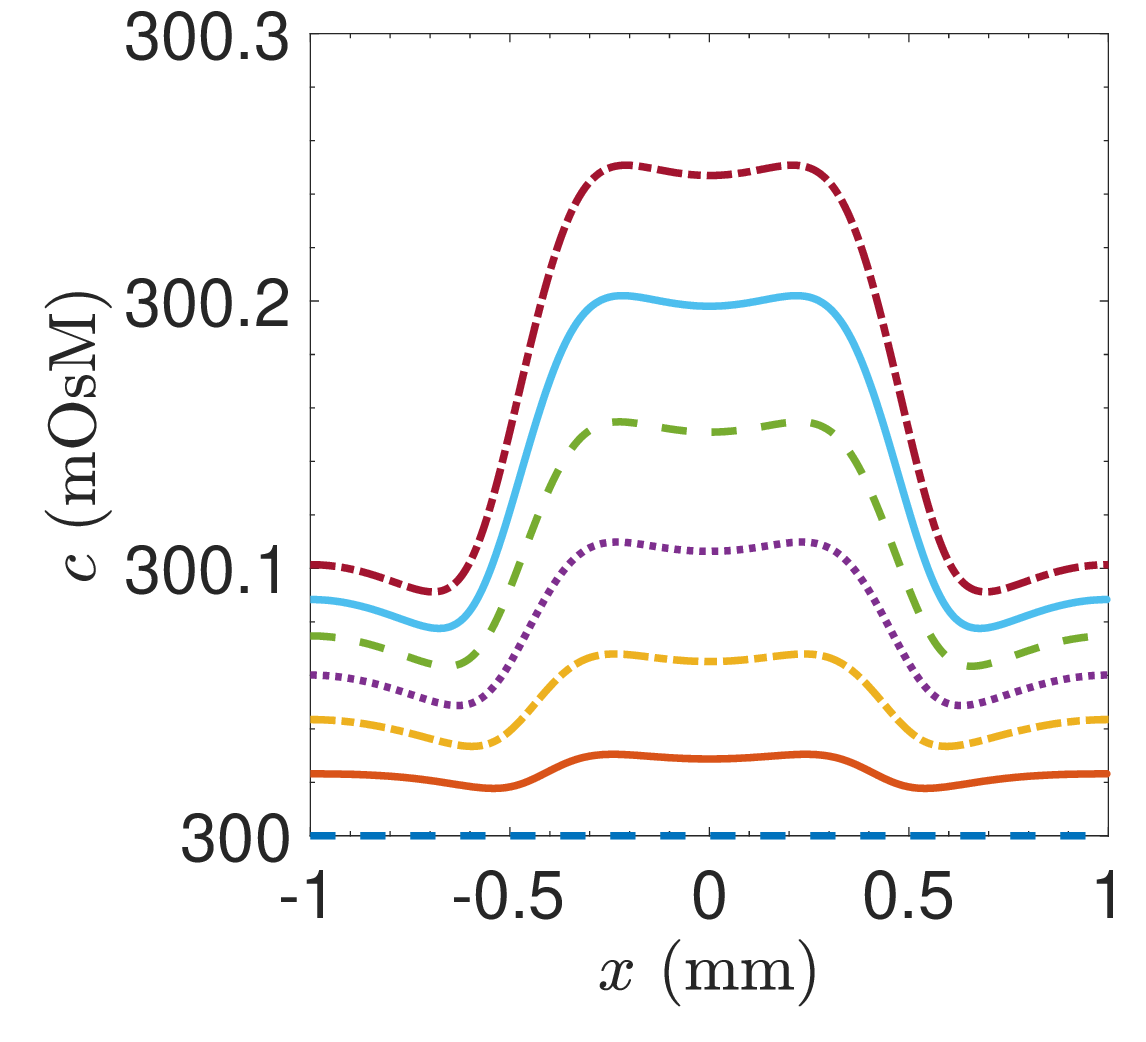}
\includegraphics[width=0.49\textwidth]{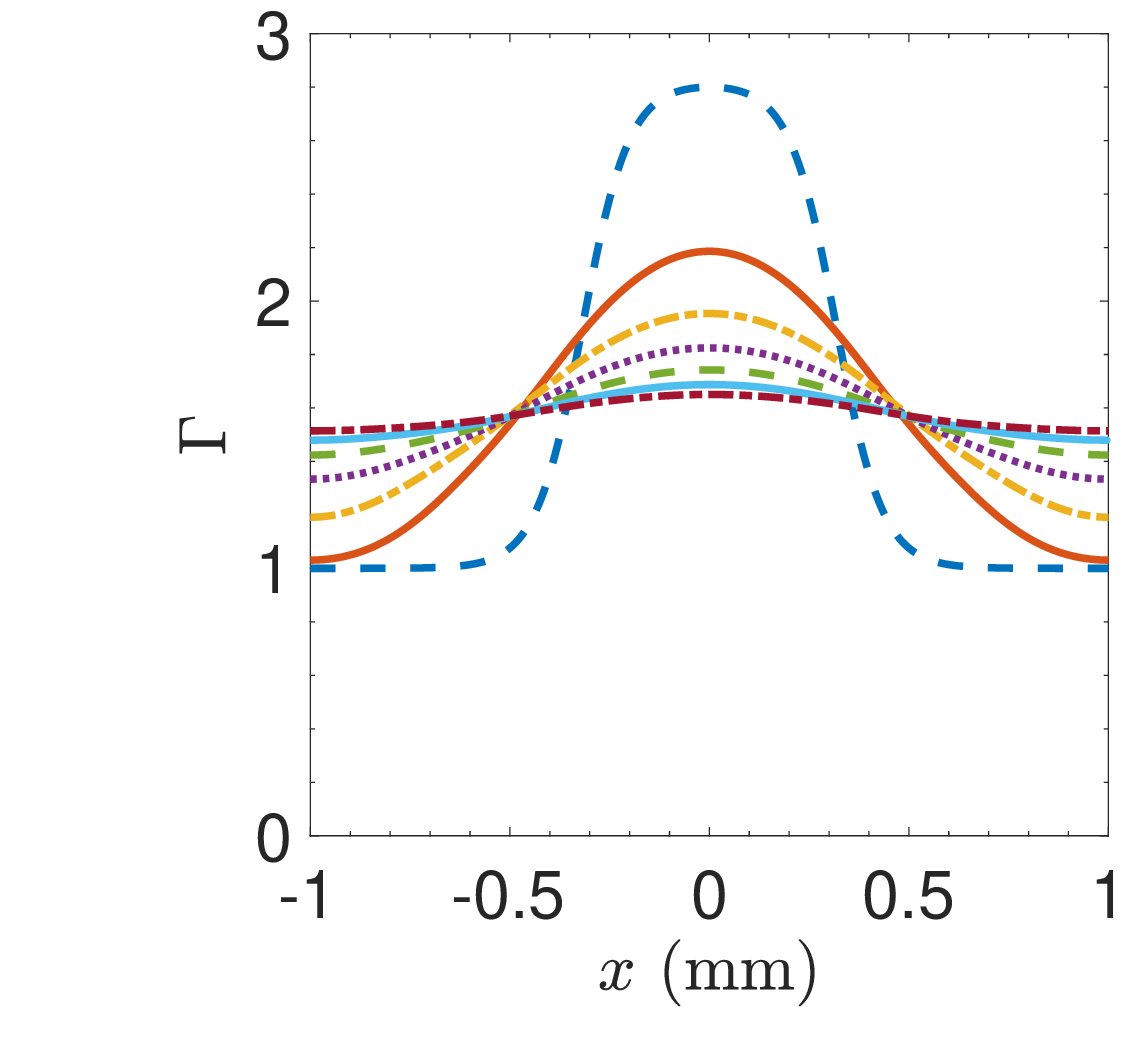}
\caption{Solutions on a short time scale for spatially-uniform initial conditions except for a local excess of surfactant. The director angle is $\theta_B=\pi/2$.  This figure illustrates typical dynamics of the dependent variables in space and time. The surface gradients of surfactant drive all the dynamics.}\label{fig:shorttime}
\end{figure}

\begin{figure}[htbp]%
\centering
\includegraphics[width=0.49\textwidth]{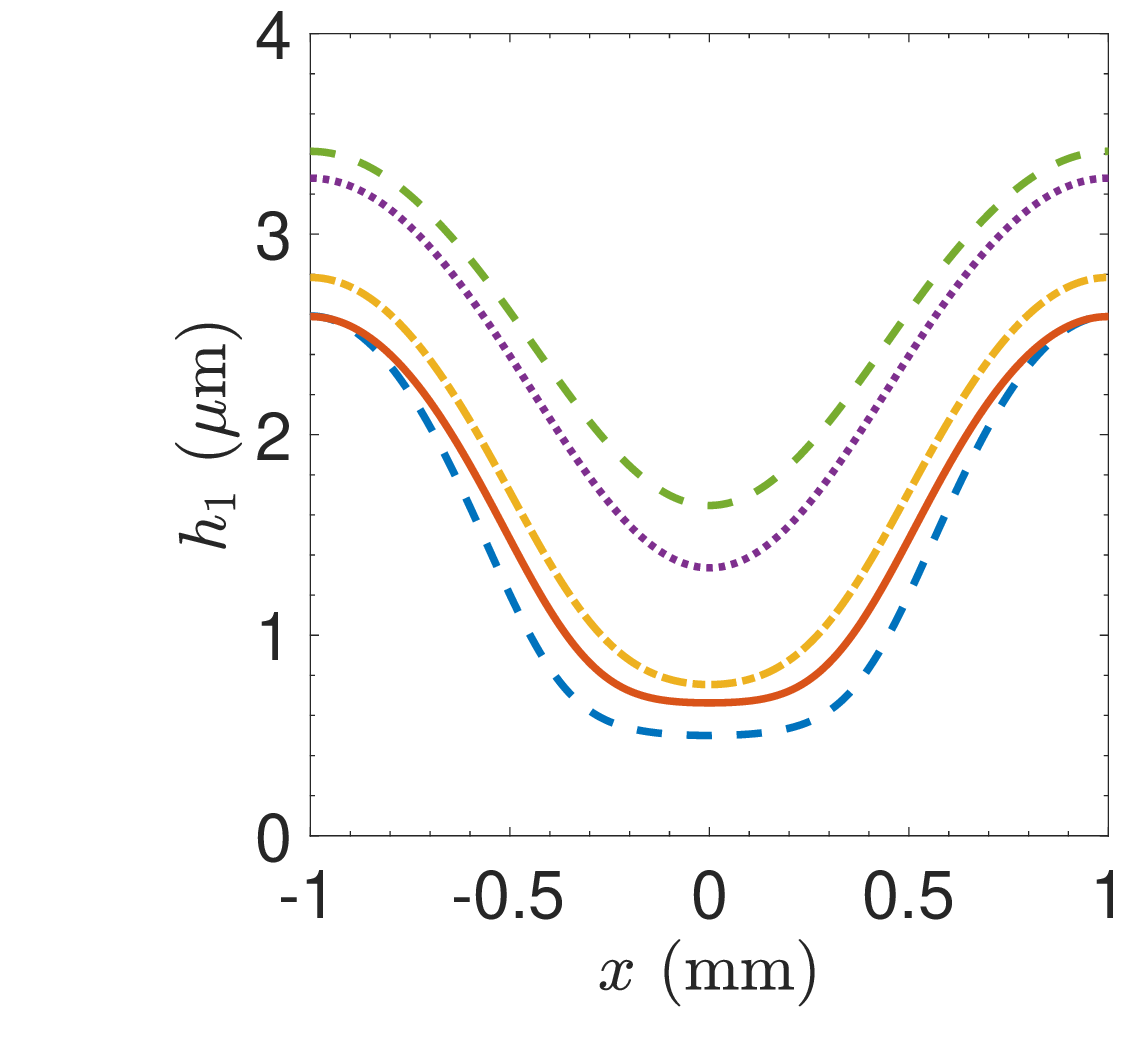}
\includegraphics[width=0.49\textwidth]{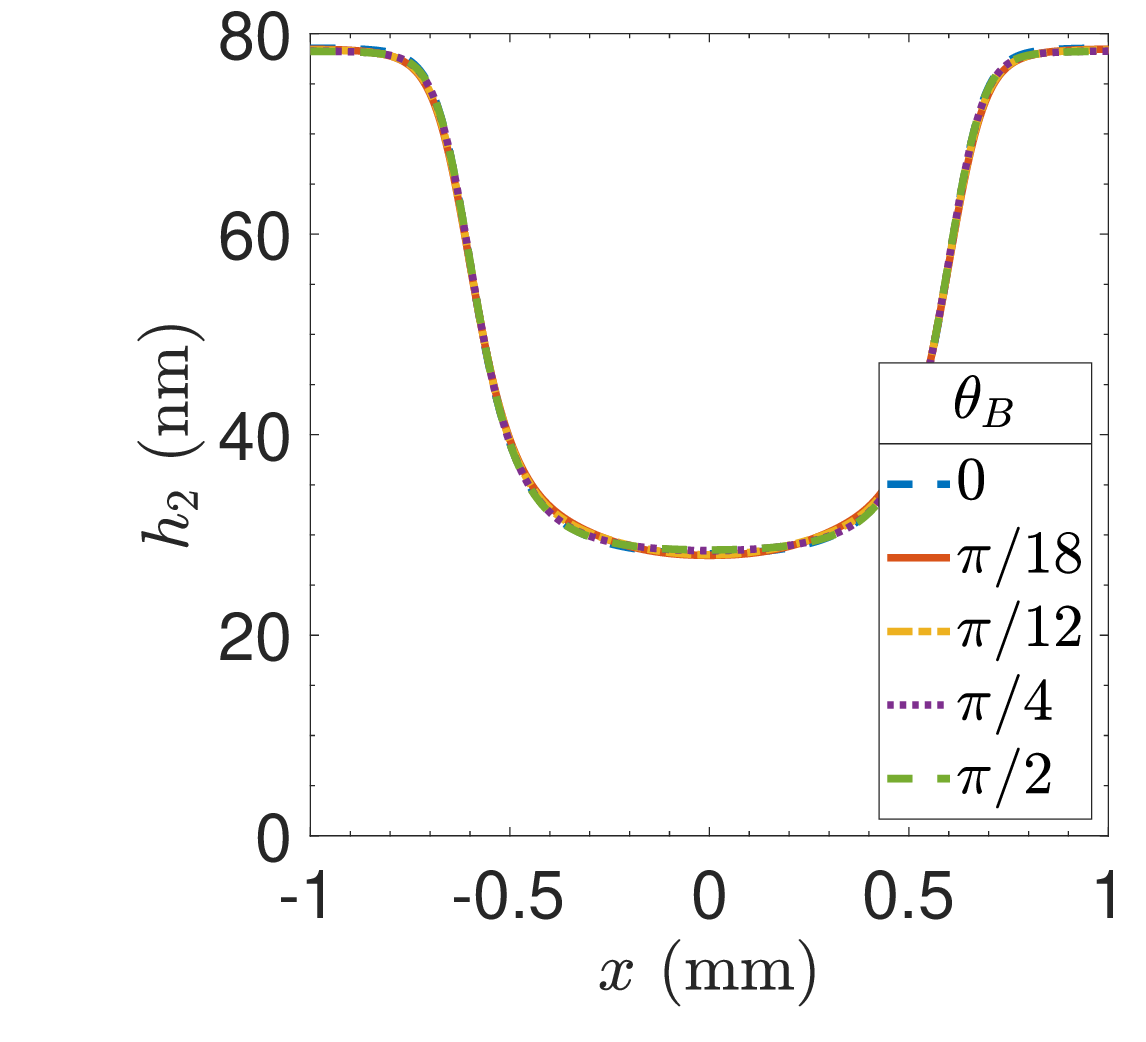}
\includegraphics[width=0.49\textwidth]{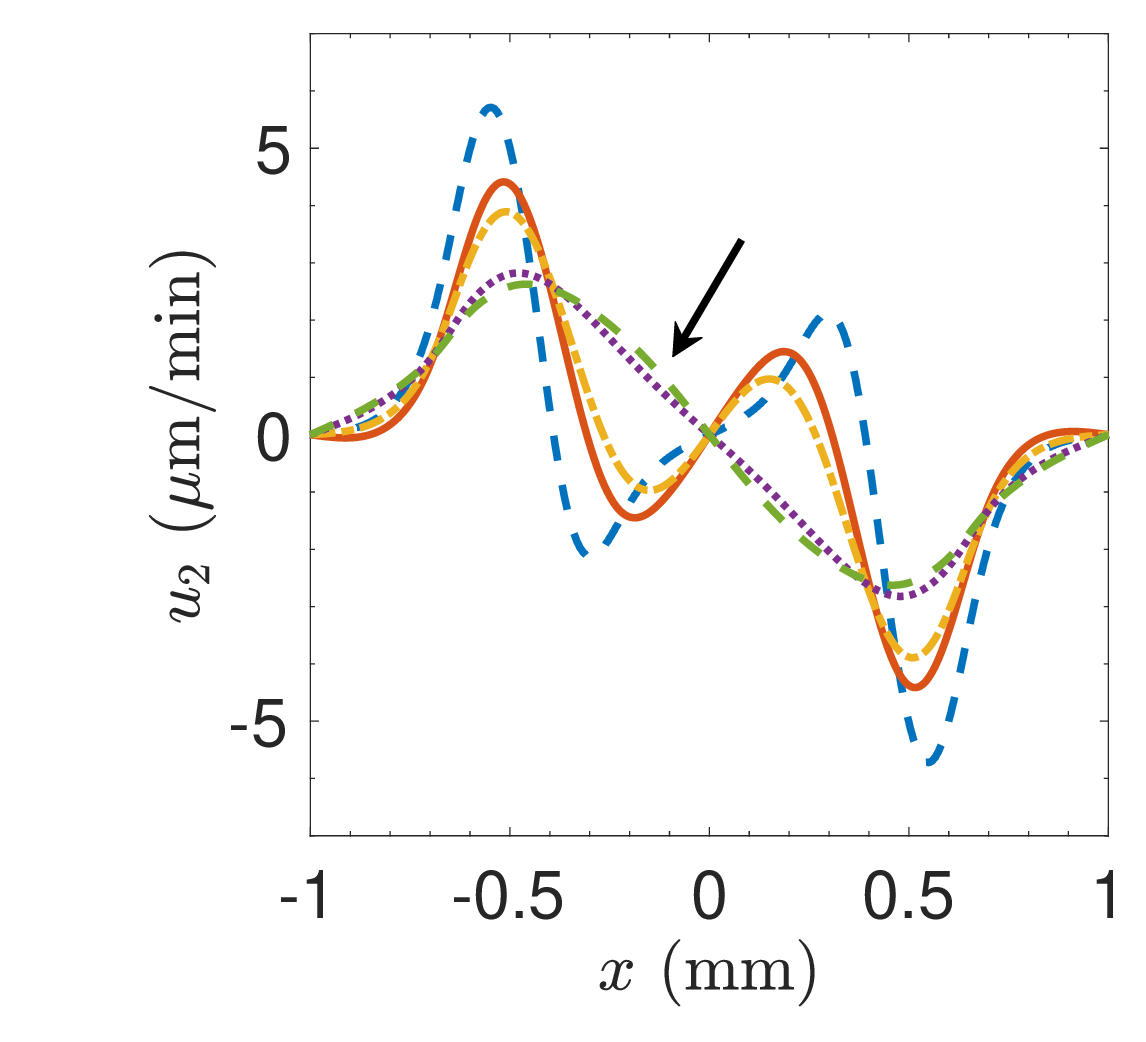}
\includegraphics[width=0.49\textwidth]{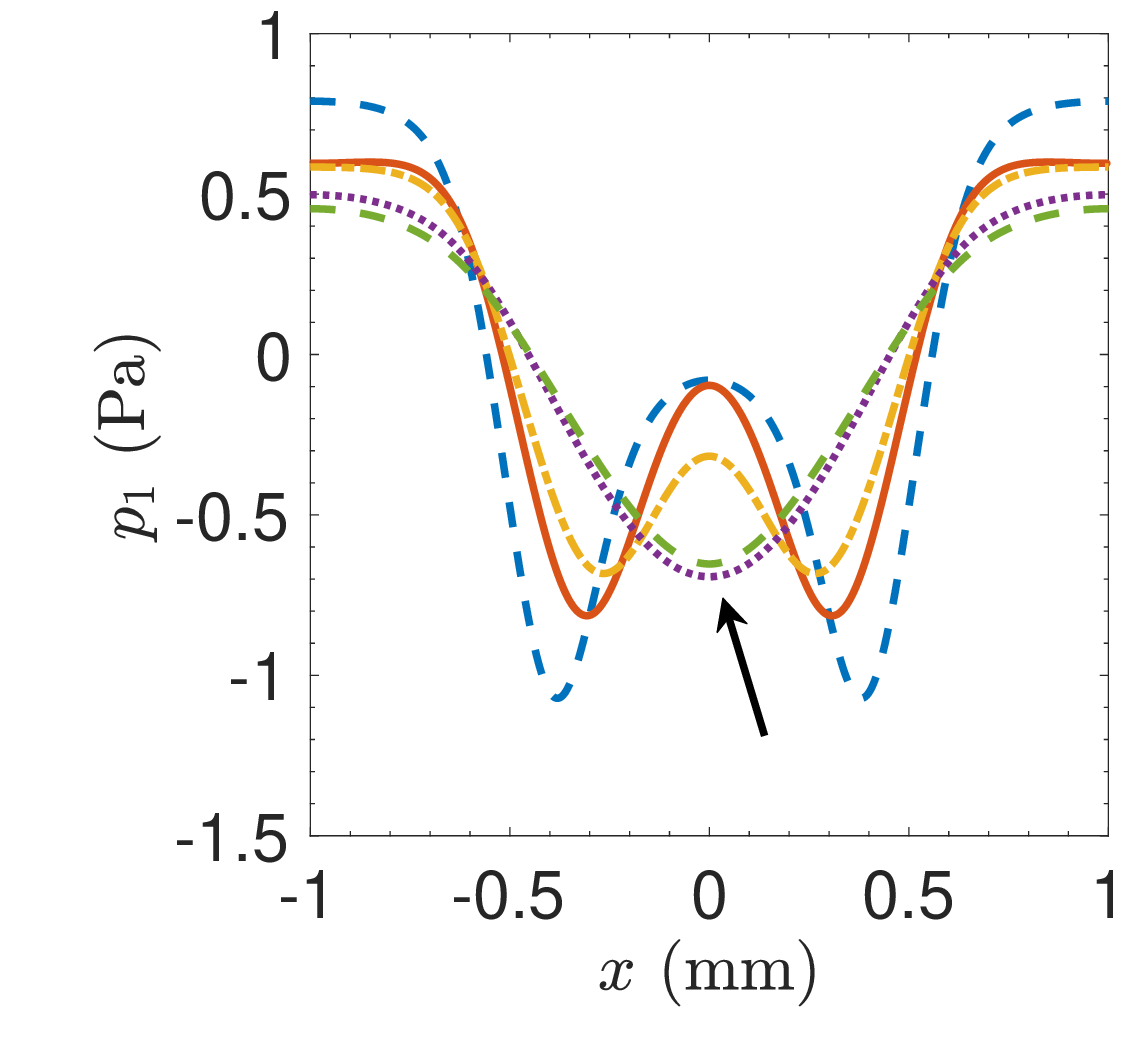}
\includegraphics[width=0.49\textwidth]{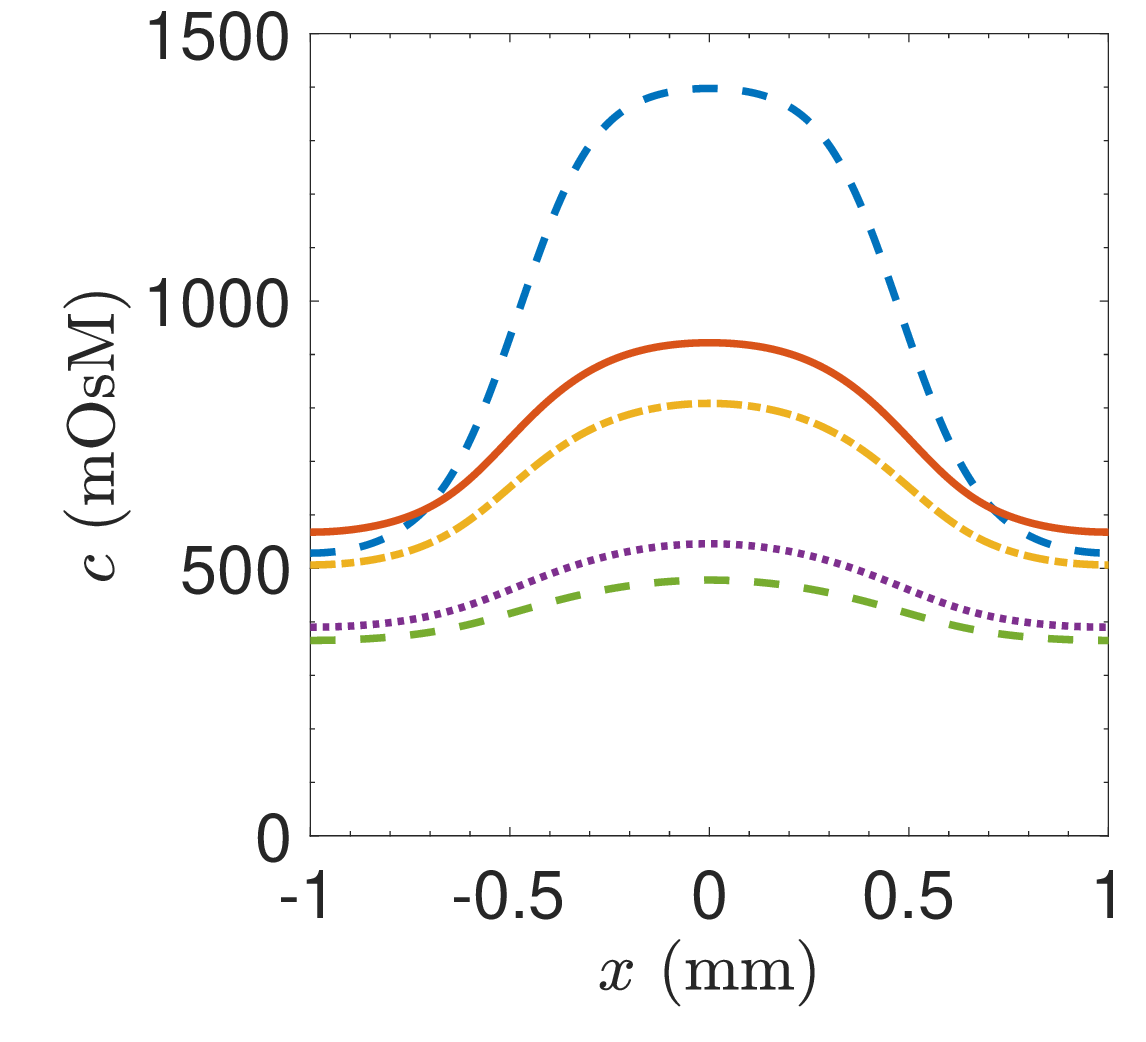}
\includegraphics[width=0.49\textwidth]{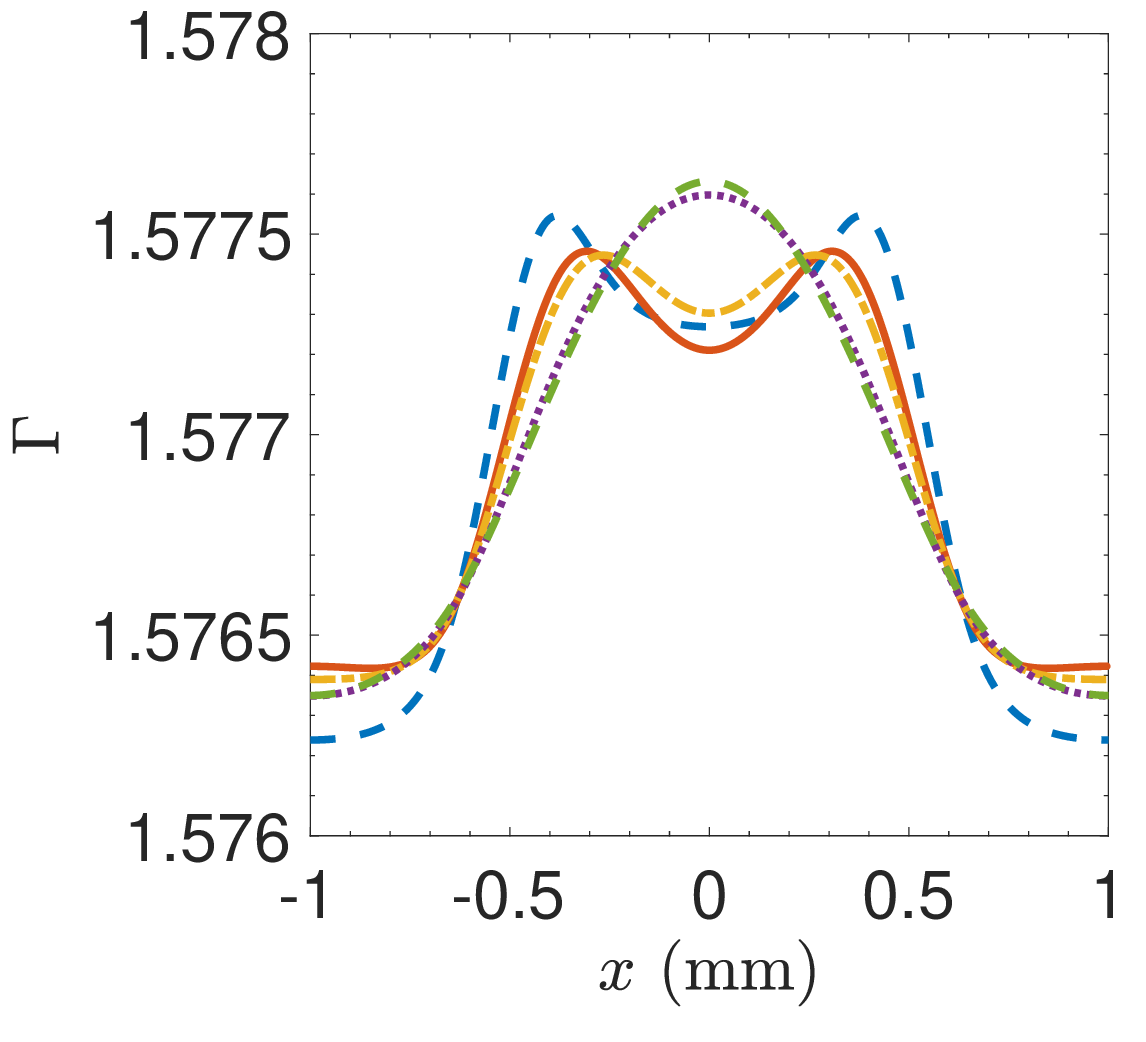}
\caption{Solutions at the final time for spatially-uniform initial conditions except for a local excess of surfactant. When $\theta_B=0$, breakup occurs at 16.6 s; for all other values of $\theta_B$ the final time is 60 s. The LL profiles (upper right) are nearly identical for all $\theta_B$.}\label{fig:longtime}
\end{figure}

We now turn to the initial conditions of Section \ref{sec:maragoni_ic}, in which concentration gradients drive the tear film dynamics. The Marangoni effect results in rapid tangential flow; thus we consider solutions on a short timescale, shown in Figure \ref{fig:shorttime}, and a longer timescale, shown in Figure \ref{fig:longtime}. Figure \ref{fig:shorttime} shows the evolution of the variables over the first 0.06 s, when $\theta_B=\pi/2$. These solutions are representative of the dynamics of tear film thinning due to surfactant concentration gradients regardless of the orientation of the liquid crystal molecules; the director angle $\theta_B$ has little effect on the short timescale. The large surfactant gradients immediately cause rapid thinning in both the lipid and aqueous layers. As the local excess of surfactant flattens, the Marangoni effect concludes. The LL thins dramatically, with the valley thickness roughly half of the initial value of 50 nm. Fluid is pulled away from the excess surfactant, and builds up on the edges of the domain. The pattern of thinning is very similar in the AL, although to a lesser degree. During this initial redistribution of fluid, the LL fluid velocity is very large in magnitude (around $3\times 10^5\,\mu$m/min), but quickly slows to the range of $\pm 5\,\mu$m/min throughout the domain (not shown). Similarly, AL pressure (not shown) drops in magnitude from a range of $\pm 3$ Pa during the first 1 s, to $\pm 1$ Pa, and thinning continues over a longer timescale. 

Figure \ref{fig:longtime} shows solutions at the final time for fives values of $\theta_B$ ranging from 0 to $\pi/2$. Breakup conditions are reached when $\theta_B=0$ by 16.6 s; for all other values of $\theta_B$, the final time of 60 s is reached. The surfactant concentration  remains effectively uniform around 1.58 times the original background concentration, and over the longer timescale it is the evaporative resistance in the LL that influence thinning dynamics. When $\theta_B = \pi/2$ or $\pi/4$ thinning becomes more localized in the center of the domain, while smaller values of $\theta_B$ maintain a wider valley of thinning through time. For larger values of $\theta_B$, we observe that negative pressure in the center of the AL domain pulls fluid in, and helps to reduce thinning. This results in an overall thicker AL at the final time. Arrows in the velocity and pressure plots of Figure \ref{fig:longtime} point to the difference in behavior when $\theta_B=\pi/4$ or $\pi/2$ as compared to smaller director angles.

 On the other hand, the LL profiles are indistinguishable from each other at the final time and look very similar to the solution at 0.06 s shown in Figure \ref{fig:shorttime}. Virtually all of the movement in the LL happens within the first second; this immobilization of the fluid surface due to the Marangoni effect has been studied in other contexts such as vertical drainage in \citet{braun2000limiting} and remobilization of droplet surfaces in \citet{stebe1994remobilizing}. 
 
 Osmolarity in the AL increases for all values of $\theta_B$ over the original value of 300 mOsM; however, it is most pronounced when $\theta_B=0$. While the solutions in the AL are not dissimilar for $\theta_B=0$ and $\theta_B=\pi/18$, the osmolarity is very different. When $\theta_B=0$, peak osmolarity reaches nearly 1400 mOsM by 16.6 s, more than one and a half times the peak osmolarity at 60 s when $\theta_B=\pi/18$. When $\theta_B=0$, thinning and TBU happen so rapidly that the solutes do not have enough time to diffuse through the AL. 

\subsubsection{Varying the Marangoni number}\label{sec:maragoni_surf}

The second row of Figure \ref{fig:thetaB_varyM} shows extreme values for AL thickness, osmolarity, and LL thickness as the director angle varies from 0 to $\pi/2$ for three values of the Marangoni number: the default $\mathcal{M}$ (187.7), $\mathcal{M}/10$, and $10\mathcal{M}$. In general, compared to the initial condition with a valley in the LL which are discussed in Section \ref{sec:gaussian_marangoni}, varying the Marangoni number has a much smaller effect on the variables. This is because with localized excess surfactant in the initial condition, the Marangoni effect is already driving dynamics, and our default Marangoni number is strong enough that increasing or decreasing by a factor of 10 does not have a dramatic effect. 

For all three values of the Marangoni number, TBU occurs for $\theta_B<\pi/50$ when the initial condition has a local excess of surfactant. For larger director angles, there is no breakup within 60 s, but the AL thins much more than in the case with the lipid valley, with minimum AL thickness ranging from 1.62 to 1.84 $\mu$m when $\theta_B=\pi/2$. After the initial rapid redistribution of fluid due to the Marangoni effect, the valley formed in the LL has a minimum value of 28 nm when $\theta_B=0$ for all three cases. For $\mathcal{M}$ and $10\mathcal{M}$ the minimum remains the same for all values of the director. For $\mathcal{M}/10$, the minimum increases slightly as the director angle increases, with a minimum value of 31.6 nm when $\theta_B=\pi/2$. Osmolarity between the three cases is very similar, with a maximum difference of 28 mOsM; in fact, the osmolarity curve for $10\mathcal{M}$ is indistinguishable from that of $\mathcal{M}$ in Figure \ref{fig:thetaB_varyM}. 

\subsubsection{Varying the evaporative resistance parameter}\label{sec:r0_surf}
\begin{figure}[htbp]%
\centering
\includegraphics[width=0.9\textwidth]{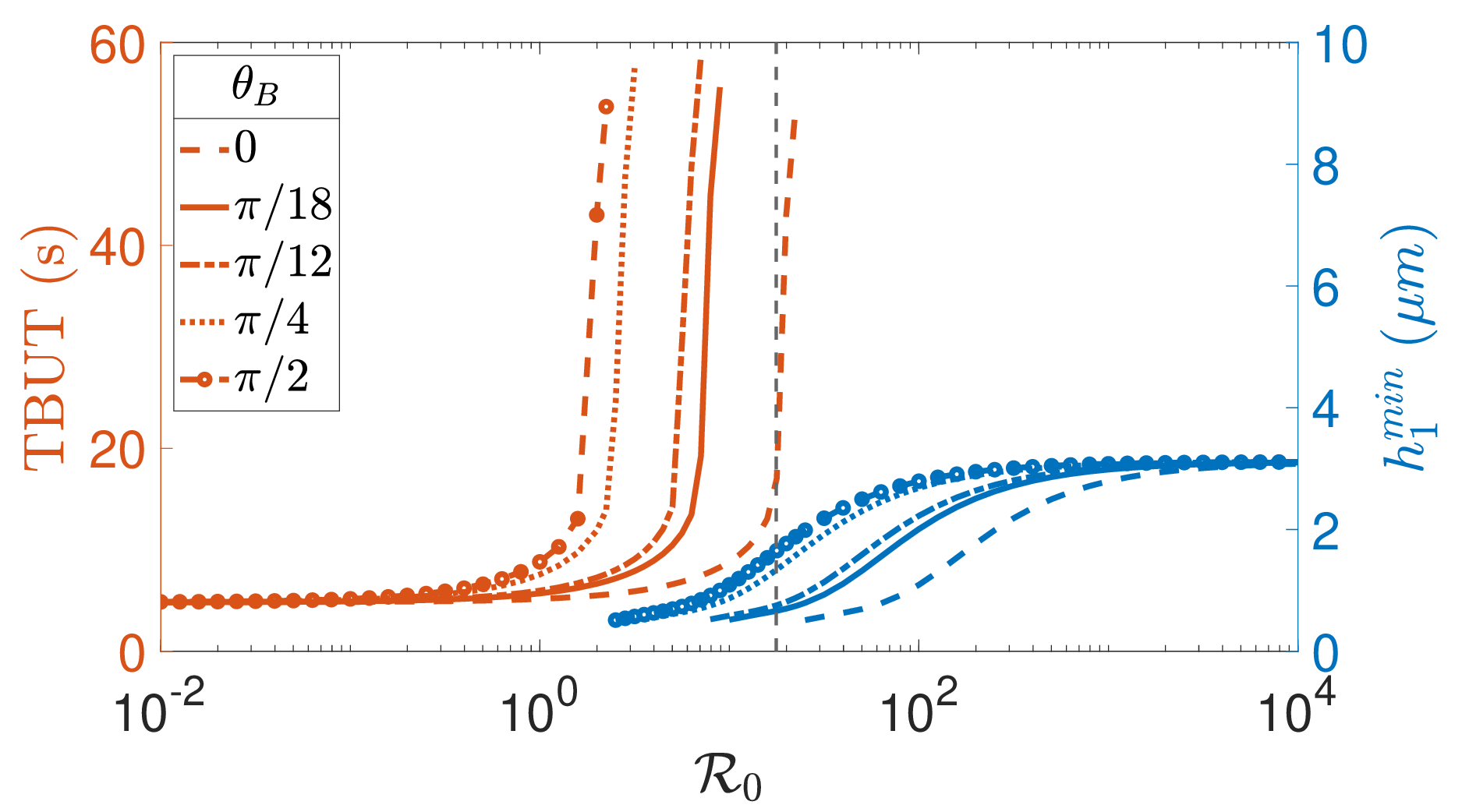}
\caption{Tear film breakup time or the minimum thickness of the AL after 60 s as a function of the evaporation resistance parameter $\mathcal{R}_0$ for the initial condition that has a local excess of surfactant. The black dashed vertical line denotes the default value of $\mathcal{R}_0=17.68$. }\label{fig:Rbreakup}
\end{figure}

We now explore the effect of the evaporative resistance parameter $\mathcal{R}_0$. 
Figure \ref{fig:Rbreakup} shows the TBU conditions as $\mathcal{R}_0$ varies for various orientations of the director. A small evaporative resistance parameter results in rapid thinning of the AL with breakup just under 5 s for all values of $\theta_B$. As $\mathcal{R}_0$ increases, breakup time increases, until eventually the final time of 60 s is reached before breakup occurs. The value of $\mathcal{R}_0$ at which this occurs varies from 2.5 for $\theta_B=\pi/2$ to 25.1 for $\theta_B=0$. The default value, denoted by the dashed vertical line, falls into this range where thinning may result in breakup before 60 s, depending on the director angle. See Figure \ref{fig:longtime} for solutions at the final time.  As $\mathcal{R}_0$ becomes very large, evaporation is diminished, and the minimum value tends toward 3.1 $\mu$m for all values of $\theta_B$. 

\begin{figure}[htbp]%
\centering
\includegraphics[width=0.7\textwidth]{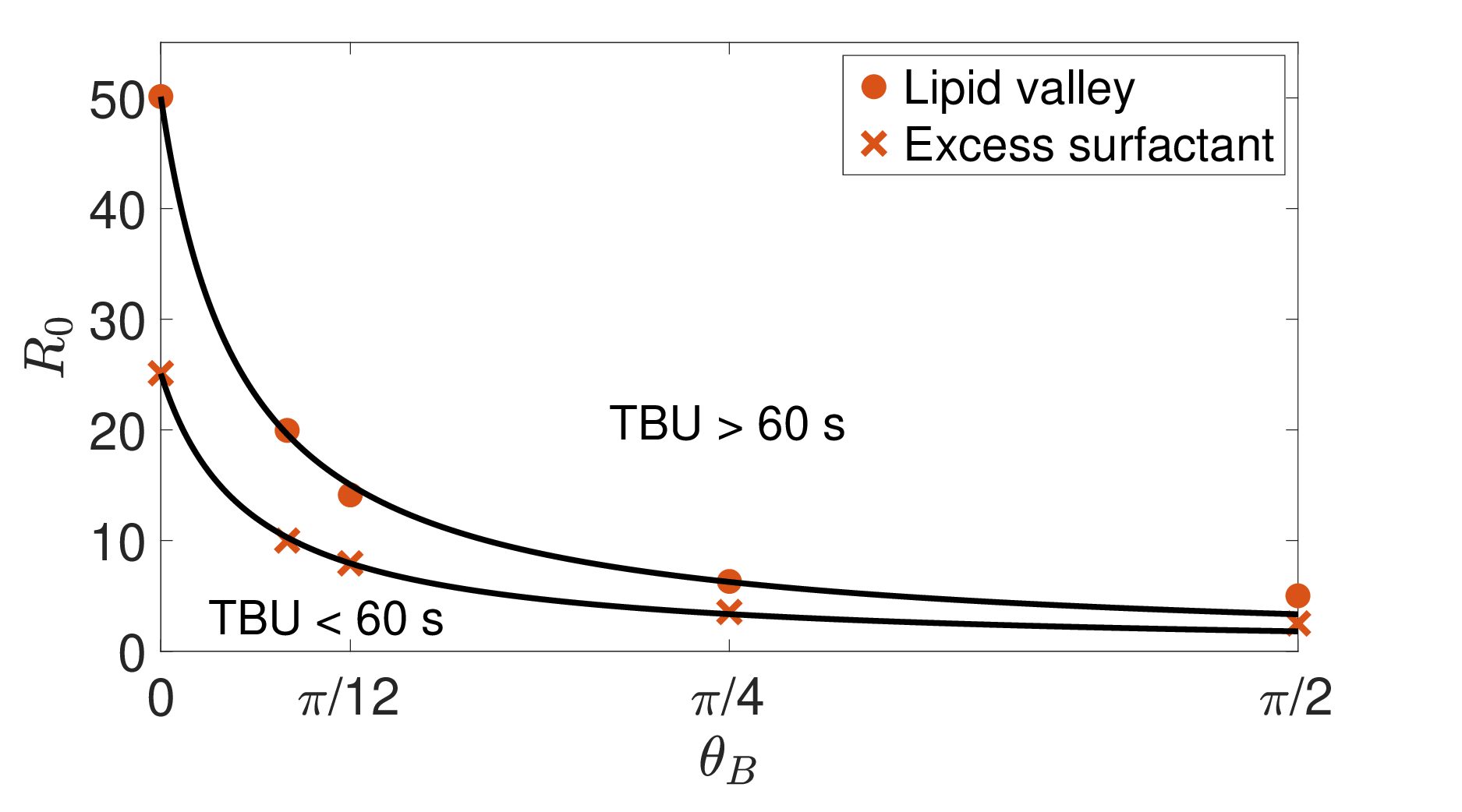}
\caption{A comparison of the $\mathcal{R}_0$ value which results in TBU at 60 s for two sets of initial conditions: one describing evaporation-driven flow (red circles), and one describing flow driven by the Marangoni effect (red crosses), as a function of the director angle of the liquid crystal molecules in the LL. The black lines are a fit to the respective data. To the left of each black line, TBU occurs before 60 s; to the right of each black line, TBU may occur after 60 s.  } \label{fig:varyR}
\end{figure}

Figure \ref{fig:varyR} shows the relationship between the evaporative resistance parameter and the angle of the director for the two sets of initial conditions. The value of $\mathcal{R}_0$ which results in breakup conditions at 60 s is shown in red for $\theta_B=0,\,\pi/18,\,\pi/12,\,\pi/4,$ and $\pi/2$. The circles represent the solution from the evaporation-driven flow caused by a valley in the LL; the crosses represents solutions from flow driven by the Marangoni effect caused by localized excess surfactant. The black lines are rational functions fit to the respective data delineating whether or not breakup conditions are reached by the final time of 60 s. If the evaporative resistance parameter is decreased by a factor of 10 from the default value of 17.68, then TBU will occur for all angles of the director for both sets of initial conditions, although TBUT will vary. 

If we compare the two curves, we see that the $\mathcal{R}_0$ which results in TBU at 60 s for the case of evaporation-driven thinning is roughly double that of the case of Marangoni-driven flow. In the evaporative case, when $\theta_B=0$, the evaporative resistance parameter $\mathcal{R}_0=50$, whereas for the Marangoni-driven thinning, $\mathcal{R}_0=25$. When $\theta_B=\pi/2$, $\mathcal{R}_0$ is 5 and 2.5, respectively. This trend roughly holds for all values of the director angle. In general, this is expected behavior; the LL acts as a barrier to evaporation in part due to its thickness. In the first set of initial conditions, a deficiency in the LL leads to increased evaporation. Thus, it cannot tolerate an evaporation resistance parameter value of any less than 50 without TBU occurring before 60 s when $\theta_B=0$. On the other hand, in the Marangoni-driven flow, the LL begins with a uniform thickness of 50 nm which helps to prevent TBU even though other aspects of the barrier may be compromised. 

\clearpage
\subsubsection{Reducing initial lipid layer thickness}
\begin{figure}[htbp]%
\centering
\includegraphics[width=0.49\textwidth]{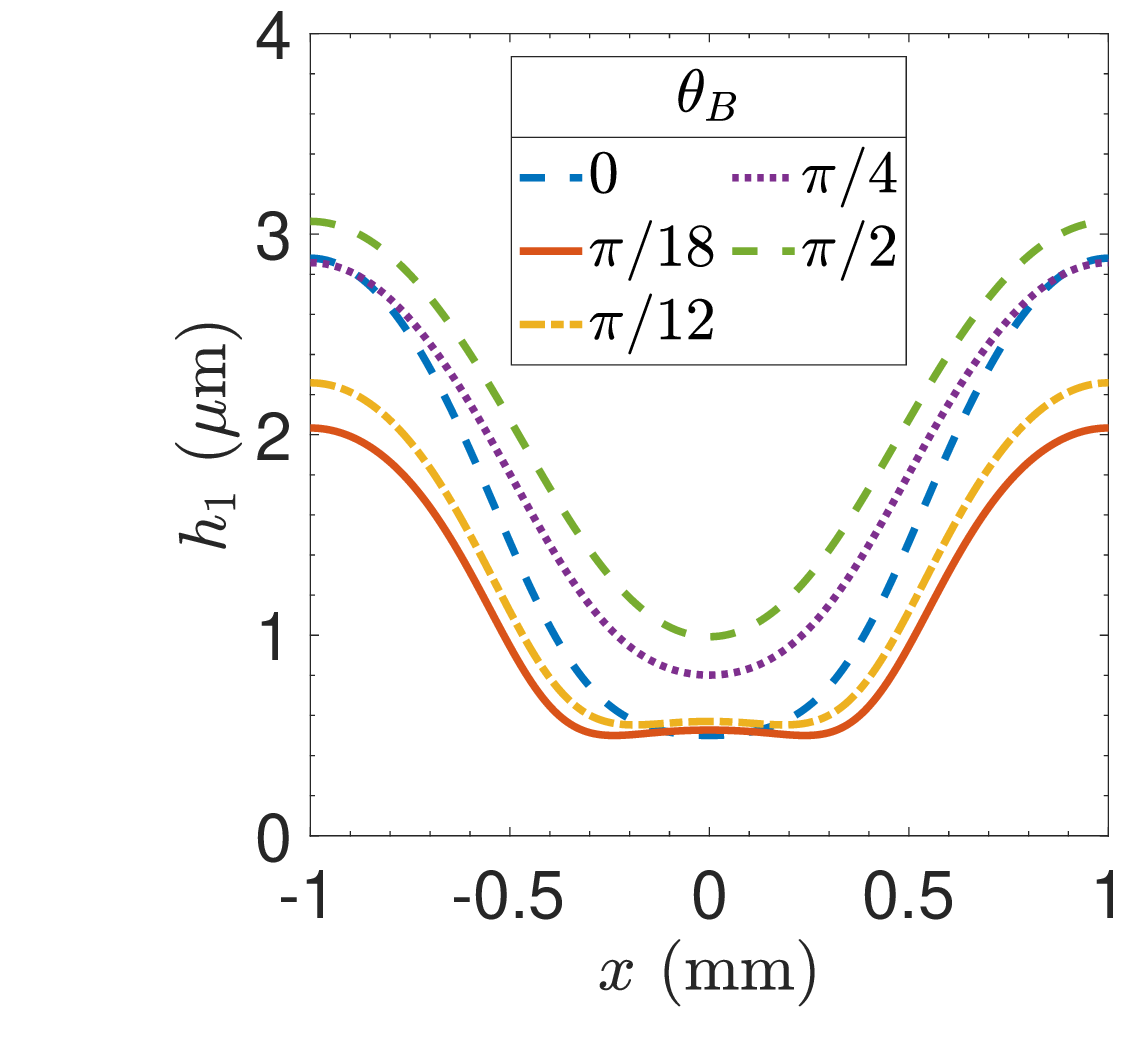}
\includegraphics[width=0.49\textwidth]{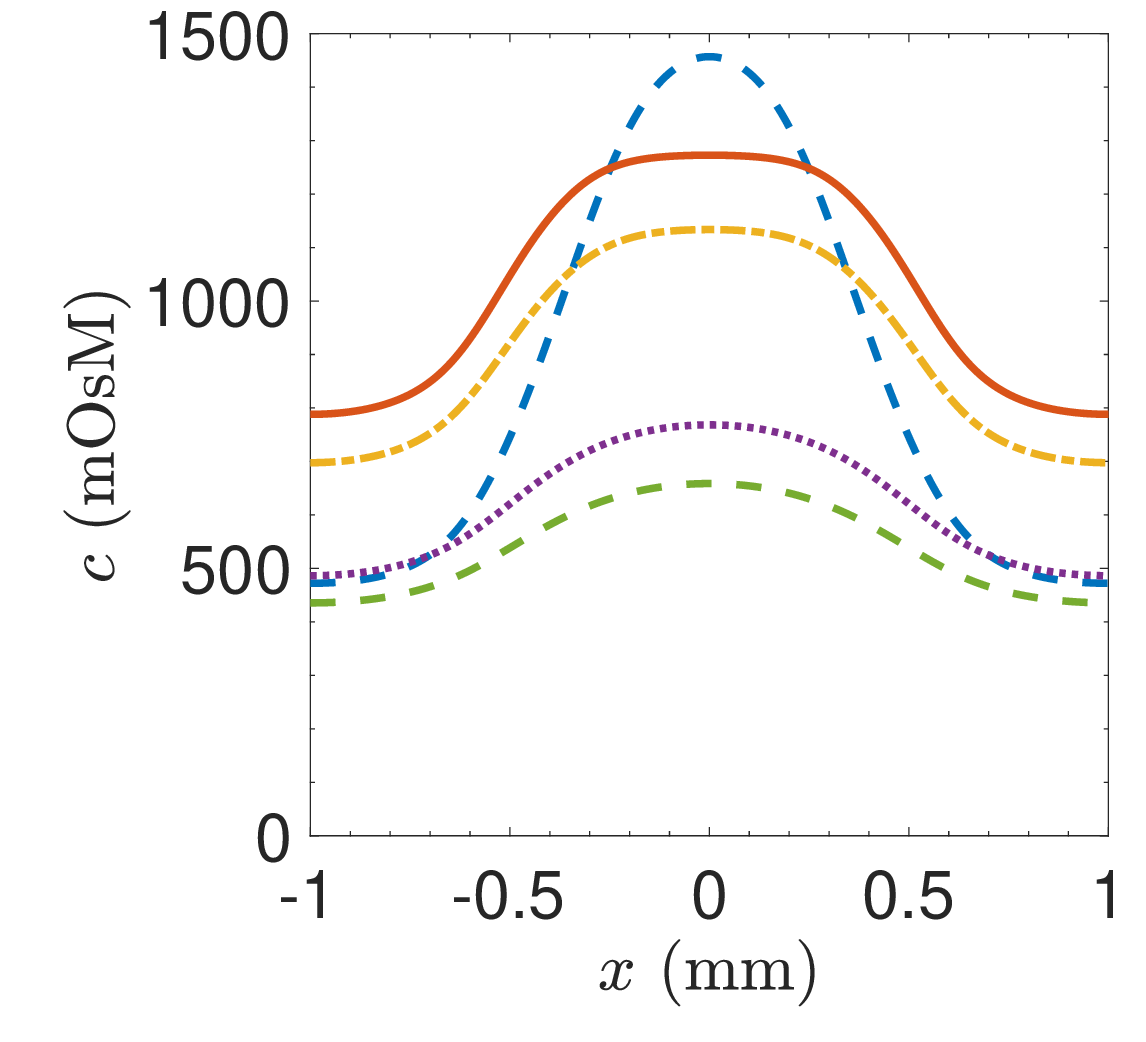}
\caption{Solutions at the final time for spatially-uniform initial conditions except for a local excess of surfactant. The initial thickness of the LL is 25 nm (half the default value). When $\theta_B=0$, breakup occurs at 8.2 s; when $\theta_B=\pi/18$, breakup occurs at 54.8 s. For all other values of $\theta_B$, the final time is 60 s.}\label{fig:h2_finaltime}
\end{figure}

The thickness of the LL ranges between 20 - 120 nm, with more recent measurements favoring a thinner LL \cite{KSHinNic10, king2011high, goto2003kinetic}. We halve the initial LL thickness from 50 nm (a ``normal" tear film thickness), to 25 nm (a ``thin" thickness). Figure \ref{fig:h2_finaltime} shows solutions for the AL and osmolarity at the final time for a few values of $\theta_B$. The solutions can be compared with those of Figure \ref{fig:longtime} which show solutions for a 50 nm thick initial LL. Qualitatively, solutions for large values of $\theta_B$ are similar, although the AL is thinner, especially at its minimum. However, for $\theta_B=\pi/12$ or $\pi/18$, there are two minima, located symmetrically about the center. For $\theta_B=0$, the region of thinning is steeper and more narrow than when the LL is thicker. In all cases, the solutions for osmolarity show higher concentrations with a thinner LL. 

Table \ref{tab:h2_comp} shows that a 50\% decrease in LL thickness results in a 40.1\% decrease in the minimum AL thickness when $\theta_B=\pi/4$; for other values of the director, the percent decrease is even less. The factor increase in peak osmolarity over the ideal osmolarity is also shown. These numbers are highest when TBU occurs; however, with a thinner LL, peak osmolarity is well over twice the ideal value for all angles of the director.

\begin{table}
    \centering
\caption{Comparison of minimum AL thickness and maximum osmolarity when the initial LL thickness is reduced from 50 nm to 25 nm. The factor increase in peak osmolarity is compared to the ideal osmolarity of 300 mOsM.  }
\label{tab:h2_comp}
    \begin{tabular}{lllclclc}\toprule
 & \multicolumn{3}{c}{$h_{1min}$ }& \multicolumn{4}{c}{$c_{max}$ }\\
 \cmidrule(lr){2-4}\cmidrule(lr){5-8}
         $\theta_B$ &  50 nm &  25 nm &  \% decrease&  50 nm&  factor $\uparrow$&  25 nm&  factor $\uparrow$\\
         &(TBUT)& (TBUT)& &&&  &  \\\midrule
         0&  0.5 (16.6 s)&  0.5 (8.2 s)&  0&  1398&4.66&  1457&4.86  \\
         $\pi/18$&  0.6626&  0.5 (54.8 s)&  24.5&  922&3.07&  1273&4.24  \\
         $\pi/12$&  0.7538&  0.5528&  26.7&  809&2.70&  1134&3.78  \\
         $\pi/4$&  1.3361&  0.8010&  40.1&  546&1.82&  768&2.56  \\
         $\pi/2$&  1.6472&  0.9926&  39.7&  478&1.59&  659& 2.20 \\\botrule
    \end{tabular}
\end{table}

\begin{figure}[htbp]%
\centering
\includegraphics[width=0.49\textwidth]{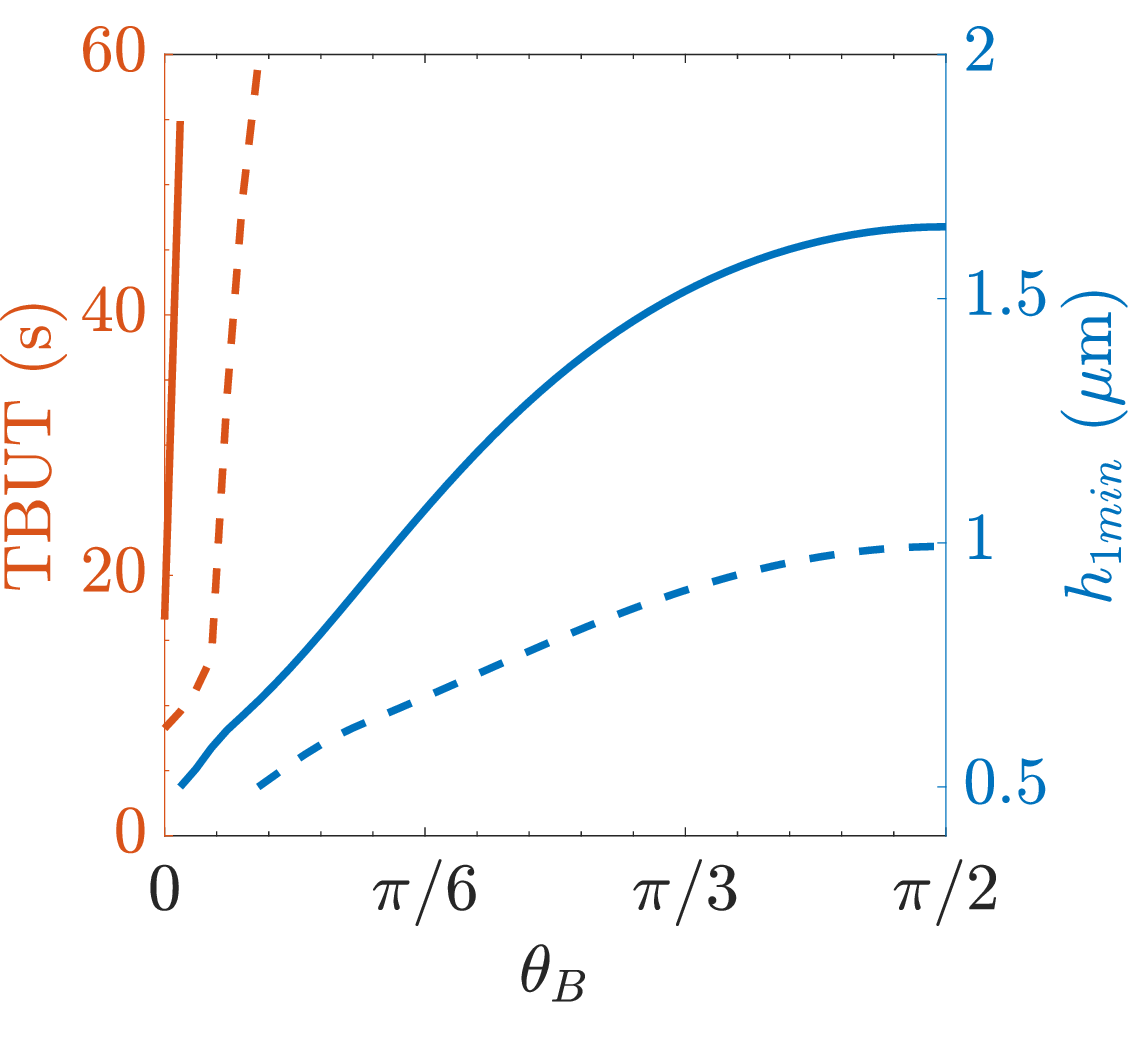}
\includegraphics[width=0.49\textwidth]{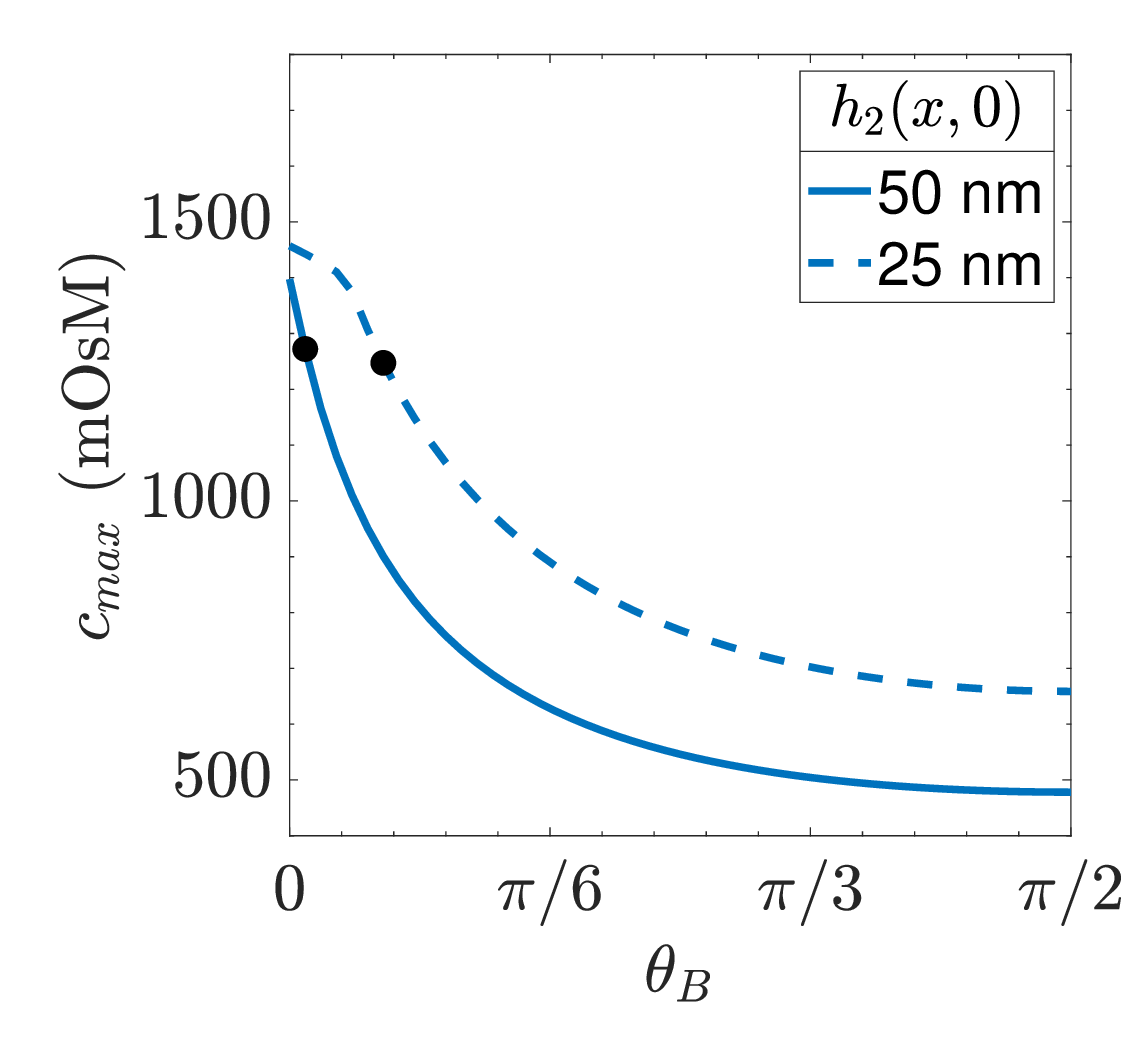}
\caption{Extreme values of AL thickness, osmolarity, and LL thickness at the end of each simulation as $\theta_B$ is varied from $0$ to $\pi/2$. To the left of the black dot, TBUT occurs before 60 s; to the right of this dot, the simulations end at 60 s.  }\label{fig:h2comp}
\end{figure}

Figure \ref{fig:h2comp} shows the effect of varying the director angle by plotting parametric changes to the minimum value of the AL thickness and the maximum value of osmolarity at the final time as $\theta_B$ varies from 0 to $\pi/2$ for both a normal and thin LL. When the LL thickness is 50 nm, TBU occurs before 60 s only for values of $\theta_B<\pi/50$; when the LL thickness is 25 nm, TBU occurs for $\theta_B<\pi/14$. The black dots on the plot mark this change. Minimum LL thickness as $\theta_B$ varies is not shown, as changes are minimal. When the initial LL thickness is 50 nm, minimum values hovers near 28 nm; when the initial thickness is 25 nm, the minimum values remain around 14 nm for all values of the director angle. 

\clearpage
\subsection{Combining lipid layer and surfactant perturbations}

\begin{figure}[htbp]%
\centering
\includegraphics[width=0.49\textwidth]{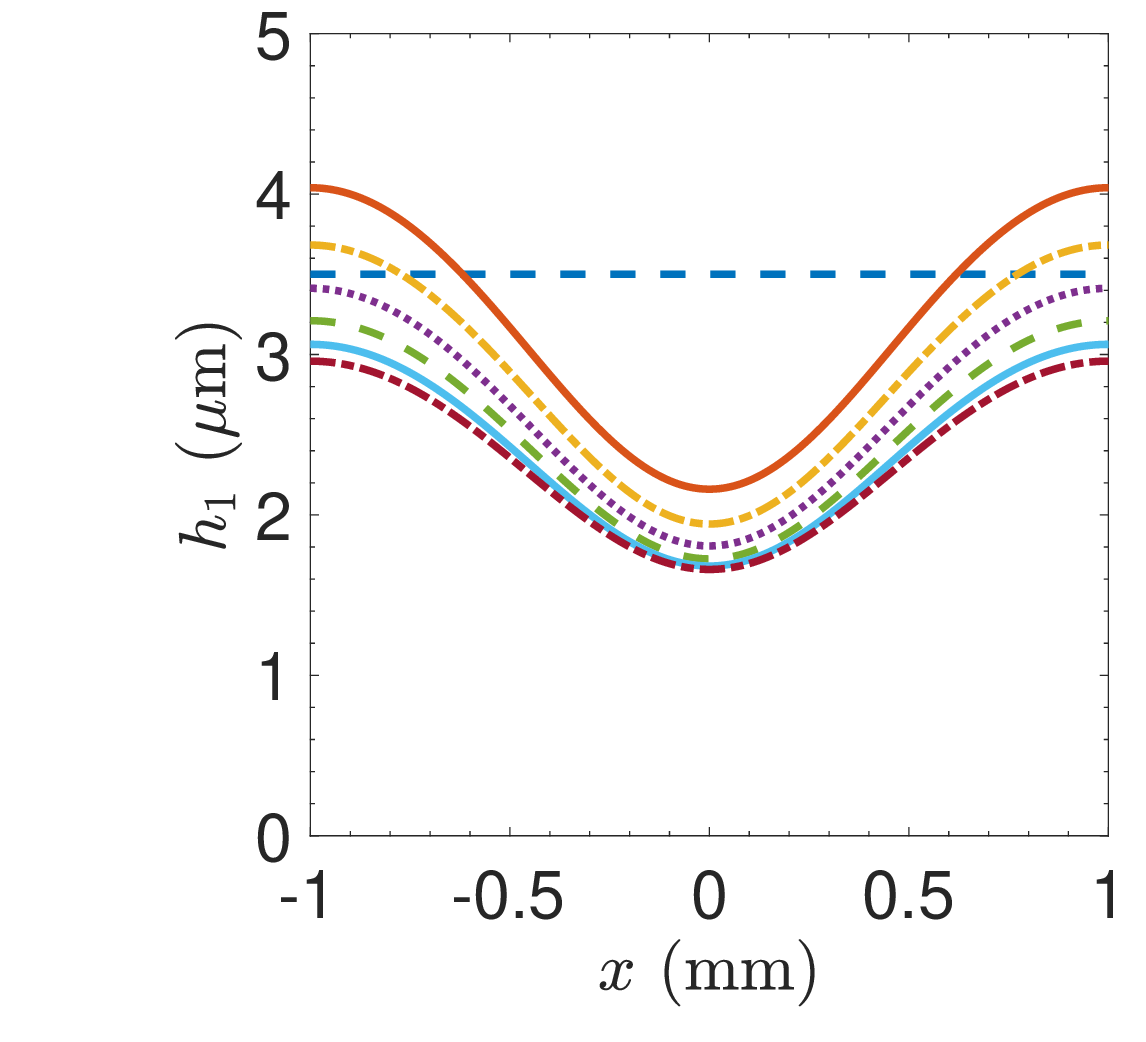}
\includegraphics[width=0.49\textwidth]{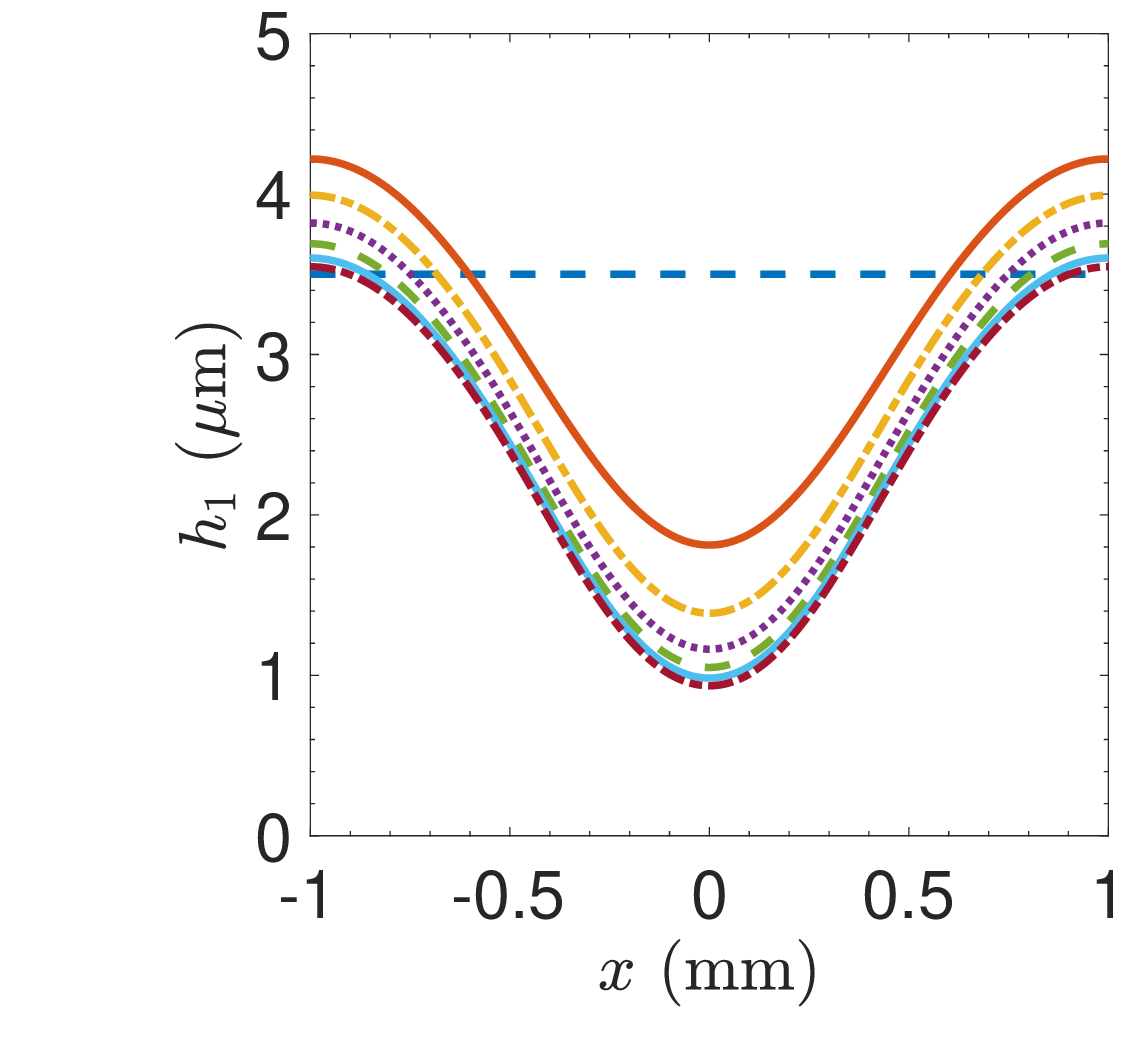}
\includegraphics[width=0.49\textwidth]{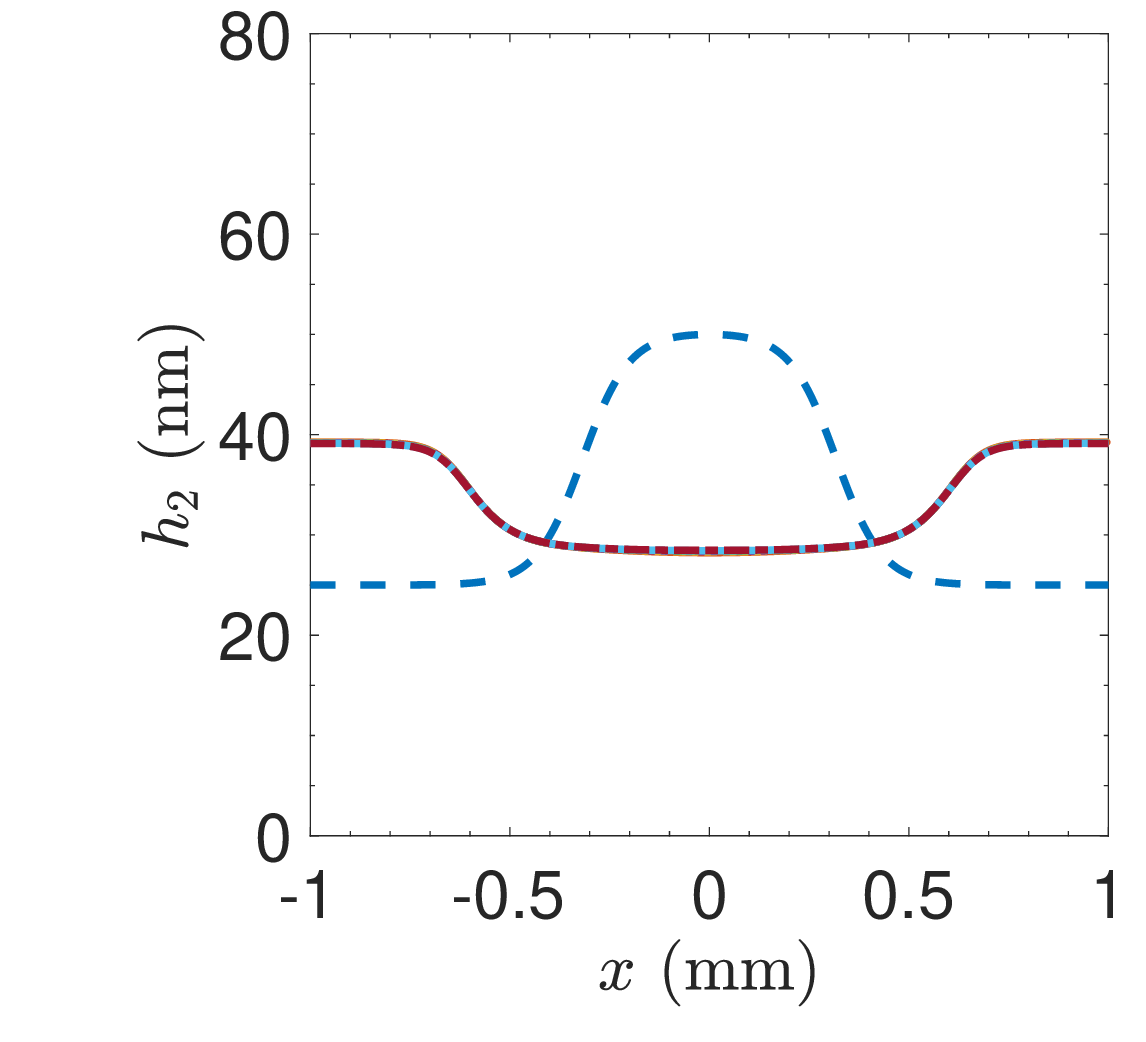}
\includegraphics[width=0.49\textwidth]{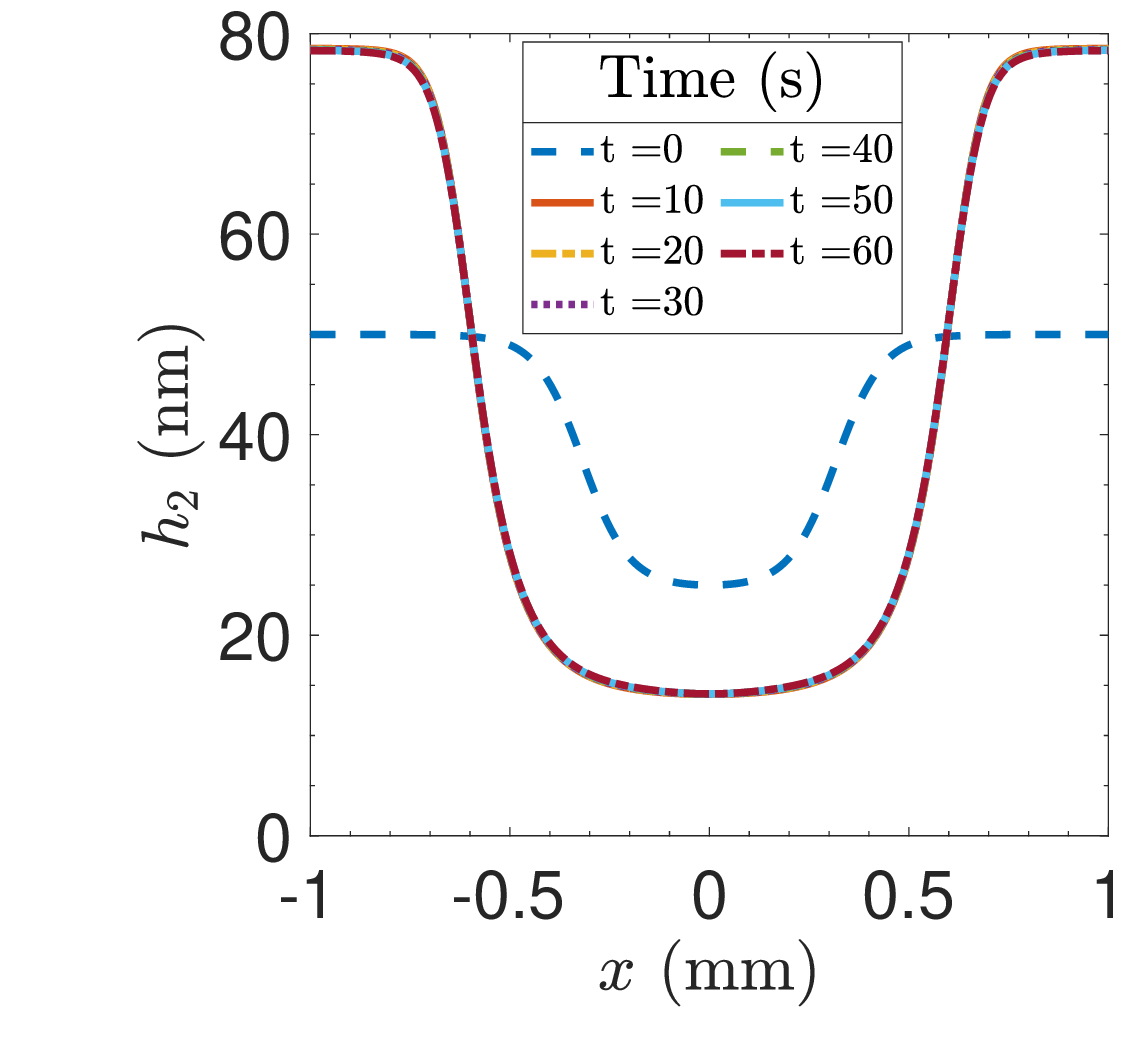}
\includegraphics[width=0.49\textwidth]{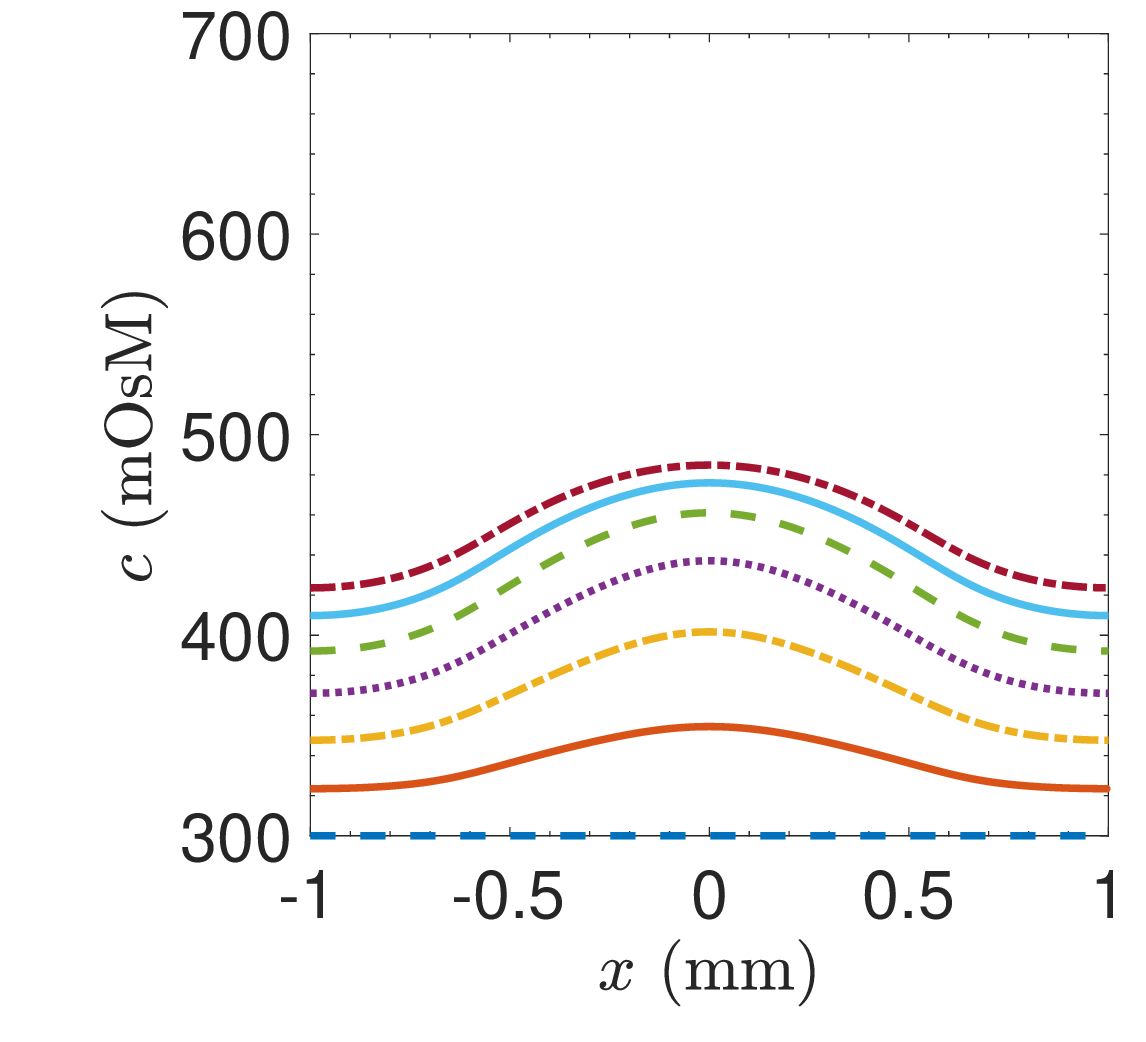}
\includegraphics[width=0.49\textwidth]{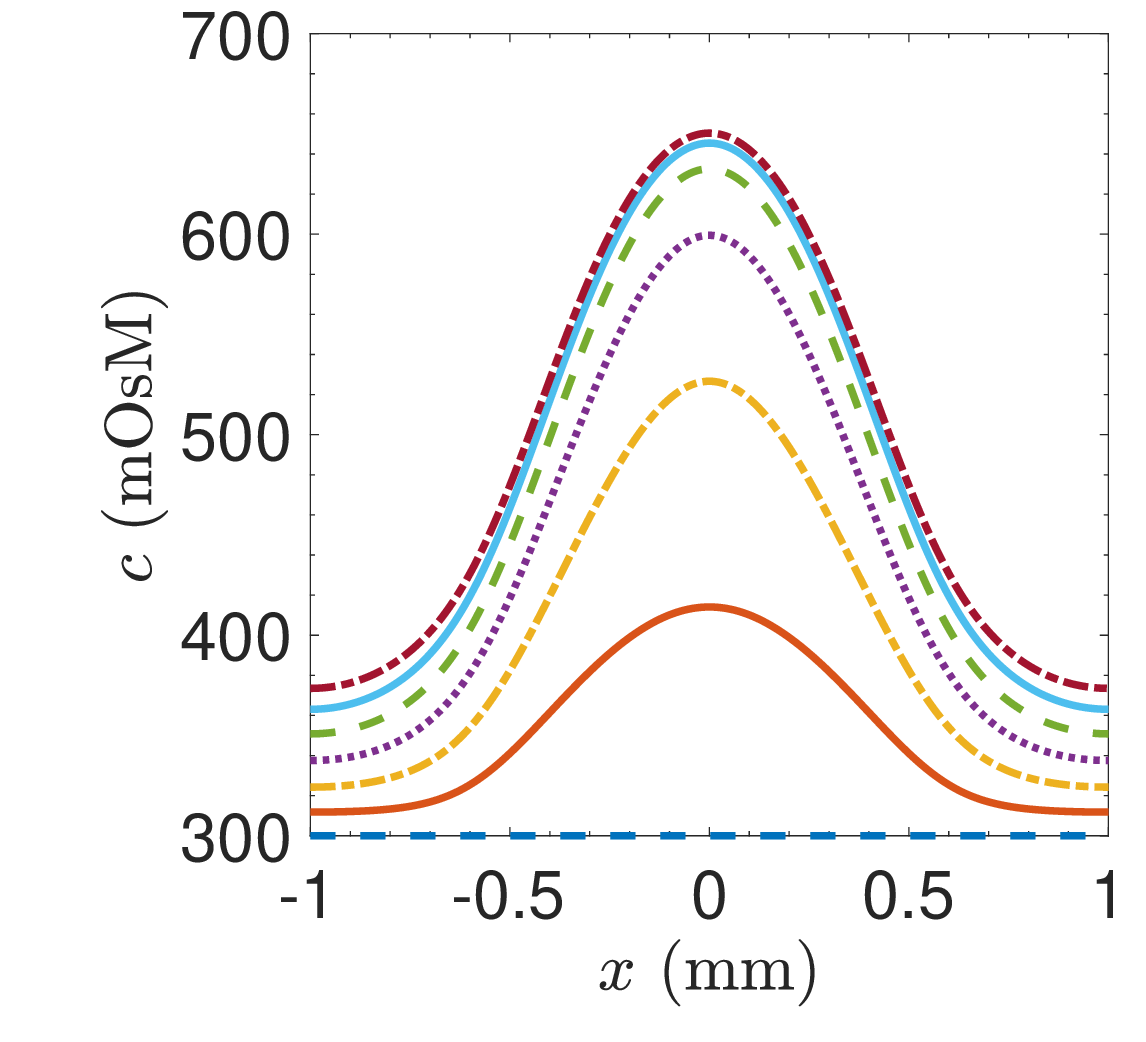}
\caption{Left column: Solutions when $\theta_B=\pi/2$ for a bump $h_2$ combined with a local excess of $\Gamma$. Right column: Solutions when $\theta_B=\pi/2$ for a valley in $h_2$ combined with a local excess of $\Gamma$. In both cases $\Gamma(x,0)$ is given by Equation (\ref{eq:surfactant_IC}). The different $h_2(x,0)$ appears in the middle row. }\label{fig:vb_bb_comp_pi2}
\end{figure}

We now compare the two scenarios which combine perturbations in both the LL thickness and the surfactant concentration. The first combination is a bump in $h_2$ with a local excess of $\Gamma$, and the second is a valley in $h_2$ with a local excess of $\Gamma$. Figure \ref{fig:vb_bb_comp_pi2} compares solutions for AL and LL thickness, as well as osmolarity for these two cases with $\theta_B=\pi/2$. While neither case reaches breakup for this angle of the director, thinning in the AL is more pronounced when there is a valley in the LL, with a minimum value of 0.94 $\mu$m, as compared to 1.66 $\mu$m when the LL contains a bump. The LL behavior is similar in both cases; it undergoes rapid redistribution of fluid initially, and then remains tangentially immobile for the remainder of the time. When the LL contains a bump, this extra fluid is quickly redistributed throughout the domain; an area of thinning results, with the minimum thickness of 28.46 nm. When the LL contains a valley, the Marangoni effect magnifies this profile resulting in a solution similar to those of Figure \ref{fig:longtime}, but with a deeper valley: the depth is now 14.15 nm versus 28.1 nm (when $\theta_B=\pi/2$). 

Osmolarity in both cases reflects the respective changes in the AL. When the LL contains a valley, the AL thinning is more pronounced, and thus osmolarity reaches a peak of 650 mOsM, with little increase in concentration at the edges of the domain. When the LL contains a bump, the AL thins to a lesser degree, and the osmolarity steadily rises throughout the domain, with a smaller peak in the center. Peak osmolarity at the final time reaches 485 mOsM. In all scenarios that contain a local excess of surfactant initially, the Marangoni effect results in a rapid flattening of the surfactant concentration to a more or less uniform value of 1.58. 

\begin{figure}[htbp]%
\centering
\includegraphics[width=0.49\textwidth]{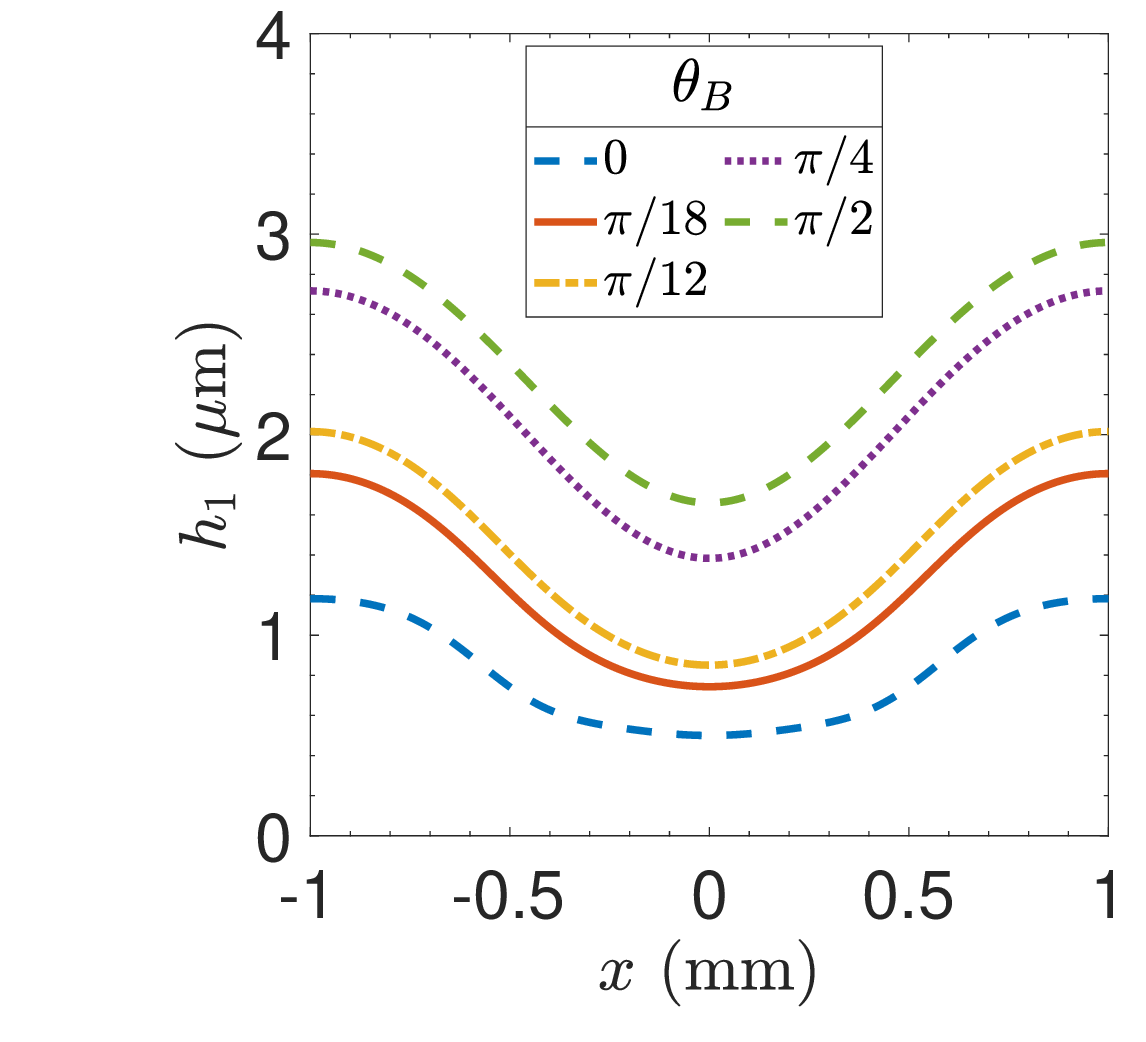}
\includegraphics[width=0.49\textwidth]{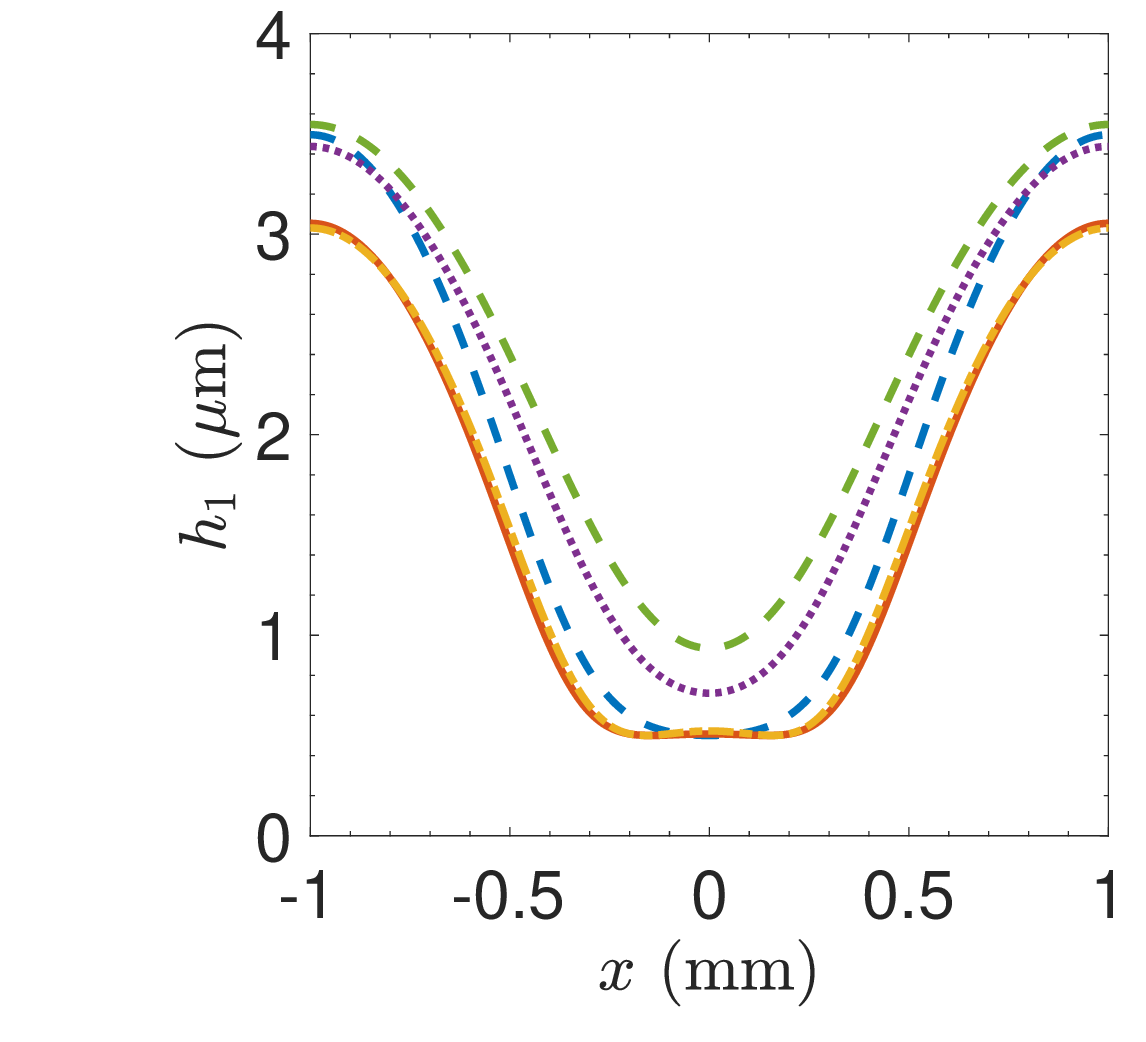}
\includegraphics[width=0.49\textwidth]{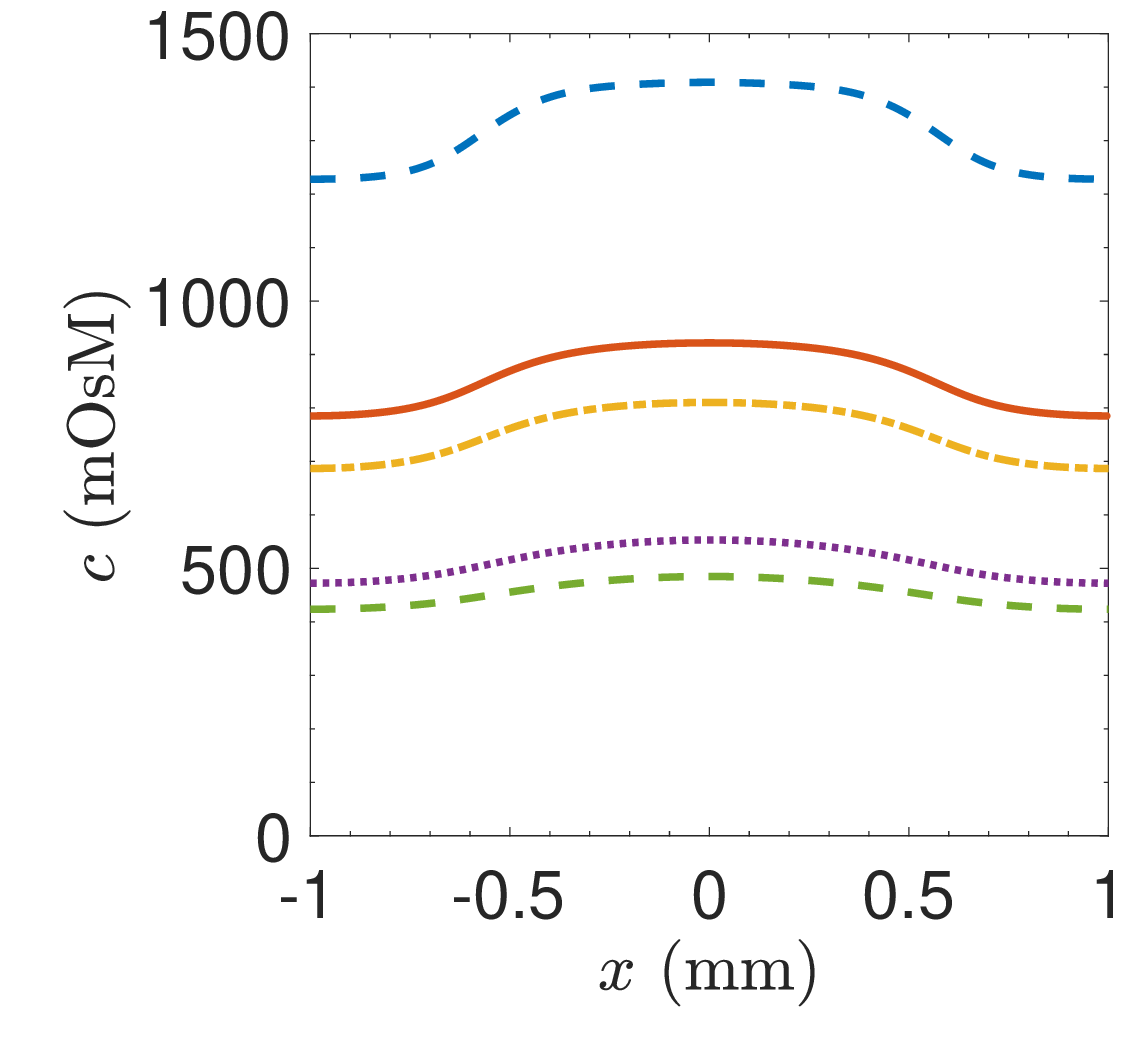}
\includegraphics[width=0.49\textwidth]{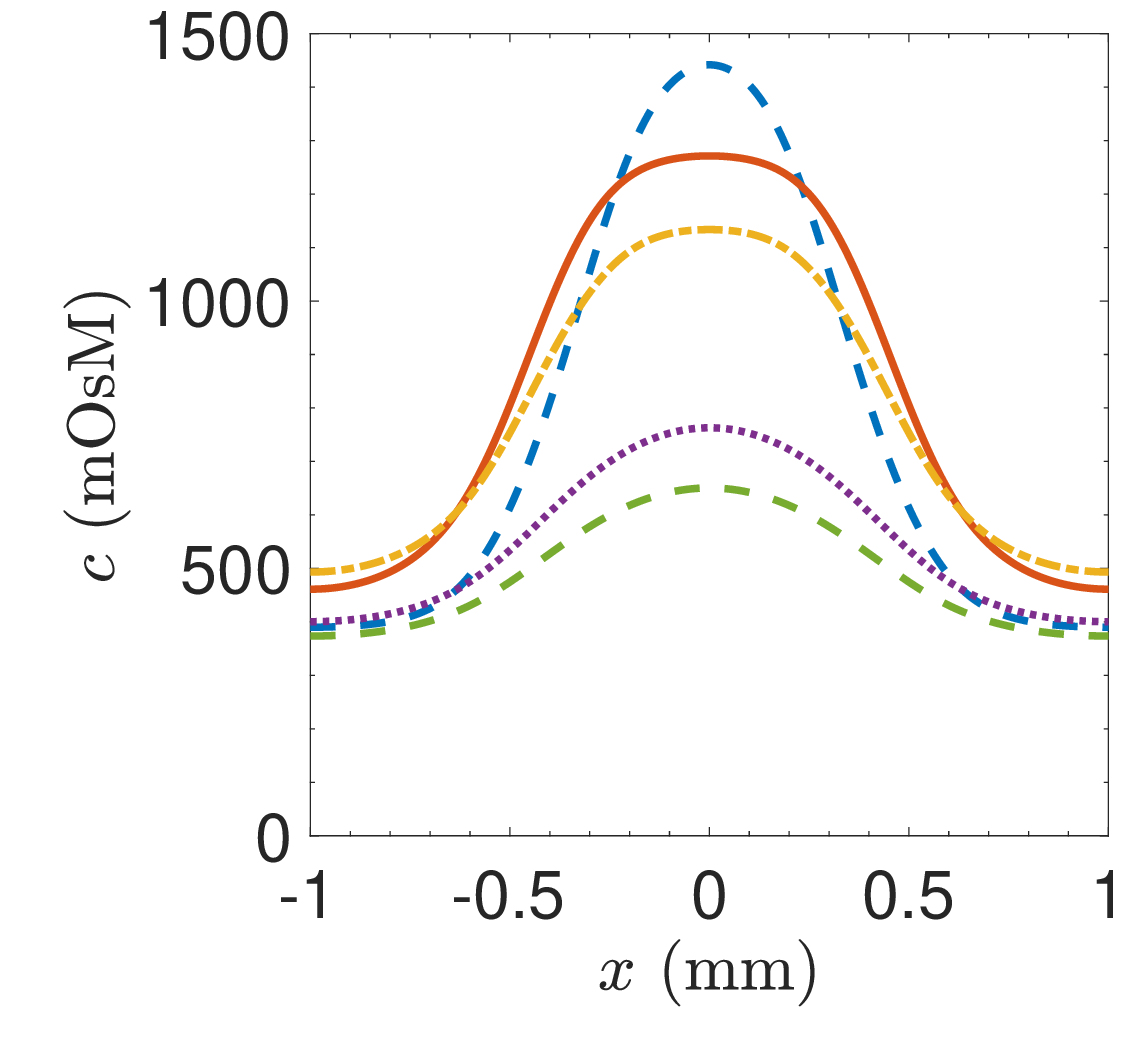}
\caption{Left column: Solutions at the final time for a bump in $h_2$ combined with a local excess of $\Gamma$. Right column: Solutions at the final time for a valley in $h_2$ combined with a local excess of $\Gamma$.}\label{fig:vb_bb_finaltime}
\end{figure}

We can consider the influence of the director angle, by observing the final time solutions for various angles as shown in Figure \ref{fig:vb_bb_finaltime}. When there is a bump in the LL, we see more variation resulting from changes in the director angle. When $\theta_B=0$, TBU occurs at 46.8 s, but no other angles of the director result in breakup. When there is a valley in the LL, the director angle has less of an impact because rapid thinning in the AL leads to TBU for $\theta_B=0,\,\pi/18,$ and $\pi/12$. Breakup times are detailed further in Table \ref{tab:TBU_comp} which is discussed more below. Osmolarity at the final time is strongly influenced by the director angle. When there is a bump in the LL, peak osmolarity is almost three times higher for $\theta_B=0$ (1409 mOsM) than for $\theta_B=\pi/2$ (485 mOsM). The difference is just over double when there is a valley in the LL (1442 versus 650 mOsM).

Thus far we have considered a valley in the LL with depth 25 nm in a 50 nm LL. We now vary this profile  keeping a valley of half thickness. Table \ref{tab:TBU_comp} compares the TBUT for three different LL background thicknesses: 30, 40, and 50 nm. We see that the dynamics are not sensitive to LL thickness when $\theta_B=0$. This is because evaporative resistance is at its lowest for this angle of the director; for all of the scenarios we have considered, $\theta_B=0$ results in TBU before 60 s. LL thickness has a stronger effect for intermediate angles of the director. For example,  when $\theta_B=\pi/18$, decreasing the LL thickness by 20\%, from 50 nm to 40 nm, results in TBU in half the time. When $\theta_B$ is large enough, TBU does not occur; for all three LL thicknesses TBU does not occur for $\theta_B=\pi/4$ or $\pi/2$.

\begin{table}
    \centering
\caption{TBU time comparison in seconds for various initial LL thicknesses when a valley in the LL is combined with a local excess of surfactant.  }
\label{tab:TBU_comp}
    \begin{tabular}{cccccc}\toprule
   & \multicolumn{5}{c}{$\theta_B$ }\\
   \cmidrule(lr){2-6}
      LL max/min (nm)  &    0 &  $\pi/18$&  $\pi/12$&  $\pi/4$&  $\pi/2$\\\midrule
   50/25 & 8.2  & 29.6 & 48.4 & - & - \\
   40/20 & 7.4  & 15.0 & 32.2 & - & - \\
   30/15 & 6.7  & 10.7 & 14.3 & - & - 
 \\\botrule
    \end{tabular}
\end{table}

\clearpage
\section{Discussion and conclusion}\label{sec:discussion}

 In this paper we explore a new model for the tear film of the eye that was derived in \citet{taranchuk24}. This two-layer model combines a Newtonian AL with a nematic liquid crystal LL, and links molecular orientation of the LL with evaporative resistance that depends on the director angle. The LL acts as a barrier to evaporation of the AL, and thickness alone does not explain its role in preventing evaporation \cite{fenner2015moretostable, king-smith2015moretostable}. By representing the LL as nematic liquid crystal, we are able to consider possible connections between the structure and composition of the LL and its affect on tear film thinning and breakup. 

 We examined two key drivers of TBU: evaporation and the Marangoni effect. We consider several scenarios which may lead to TBU. The first is a narrow valley in the LL, which leads to thinning of the AL due to increased evaporation. The second is a local excess of surfactant on the AL/LL interface, for which we see thinning on two timescales. Initially, surfactant gradients lead to rapid outward flow in both the AL and LL. The surfactant concentration levels out within the first second, and then evaporation takes over, continuing tear film thinning over a longer timescale of 60 s. Finally, we considered combinations of perturbations in both the LL and surface surfactant; one in which there is a bump in the LL combined with a local excess of surfactant, and one in which there is a valley in the LL combined with a local excess of surfactant. This final case depicts a worst case scenario, where both excess surfactant and a deficiency in the LL drive thinning, and breakup conditions are reached for $\theta_B\lesssim\pi/12$.

The evaporative resistance parameter $\mathcal{R}_0$ directly influences how quickly the AL thins. However, as shown in Figure \ref{fig:varyR}, the minimum value of $\mathcal{R}_0$ that the system can tolerate while withstanding breakup also depends on the angle of the director, $\theta_B$, and the initial condition for the LL. The evaporation function, given in Equations (\ref{eq:je}) and (\ref{eq:r0}) and derived in \ref{sec:osm_evap}, shows one way these quantities may be related. While we have chosen to introduce director angle dependence in the evaporation function \cite{king2018,King-SmithOS13, paananen2020CEs}, we also see that the director angle has an effect on flow dynamics. Small angles of $\theta_B$ cause the LL to become more mobile, which results from liquid crystal properties in the LL.

In addition, we find that the default Marangoni number has a strong effect on dynamics, regardless of the initial conditions or other parameter values used. Thus, in order to observe the effect of the liquid crystal parameters, especially the Leslie viscosities, we reduce the Marangoni number by a factor of ten. In the absence of any measurements in the eye, we use the properties of 5CB as a starting point for the Leslie viscosities of the liquid crystal lipid layer. One unexpected finding is that varying the Leslie viscosities from the 5CB values has little effect on the dynamics. Not only that, but we are constrained in how much we can vary them, at least when using constant scalings of 5CB. On the other hand, the other main liquid crystal parameter $\theta_B$, which describes the orientation of the molecules, affects both evaporation (by design), and flow. This is especially true for small angles. 

Rather than varying all of the Leslie viscosities, we focused on varying $\alpha'_4$, which controls the dynamic viscosity of the LL (because $\mu_2=\alpha'_4/2$). These results are very similar to the Newtonian case, explored in \citet{stapf2017duplex}, in that it takes a very large increase in $\mu_2$ to effect the dynamics of the AL. The director angle does have an effect on the mobility of the LL, until $\mu_2$ become very large (on the order of $10^5$). After this point, the LL becomes virtually tangentially immobile for all angles of the director.  

Finally, we consider two initial conditions which combine perturbations in both the LL and the surface surfactant. The first is an undesirable case where  a valley of lipid is combined with a local excess of surfactant. As expected, we see more breakup cases; for all angles of $\theta_B\lesssim10$, TBU occurs within 60 s as a result of both the Marangoni effect and increased evaporation due to the deficiency in the LL. 

Alternatively, we consider an initial condition which combines a bump in the LL with a local excess of surfactant. This scenario corresponds to Section 5.6 in \citet{bruna2014influence}, where in a limit $h_2$ and $\Gamma$ can both be governed by the same equation. While we do not take that limit here, we model an initial condition in which the LL follows the surfactant. This area of increased thickness in the LL seem to be protective by creating a buffer against the Marangoni effect, and thus helping to prevent excessive thinning in the AL. Imaging of the LL has shown areas of varying thickness on occasion \cite{GotoTseng03}. This type of initial condition could resemble this scenario, but more work is needed to link them.

\citet{choudhuryMembraneMucin2021} modeled TBU driven by dewetting due loss of mucins in the glycocalyx using a single layer model. They consider 119 s to be a normal time to breakup in a healthy eye, and their simulation resulted in TBU as short as 3.3 s when there is mucin loss over a third of the corneal surface.  \citet{deyContinuousMucinProfiles2019,deyContinuousMucinCorrection2020} studied the same mechanism for TBU, but used a continuous-viscosity model which combined the mucus and aqueous layers. Their model predicted breakup in 2.4 to 135 s for scenarios representing pathological to healthy conditions, respectively. \textsl{in vivo} measurements for subjects with dry eye or other eye conditions have shown TBU times well under 10 s. In healthy subjects, TBU times may be longer, but would generally occur in less than two minutes \cite{begleyQuantitativeAnalysis2013, yokoi2019TForiented, King-SmithIOVS13a}.

While our model focuses on different mechanisms for thinning and breakup, TBU times from our simulations fall within a similar range as previous models and experiments. We consider that a healthy eye has high evaporative resistance, corresponding to a large director angle. For the scenarios we considered, a director angle of $\theta_B=\pi/4$, or $\pi/2$ did not result in breakup within 60 s. We did not compute time to TBU; however, at the end of 60 s, minimum AL thicknesses were well above the threshold value of 0.5 $\mu$m for these larger director angles. For small angles of the director in which evaporative resistance was low and there was a deficiency in the LL or an excess of surfactant, our model predicted breakup in as little as 6 to 10 s. These predicted breakup times are not as rapid as the shortest TBU times observed in eyes \cite{King-SmithIOVS13a,yokoi2019TForiented}. 

Thinning and breakup due to a glob surfactant was studied in \citet{ZhongJMO18,ZhongBMB19}. The glob consisted of an interval of fixed concentration as well a part that was transported outside this interval. This area of fixed concentration, which acted as a source, was varied in size, and found to cause TBU either under or at the edge of the glob depending on size. TBU occurred in $< 1$ s to 15 s, and it was noted that evaporation aided TBU. Our results have central breakup and evaporation aligns with the Marangoni effect to promote TBU. Although our TBU time is on the longer end of the range they predicted, our treatment of surfactant is simpler. We conserve surfactant, and the local excess of surfactant does not act as a source; nor did we experiment with larger or smaller concentrations. 

Thinning due to evaporation has been modeled previously both with a fixed evaporation distribution \cite{BraunDrisTBU17}, and with a static LL \cite{peng2014evaporation}. Alternatively, \citet{lukeParameterEstimation2020} developed an evaporation-driven model that was fit to experimental data. We have taken a mechanistic approach, and incorporated LL thickness, permeability, and structure in our evaporation function. 

We have studied a variety of initial conditions and explored the effects of several key parameters on tear film dynamics when modeling the LL as nematic liquid crystal. Now that we have a sense of the key features and limitations of this model, future work could include a moving end to imitate the eyelid during a blink. Tear breakup is often related to the LL pattern \cite{King-SmithIOVS13a,braun2015dynamics}; more fidelity would require a moving domain \cite{JonesEtal06}. In addition, it would be valuable to determine more rheological properties for this kind of model \cite{rosenfeld2013structural, Leiske11, Leiske12}.

\backmatter



\section*{Acknowledgements}
This material is based upon work supported by the University of Delaware Graduate
College through the Doctoral Fellowship for Excellence (MJT). Any opinions, findings, and
conclusions or recommendations expressed in this material are those of the authors.  RJB is grateful for SHD's mentoring and many examples of the right way to do things.  Remembering how he got his unofficial middle name, we have joined words wherever a possibility of using a hyphen has occurred in this paper.

\section*{Declarations}
\subsection*{Competing interests}
The authors have no competing interests to declare that are relevant to the content of this article.

\begin{appendices}

\section{Dimensional equations}\label{sec:appendix_derivation}
 We give the dimensional equations (after \cite{taranchuk24}) and a very brief outline of the derivation. 
\subsection{Governing equations}
Inside the AL, $0<y'<h'_1$, we have conservation of mass and momentum using the Navier-Stokes equations. The advection-diffusion equation governs osmolarity transport within the AL. The equations in dimensional form with a prime denoting a dimensional quantity, respectively, are:
\begin{align}
    \nabla' \cdot {\bf u'}_1=0,\\
    \rho_1 ({\bf u'}_{1t}+{\bf u'}_1\cdot \nabla' {\bf u'}_1)= -\nabla' p'_1 + \mu_1 \Delta' {\bf u'}_1,\\
    c'_{t'} + {\bf u'}_1\cdot\nabla' c' - D_1 \Delta' c'=0,
\end{align}
where the divergence of ${\bf u'_1}$ is $\nabla'\cdot{\bf u'_1}=u'_{1x'}+v'_{1y'}$, the gradient is $\nabla'{\bf u'_1}=\begin{bmatrix} u'_{1x'} &v'_{1x'}\\u'_{1y'}& v'_{1y'}\end{bmatrix}$, and the Laplacian of the first component of ${\bf u'_1}$ is $\Delta' u'_1=u'_{1x'x'}+u'_{1y'y'}$. 

Inside the LL, $h'_1<y'<h'$, we have conservation of energy, mass, and momentum using the Ericksen-Leslie equations for liquid crystals.  
\begin{align}
    \frac{\partial }{\partial x'}\left( \frac{\partial W'}{\partial \theta_{x'}}\right)+\frac{\partial }{\partial y'}\left( \frac{\partial W'}{\partial \theta_{y'}}\right)-\frac{\partial W'}{\partial \theta}+\tilde{g}'_x\frac{\partial n_x}{\partial \theta}+\tilde{g}'_y\frac{\partial n_y}{\partial \theta}&=0,\\
    -\frac{\partial \pi'}{\partial x'}+\tilde{g}'_x\frac{\partial n_x}{\partial x'}+\tilde{g}'_y\frac{\partial n_y}{\partial x'}+\frac{\partial \tilde{t}'_{xy}}{\partial y'}+\frac{\partial \tilde{t}'_{xx}}{\partial x'}&=0,\\
    -\frac{\partial \pi'}{\partial y'}+\tilde{g}'_x\frac{\partial n_x}{\partial y'}+\tilde{g}'_y\frac{\partial n_y}{\partial y'}+\frac{\partial \tilde{t}'_{yx}}{\partial x'}+\frac{\partial \tilde{t}'_{yy}}{\partial y'}&=0,\\
    \frac{\partial u_2'}{\partial x'}+\frac{\partial v_2'}{\partial y'}&=0,
\end{align}
where, working in two dimensions, 
\begin{align}
    \tilde{g}'_i=&-\gamma_r N'_i-\gamma_t e'_{ik}n_k, \hspace{25pt} e'_{ij}=\frac{1}{2}\left(\frac{\partial u'_i}{\partial x'_j}+\frac{\partial u'_j}{\partial x'_i}\right), \\
    N'_i=&\;\dot{n}'_i-\omega'_{ik}n_k, \hspace{55pt} \omega'_{ij}=\frac{1}{2}\left(\frac{\partial u'_i}{\partial x'_j}-\frac{\partial u'_j}{\partial x'_i}\right),\hspace{25pt}\pi'=\;p_2'+W',\\
    W'=& \;\frac{1}{2}\bigg[K_1(\nabla' \cdot {\bf n})^2+K_3(({\bf n}\cdot \nabla') {\bf n})\cdot (({\bf n}\cdot \nabla') {\bf n})\bigg],\\
    \tilde{t}'_{ij}=&\;\alpha_1'n_kn_pe'_{kp}n_in_j+\alpha'_2N'_in_j+\alpha'_3 N'_jn_i+\alpha'_4e'_{ij}+\alpha'_5e'_{ik}n_kn_j+\alpha'_6e'_{jk}n_kn_i .
\end{align}
We use summation notation here as it is standard when working with the Ericksen-Leslie equations, with $i,j,k=1,2$, and $\dot{n}_i$ denotes the convective derivative of the $i$th component of ${\bf n}$. We further assume that the elastic constants are equal; that is $K=K_1=K_3$. These liquid crystal quantities are defined further in Table \ref{tab:lc_param}. 

\begin{table}[htbp]
\caption{Liquid crystal variables and parameters \cite{stewart2019static}. }
\label{tab:lc_param}%
\begin{tabular}{ll}
\toprule
Quantity  & Description\\
\midrule
${\bf n}= (\sin\theta,\cos\theta)$& director field\\
$\theta(x',y',t')$&angle the director angle makes with the $y$-axis\\
$p'_2$      & pressure in LL \\
$W'$      & bulk energy density \\
$\pi'=p'_2+W'$& modified pressure\\
$\tilde{g}'_i$& viscous dissipation \\
$e'_{ij}$ & rate of strain tensor\\
$N'_i$ & co-rotational time flux of the director \bf{n}\\
$\omega'_{ij}$ & vorticity tensor \\
$\tau^{V'}_{2ij}$& viscous stress tensor\\
$\tau^{E'}_{2ij}$& elastic stress tensor\\
$\tau_2'$ & stress tensor\\
$\alpha'_i,\,i=1,...,6$ & Leslie viscosities (Newtonian: $\mu'=\alpha'_4/2$, all other $\alpha'_i=0$)\\
$\gamma'_r=\alpha'_3-\alpha'_2$ & rotational/twist viscosity \\
$\gamma'_t=\alpha'_6-\alpha'_5$ & torsion coefficient \\
$K_1,\;K_3$      & elastic constants for splay and bend respectively    \\
\botrule
\end{tabular}
\end{table}

\subsection{Boundary conditions}
We now list the boundary conditions, proceeding from the cornea out towards the air. 
At $y'=0$, we have velocity continuity and a no flux boundary condition for osmolarity transport. These are, respectively:
\begin{align}
    {\bf u'_1} = {\bf J_0},\\
     c'\,({\bf u'}_1 - {\bf s'}_{0t'})-D_1 \nabla' c'=0.
\end{align}
Note that as we assume the cornea is flat, ${\bf s'}_{0t'}={\bf 0}$.

At $y'=h'_1$, we have velocity continuity, aqueous and lipid mass conservation, and the stress balance. In addition, we have a no flux boundary condition for osmolarity transport, and the anchoring boundary condition for the LL. The equations, respectively, are:
\begin{align}
    ({\bf u'}_1 - {\bf u'}_2)\cdot \hat{t}'_1 = 0,\\
    \rho_1 ({\bf u'}_1 - {\bf s'}_{1t'}) \cdot \hat{n}'_1 = J'_e,\\
    \rho_2 ({\bf s'}_{1t'}-{\bf u'}_2) \cdot \hat{n}'_1 = 0,\\
    ({\bf \tau'}_1 - {\bf \tau'}_2)\cdot \hat{n}'_1 = -\gamma'_{s1}\hat{n}'_1 \nabla' \cdot \hat{n}'_1 + \nabla'_{s_1} \gamma'_{s1},\\
    c'\,({\bf u'}_1 - {\bf s'}_{1t'})-D_1 \nabla' c'=0,\label{eq:ALLLnoflux}\\
    \theta=\theta_B.
\end{align}
To resolve the stress balance into tangential and normal stress components, we take the dot product of the stress balance and $\hat{t}'_1$ or $\hat{n}'_1$ respectively. Here $\nabla'_{s_1}=({\bf I}-\hat{n}'_1\hat{n}'_1)\cdot \nabla'$ is the surface gradient, where ${\bf I}$ is the identity matrix \cite{stone1990simple}. 

At $y'=h'$, we have lipid mass conservation, stress balance, and anchoring respectively:
\begin{align}
    \rho_2 ({\bf u'}_2 - {\bf s'}_{2t'})\cdot \hat{n}'_2=0,\\
    {\bf \tau'}_2 \cdot \hat{n}'_1=-\gamma_2 \hat{n}'_2 \nabla' \cdot \hat{n}'_2,\\
    \theta=\theta_B.
\end{align}
The stress balance can be resolved into components in the same way as above, using the unit tangent and normal vectors at this interface, $\hat{t}'_2$ and $\hat{n}'_2$.

At the aqueous-lipid interface $y'=h'_1$, we have surfactant transport and the linear equation of state for the surface tension, respectively:
\begin{align}
    \Gamma'_{t'}+\nabla'_{s_1}\cdot(\Gamma' u'_1) = D_s \nabla^{'2}_{s_1}\Gamma',\\
    \gamma'_{s1}=\gamma_1 - RT_0\Gamma'.\label{eq:eqofstate}
\end{align}

\subsection{Osmosis and evaporation}\label{sec:osm_evap}

The effect of osmosis is determined by the difference in concentration on either side of the aqueous-corneal interface $y'=0$,
\begin{align}
    {\bf J'}_0 \cdot \hat{n}'_0 &= P_c\,(c'-c_0).
\end{align}
Here we assume an ideal osmolarity ($c_0=300$ mOsM) on the corneal side of the interface.

To account for the effect of evaporation at the aqueous-lipid interface $y'=h'_1$, we modify a boundary layer model derived thoroughly in \citet{stapf2017duplex}
 \begin{align}
    J'_e&=\frac{E_0}{1+\frac{k_m}{Dk}h'_2}.
\end{align}
Here, $Dk$ represents resistance to evaporation provided by the LL based on its thickness and permeability. Its value comes from a nonlinear least squares fit to clinical data \cite{ king2010application} performed by \citet{stapf2017duplex}, which is very similar to that of \citet{cerretani2013water}. We further modify $Dk$ to include a dependence on the director angle of liquid crystals in the LL. We suggest the orientation of the molecules affects evaporation through 
\begin{align}
    J'_e&=\frac{E_0}{1+\frac{k_m}{Dk}(0.1+0.9\sin \theta_B)h'_2}.\label{eq:evap_function}
\end{align}
The maximum evaporation rate occurs when $\theta_B=0$, which is ten times that of the rate for $\theta_B=\pi/2$.  The minimum rate matches that used in \cite{stapf2017duplex}.

\subsection{Deriving the leading order system}
We nondimensionalize using the scalings in (\ref{eq:scalings}), along with 
\begin{align}
    W'=\frac{K}{\delta^2 L^2}W, \nonumber\hspace{25pt}\alpha_i'=\mu_2\alpha_i,\hspace{25pt} \gamma'_t=\mu_2\gamma_t,\hspace{25pt}\gamma'_r=\mu_2\gamma_r.
\end{align}
Then, we asymptotically expand all the dependent variables in powers of the small parameter $\epsilon \ll 1$. For example,
\[\Gamma(x,t) \sim \Gamma_0(x,t)+\e\, \Gamma_1(x,t)+\e^2 \Gamma_2(x,t)+\cdots\]
We substitute these expansions into the nondimensionalized equations and collect like powers of $\e$. At leading order, we are able to determine the director angle, conservation of mass in both the AL and LL, and conservation of surfactant concentration. At $O(\e)$, we are only able to determine intermediate variables, so we proceed to $O(\e^2)$. We are then able to determine the AL depth averaged velocity, AL pressure, the force balance equation for the axial velocity in the LL, and conservation of osmolarity. This gives us the closed system of equations from Section \ref{sec:governingeqs}. This progression is typical of problems like this \cite{stapf2017duplex, bruna2014influence}. 

\end{appendices}


\bibliography{sn-bibliography-rb}

\end{document}